\documentclass[fleqn,usenatbib,useAMS]{mnras}

\usepackage[T1]{fontenc}
\usepackage{ae,aecompl}
\usepackage{longtable}
\usepackage{graphicx}	
\usepackage{amsmath}	
\usepackage{amssymb}	
\usepackage{txfonts}

\title[Structure-dynamics relations]{Structure-dynamics relations for late-type spiral and dwarf irregular galaxies revisited}
\author[B.~R. Parodi]{Bernhard R. Parodi$^{1}$\thanks{E-mail:	bernhard.parodi@gibz.ch}
\\
$^{1}$Gewerblich-industrielles Bildungszentrum Zug (GIBZ), Baarerstr. 100, CH-6300 Zug, Switzerland}
\date{}
\pubyear{2018}

\begin{document}
	
\label{firstpage}
\pagerange{\pageref{firstpage}--\pageref{lastpage}}
\maketitle

\begin{abstract}
Scaling relations among structural and kinematical features of 79 late-type spiral and dwarf irregular galaxies of the SPARC sample are revisited or newly established. The mean central surface brightness $<$$\mu_{0,[3.6]}$$>$$= 19.63\pm0.11\, \mathrm{mag\,arcsec^{-2}}$ allows for a clear-cut distinction between low and high surface brightness galaxies. At a given luminosity, LSB galaxies are more extended than HSB galaxies and the rotation curves have smaller inner circular velocity gradients $dv(R_d)/dr$ at one disk scale length $R_d$. Irrespective of luminosity, the geometry of rotation curves is characterized by the relation  $dv(R_d)/dr \approx v_\mathrm{max}/R_\mathrm{max}$, with $v_\mathrm{max}$ being the maximum circular velocity reached at $R_\mathrm{max}$. For the rotation curve decompositions disk mass-to-light ratios are restricted to have constant, but semi-free best-fit values 0.2, 0.5, or 0.8 $M_\odot/L_\odot$ at [3.6]; they exhibit an asymmetric bimodal distribution with the dominant peak located at the median value of 0.2 (minimum disks) and with the subdominant peak at 0.8 (maximum disks).  Assuming dark matter halos of Burkert and of pseudo-isothermal (PITS) type, the former provide better fits for about two thirds of all galaxies.  While the halo core densities $\rho_0$ are about equal, the core radii $r_0$ of PITS halos are systematically lower by a factor of about 0.6 as compared with those of the Burkert type. Focussing on the Burkert halo, the baryonic mass fraction at intermediate radii is included to address both an adjusted baryonic Tully-Fisher relation and the significance of deviations from the mean radial acceleration relation. The average radial decrease of the baryonic mass fraction within galaxies is quantified. The Burkert halo parameters obey $\rho_0$$\,\propto\,$$r_0^{-1.5\pm 0.1}$ with considerable scatter, but allowing for $v_\mathrm{max}$ as a third variable we find $\rho_0$$\,\propto\,$$r_0^{-1.84\pm0.07} v_\mathrm{max}^{2.00\pm 0.11}$ with small scatter. The halo central surface density $\rho_0 r_0$, with a sample median $<$$\rho_0 r_0$$>$$\,\approx 121\, \mathrm{M_\odot pc^{-2}}$ ($\sigma = 112$), weakly correlates with $\mu_0$ and with compactness C and strongly correlates with the observed radial acceleration $g_{\mathrm{obs}}=v^2_{\mathrm{obs}}(r)/r$ at different galactocentric radii. Consequently, because $R_\mathrm{max} \propto r_0$, we have a tight central halo column density versus maximum circular velocity relation $v_\mathrm{max}^2 \propto \rho_0 r_0^{2-\epsilon}$. Halo cores barely extend over the luminous disk, but their sizes do not correlate with the optical radii. We introduce an alternative to a prominent conventional universal rotation curve (URC); it is based on the non-singular total matter density profile $\rho_\mathrm{total}(r)= (v_\mathrm{max}^2/4\pi G r^2)\left( 1 - (1-r/r_c) \exp(-r/r_c) \right)^2$, with the scaling parameter $r_c$ correlating with the halo core size $r_0$. Fitting our synthetic URC to a selection of galaxies, the co-added doubly-normalized rotation curves exhibit a high degree of similarity. A couple of analytic URC decompositions into a baryonic disk and a dark matter component is accomplished. 
\end{abstract}
\begin{keywords}
	Galaxies: spiral -- Galaxies: irregular -- Galaxies: dwarf --  Galaxies: halos -- Galaxies: fundamental parameters
\end{keywords}
\section{Introduction}

Late-type spiral and dwarf irregular galaxies are composed of, firstly, bulgeless and non-barred gaseous and luminous stellar disks, where accretion, star formation, and stellar evolution takes place. Within classical Newtonian physics, disks are recognized to be prominently surrounded by, secondly, extended nonluminous accumulations of invisible matter, called dark matter (DM) halos. The presence of DM is inferred only indirectly due to a luminous mass (LM)-deficiency or mass-degeneracy.  Historically, the first dynamical evidence for Galactic non-luminous matter goes back to Lord Kelvin who applied the kinetic theory of gases to stellar systems and in particular to a handfull of galactic-plane stars with known velocity dispersion \citep{Thomson04}. In extragalactic astronomy, the mass-degeneracy was first put on firm quantitative grounds by \citet{Zwicky33} on applying the virial theorem to the Coma cluster of galaxies, realizing an indisputable lack of luminous matter and coining the term "dark matter". For disk galaxies of all corresponding Hubble types, DM is typically inferred since the late 1950ies from the amplitude and shape of their rotation curves (RC) that disagree with Keplerian motion, in particular, the flatness observed way beyond the optical radius of the luminous disk indicates a hidden mystery. The first accurate RC was determined for M31 in the radio and goes back to \citet{vandeHulst57} and \citet{Schmidt57}. For a historical review see \citet{Bertone16} and references therein. 

One path to an understanding of the nature of DM leads over finding model relations between observed or infered features of luminous and dark matter. Luminous matter particularly in late-type spiral and dwarf irregular disk galaxies typically goes with exponential surface brightness profiles at intermediate radii (with central surface brightness and disc scale length as structural parameters) and for the dynamics rotating disks close to centrifugal equilibrium are adopted \citep{VanderKruit11}. The corresponding disk mass models predict at most radii circular velocities way below to what is observed, even if maximum disks with high mass-to-light ratios are assumed. The excess of the observed total circular velocity may be attributed to non-detected or dark matter \citep[e.g.,][]{Freeman70, Rubin70,Rubin83, vanAlbada85}. Parts of these halos are composed not of some enigmatic type of matter, but simply are baryonic due to the speedy outflow of stellar debris due to supernovae, an ongoing process within and around each disk galaxy since the early times of its formation. There is a vast literature on suggested parametrizations for total DM halo density profiles \citep[e.g.,][for double power laws and the Einasto profile]{An13}. For late-type spiral and low-luminous disk galaxies the rotation curves are particulary well described by corresponding halo density profiles with, on one hand, non-singular and non-steep ("constant") density cores, as with the \citet{Burkert95} profile, the PSS profile \citep{Persic96, Borriello01}, or pseudo-isothermal spherical (PITS) profiles \citep{Kormendy04, Kormendy16}. On the other hand, simulations of galaxies within a lambda cold dark matter ($\Lambda$CDM) universe with hierarchical clustering predict cuspier, steep core density profiles that even become singular at the galaxy center, as with the widely used three-parameter double-power law by \citet{Zhao96}. This model includes the isothermal profile (ISO), the Navarro-Frenk-White \citep[NFW,][]{Navarro96} profile and its generalization \citep[gNFW, e.g.,][]{Moore99}, and the DC14 profile by \citet{DiCintio14b}. The term "constant core" is owing to $\log \rho(r)-\log r$-diagrams wherein singular profiles show everywhere a falling straight line while cored non-singular profiles exhibit a nearly flat part in the central region. Overall, there is a prevailing degeneracy with respect to halo density profiles. A review on the cusp-core-controversy is given in \citet{deBlok10}. Comparing cored halo models,  the results do not allow to conclusively select a single superior one, \citet{Breddels13} even suspect that more than one type of parametric halo profile is necessary for realistic assessments. To give an idea, while \citet{Adams14} compares fits to the RCs of dwarf galaxies using Burkert and gNFW profiles and finds the latter to slightly perform better, \citet{Pace16} finds the PITS halo and the DC14 profiles to perform equally well and to outperform the Burkert and the NFW profiles. The result of \citet{DiCintio14b} is motivated by hydrodynamical N-body simulations and includes as special cases both the isothermal and the gNFW halos, favored within the $\Lambda$CDM scenario. These hydrodynamical numerical approaches as well as semi-analytic models manage to transform cusped DM halos into cored halos by means of galaxy evolution with supernovae feedback or due to interaction of baryon clumps with dark matter as secular dynamical processes \citep[e.g.,][]{Governato10, Oh11b, DiCintio14a, Katz17, DelPopolo18}. For the sake of analytical simplicity, halo models used for RC decompositions usually assume spherical symmetry, but most N-body simulations agree in their prediction that evolving halos in a CDM universe with hierarchical clustering are more accurately described by triaxial systems \citep[e.g.,][]{Bryan13, Despali14}. For dwarf galaxies, however, the asphericity seems to be very moderate \citep{Trachternach09}.

According to their star-formation history, late-type spirals and dwarf irregulars are capable to reproduce their stellar mass over the cosmic time, with bursty episodes \citep[e.g.,][]{Karachentsev18}. Star-forming regions within rotationally supported low-mass disk galaxies are systematically found out to the outer disk until more than two and a half optical radii (corresponding to about eight exponential disk scale lengths and about four times the size of typical halo cores) \citep{Parodi03, Hunter16}. This is more than twice as far as the onset of the flat part of typical RCs. Stellar truncation is due to reaching a critical star formation threshold density \citep{vandenBosch01b}. Disk supported galaxies of all luminosities are subject to the baryon-halo conspiracy: they obey both the \citet{Tully77} relation (TFR), linking the maximum circular velocity with disk luminosity, and the baryonic Tully-Fisher relation (BTFR), linking the total baryonic mass with the observed rotational dynamics. These structure-dynamics relations are indirectly verified by the recently established mass-deficiency  versus radial acceleration relation (MDAR) or equivalenty by the two-radial accelerations relation (RAR) of rotationally supported galaxies including dwarf disc galaxies \citep{McGaugh14a,McGaugh16}. 

Typically the observed rotation curves (RC) exhibit a linear or nearly linear increase at small galactocentric radii before they curve at intermediate radii to become flat or, in some cases, start to moderately decrease after reaching a maximum circular velocity. Further out, any RC will eventually drop. Irrespective of the maximum circular velocity the central velocity gradients may vary from galaxy to galaxy, from steep to rather flat. In general, there's a variety of possible amplitudes, from slow to fast, and of possible shapes of RCs, from narrow to extended curvatures. Despite of the diversity of observed RCs \citep{Oman15}, based on cored halo models the increasingly successful composition of universal rotation curves (URC), in particular for dwarf disc galaxies \citep{Karukes17, DiPaolo18, Lapi18}, continuously adds to the confidence concerning a deeper understanding of DM halos. 

A panopticum of other scaling relations among various variables describing structural and kinematical properties of disk galaxies are known and kept being refined, such as LM parameter correlations \citep{Giovanelli99,Courteau07}, DM correlations among halo parameters (e.g., the supposed uniformity of halo central surface density), or DM-LM relations \citep[e.g.,][and references therein]{Kormendy04, Kormendy16, Lapi18}.

In this paper we address a few known and investigate some new structural and kinematical scaling relations for a sample of 79 late-type spiral and dwarf irreguar galaxies, based on a homogenous data set. The photometry and preprocessed kinematic data are taken from the SPARC database and consistently postprocessed by us with respect to halo mass model decompositions. We proceed as follows: in Sect. 2 we present the data (Sect. 2.1), include an absolute magnitude independent distinction of high-surface and low-surface brightness at 3.6$\mu$m (Sect. 2.2), and describe the decomposition procedure applied by us to the galaxy RCs in order to obtain Burkert and PITS halo parameters (Sect. 2.3). We thereby present a peculiarity concerning the distribution of best-fit mass-to-light ratios. Section 3 presents and discusses the investigated scaling relations, which are the RAR (Sect. 3.1), the halo core density versus size relation subject to a third variable (thereby providing an original new result, Sect. 3.2), some kinematic dependencies of the central halo surface density (and refuting its supposed uniformity, Sect. 3.3), an inner circular velocity gradient (based on a new simple derivation, Sect. 3.4), an adjusted BTFR for varying radii (Sect. 3.5), the central halo column density versus maximum velocity relation (Sect. 3.6), and finally the URC in a conventional as well as in an alternative new form, the latter being based on a genuinely proposed total matter density profile (Sect. 3.7). Section 4 summarizes the results and takes a glance at further topics. Finally, the appendix contains a couple of preliminary formal decompositions of the alternative URC (App. A) as well as three tables with selected photometric, kinematic, and halo structural data (App. B).

\section{Structural and kinematical data}

\subsection{SPARC data}
The \emph{Spitzer} Photometry and Accurate Rotation Curves (SPARC) database \citep{Lelli16a} can be accessed at \href{http://astroweb.cwru.edu/SPARC/}{astroweb.cwru.edu/SPARC/}. Among the 175 galaxies of all types we select a subsample of 79 late-type spiral and dwarf irregular galaxies, i.e., those with morphologies Sdm, Sm, Im, and BCD corresponding to numerical Hubble types $T\ge8$, all of which have inclination-corrected exponential disk surface photometry at 3.6 $\mu$m and discrete rotation curve (RC) samplings with at least four data points. Two further galaxies, namely CamB and PGC1017, were omitted a posteriori because of extremly exceptional results or peculiar kinematics. We otherwise neither exclude face-on or edge-on galaxies nor galaxies with low quality HI data (i.e., those with a quality flag value $Q$=3). Tables \ref{TableB1} to \ref{TableB3} (relegated to the Appendix) list selected SPARC data (original as well as partially processed by us) and our results concerning the RC decomposition, together with some additional quantities of relevance. In particular, the columns of Table \ref{TableB1} (Selected photometry and luminous structure) are as follows:\newline 
(1)-(2) galaxy name and an alternative identifier; \newline
(3) distance (value given in SPARC database);\newline
(4) absolute magnitude, calculated from the SPARC luminosity $L/L_\odot$ by means of $M[3.6]=-2.5\log_{10}(L/L_\odot)+M_\odot[3.6]$ (with $M_\odot[3.6]$$=$$3.24$ mag as adopted from the SPARC database refering to \citet{Oh15}); \newline
(5)$\,\,$extrapolated central surface brightness, calculated from the correspon\-ding incli\-nation-corrected SPARC central surface brightness \emph{SB} (in $L_\odot \mathrm{pc}^{-2}$) by means of $\mu_0$$=$$-2.5\log_{10}(SB)$ $+21.572+M_\odot[3.6]$;\newline
(6) exponential disk scale length (SPARC value, measured at outer radii); \newline
(7) compactness parameter value (calculated according to equ. \ref{C});\newline

The columns of Table \ref{TableB2} (Dark matter halo structural parameters) are as follows:\newline
(1) galaxy name;\newline
(2)-(3) Burkert halo central density and scale length ("core radius");\newline
(4)-(5) semi-free mass-to-light ratio and (pseudo-)reduced $\chi_n^2$.\newline 
(6)-(7) PITS halo central density and scale length;\newline
(8)-(9) semi-free mass-to-light ratio and (pseudo-)reduced $\chi_n^2$.\newline 

Finally, the columns of Table \ref{TableB3} (Kinematics according to the best-fit RC applying the Burkert halo) are as follows: \newline
(1) galaxy name; \newline 
(2) velocity gradient at the radius $R_d$, calculated by means of equ. (\ref{velograd}); \newline
(3)	values for the gas, stellar disk ($v_s^\Upsilon\equiv v_s\sqrt{\Upsilon_{[3.6]}}$), and halo velocity components at the idealized stellar disk-peak radius 2.15 $R_d$ (the values are linearly interpolated between two neighboring observed data points provided by the SPARC database); \newline
(4)-(5) total calculated and observed velocity components, respectively, at 2.15 $R_d$; \newline
(6) same as (3), but at the so-called optical radius 3.2 $R_d$; \newline
(7)-(8) same as (4)-(5), but at the optical radius 3.2 $R_d$;\newline
(9)-(10) radius and corresponding observed velocity, respectively, where the flat (or declining) regime of the RC begins (the data points are selected \emph{ad hoc} from the SPARC database).

\subsection{Loci of constant central surface brightness or compactness}

\begin{figure*}
	\centering 
	\includegraphics[width=0.45\textwidth]{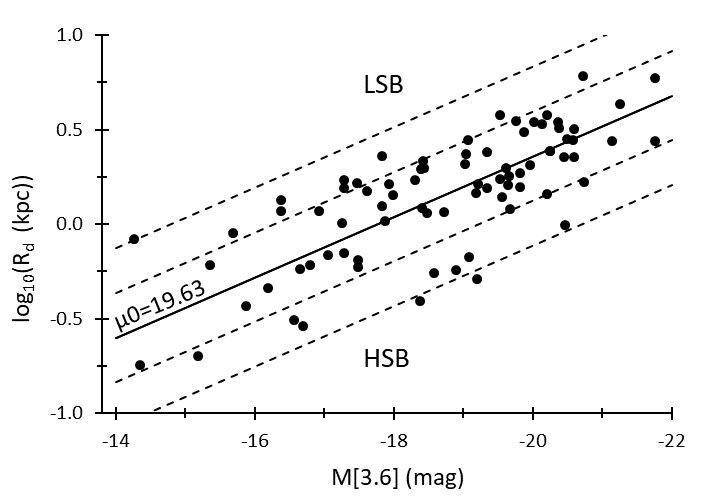}
	\includegraphics[width=0.45\textwidth]{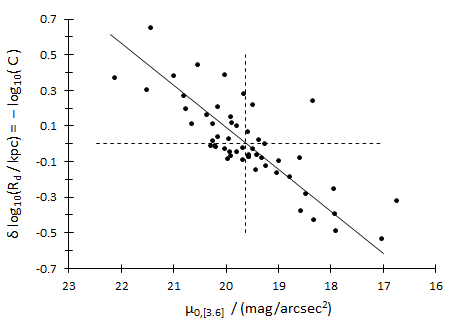}
	\includegraphics[width=0.45\textwidth]{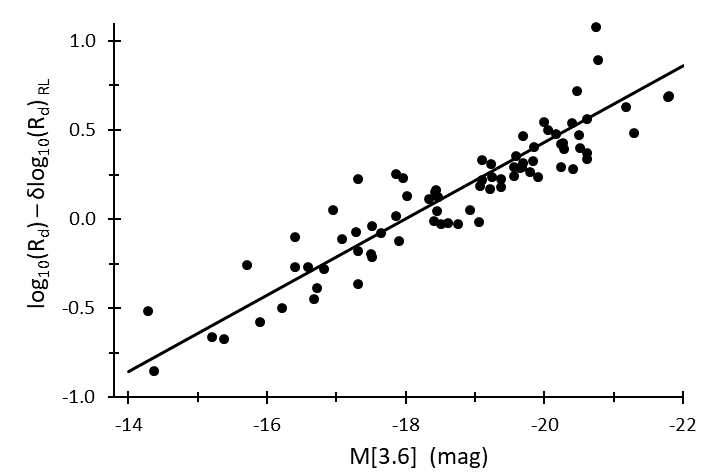}
	\includegraphics[width=0.45\textwidth]{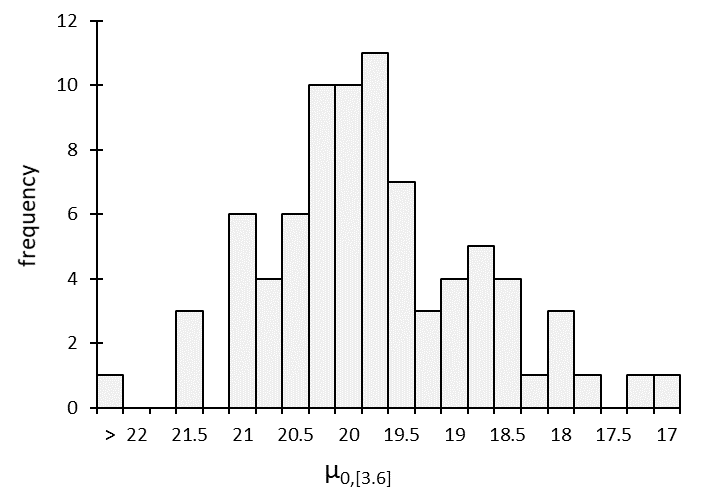}
	\caption{\small \emph{Top left:} $\log (R_d)-M$-diagram (in the 3.6-micron bandpass). The solid line is determined by means of a least-squares bisector fit to the data points. It has a slope of -0.2 (equation \ref{logRd}) and represents the locus corrresponding to the mean extrapolated central surface brightness $<$$\mu_0$$>$$ = 19.63\, \mathrm{mag/arcsec^2}$. Dashed lines are further loci of constant central surface brightness with selected values $\mu_0=21.63, 20.63, 18.63, 17.63$ mag/arcsec$^2$ (from top to bottom). 	
	\emph{Top right:} Residuals $\delta\log (R_d)=\log (R_d)\,-$$<$$\log (R_d)$$>$ as a function of extrapolated central surface brightness. We note the equivalence of these residuals with reciprocal compactness, i.e., $\delta\log (R_d)=-\log(C)=\log(1/C)$. The vertical dashed line is positioned at the sample mean $<$$\mu_0$$>$. The solid line is given by equation (\ref{dlogRd}). \emph{Bottom left:} The $\log (R_d)-M$-diagram after correction for mean residuals $<$$\delta\log (R_d)$$>$ has a considerably reduced scatter as compared to the one in the upper left figure. \emph{Bottom right:} Histogram-distribution of all extrapolated central surface brightnesses, exhibiting a peak around the mean value of $<$$\mu_{0,[3.6]}$$>$$ = 19.63\, \mathrm{mag/arcsec^2}$. \normalsize}
	\label{Fig1}
\end{figure*}

We start with recalling some basic photometric and structural model relations. Late-type disk galaxies typically have neither bulges nor bars and exhibit over a large portion of their luminous radial extent exponential intensity profiles $I(r)=I_0 \exp(-r/R_d)$, where $R_d$ is the disk scale length and $I_0$ is the central intensity. (Actually, a more concise profile that better relates to the fundamental plane (FP) of dwarf irregular galaxies is provided by the hyperbolic secant (sech) function by means of $I(r)=I_0\, \mathrm{sech}(r/R_d)$ \citep{Vaduvescu08} that is, however, not yet established; an empirical conversion among the photometric parameters of exp- and sech-model fits is available \citep{Janowiecki14}). Assuming axisymmetry and corrections for inclination and absorption included (and possibly cosmological corrections as well) central intensity translates into the (extrapolated) central surface brightness $\mu_0=-2.5\log I_0 + \mathrm{const}$. The total disk luminosity is given by $L=2\pi\int_{0}^{\infty}\,r\,I(r)\,dr=2\pi\,I_0\,R_d^2$. This either translates into some total disk mass $M_d = \Upsilon \,L$ (assuming some appropriate constant mass-to-light ratio $\Upsilon$) or into an absolute magnitude $M = -2.5\log(L)+\mathrm{const}$. This latter relation can now be expressed as a linear function in $M$, 
\begin{equation} 
\log(R_d)=-0.2 M+0.2\mu_0+\mathrm{const},   \label{M}\\
\end{equation}
with the intercept being dependent on the central surface brighness $\mu_0$. This will introduce considerable scatter in any $\log R_d\,$--$M-$diagram. For our sample of exponential disk galaxies we thus expect a linear regression line $\log(R_d)_\mathrm{RL}$ with slope -0.2 and with a scatter dominated by the central surface brightness, as illustrated in Fig. 1, upper left-hand panel. Observed vertical deviations $\delta\log(R_d)=\log(R_d)\,\, - \log(R_d)_\mathrm{RL}$ from the mean linear trend are therefore partially explained by the individual values of $\mu_0$. This indeed is observed, as shown in Fig. 1 (upper right-hand panel).  Formally, for the linear regressions we performed ordinary-least squares bisector (OLSB) fits \citep{Isobe90} in order to account for uncertainties in both coordinates; the best fits are found to be 
\begin{eqnarray} 
\log(R_d)_\mathrm{RL}&=& (-0.198\pm0.014)\,\,M_{[3.6]}-3.592 \label{logRd}\\
\delta\log (R_d)_\mathrm{RL}&=& (0.236\pm0.017)\,\,\mu_{0,[3.6]}-4.629. \label{dlogRd}
\end{eqnarray}
The first of these equations defines the mean optical scale length at some absolute magnitude, while the second gives the deviation due to the particular central surface brightness. These equations agree fairly well with the theoretical expectation as summarized by equation (\ref{M}). Correcting $\log(R_d)$ by means of the mean deviations $\delta\log (R_d)_\mathrm{RL}$ results in the $\log R_d\,$--$M-$diagram shown in the lower left-hand panel of Fig. \ref{Fig1}. The remaining scatter may be due to uncertaincies in profile fitting, inclination correction, internal/external absorption corrections, or be intrinsic (e.g., related to the halo structure). Stated differently yet, given the same value $\mu_0$ for several galaxies they are expected to lay on a mutual locus with slope -0.2 (see the dashed lines in Fig. 1, upper left panel). Hence, the diagrams shown in the upper panels of Fig. 1 allow for a distinction between low surface brightness (LSB) and high surface brightness (HSB) galaxies \emph{irrespective} of absolute magnitude.  Some intermediate surface brightness (ISB) galaxies within an interval centered at the mean SB may naturally be considered as well \citep[as done in][]{McGaugh96}, but are omitted here for brevity. Actually, for any given absolute magnitude, the mean extrapolated central surface brightness at 3.6 $\mu$m is 
\begin{equation}
<\mu_{0,[3.6]}>\,= 19.63\pm 0.11\,\,\mathrm{mag\,arcsec}^{-2},
\,\,\,\,\,\,\,\,\sigma = 1.01
\end{equation}
(corresponding to $\delta\log (R_d)$=0 in equation (\ref{dlogRd}) and corresponding to the mean of the distribution shown in the lower right-hand panel of Fig. 1). We note that the spread around the peak is rather large, one thus obviously neither can say that $\mu_{0,[3.6]}$ is approximately constant for all galaxies \citep{Freeman70} nor observe some bimodality \citep{Tully97, McDonald09}. The above \emph{mean} value is thus the watered-down form of Freeman's law at 3.6 $\mu$m. Furthermore and as is well established for other bandpasses \citep[e.g.,][]{Zwaan95}, for any \emph{given absolute magnitude} (or luminosity), larger-than-mean disk scale lengths correspond to LSB galaxies ($\mu_{0,[3.6]}$$>$$ 19.63\, \mathrm{mag\,arcsec^{-2}}$), while smaller-than-average scale lengths correspond to HSB galaxies ($\mu_{0,[3.6]}$$<$$19.63\, \mathrm{mag\,arcsec^{-2}}$). Therefore, to avoid confusion, one always should be aware that "low luminous" does not necessarily imply "LSB", and "luminous" is not necessarily related to "HSB".
\begin{figure*} 
	\centering
	\includegraphics[width=0.45\textwidth]{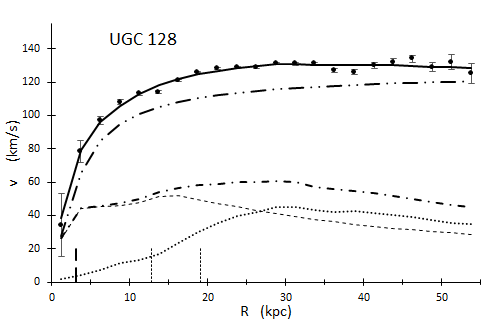}	
	\includegraphics[width=0.45\textwidth]{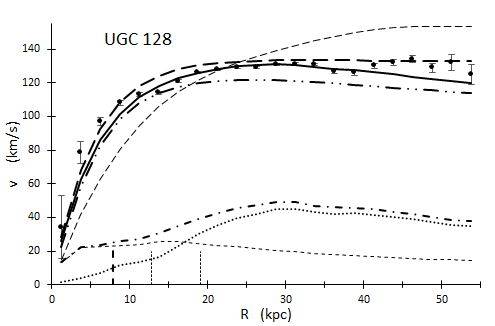}	
	\includegraphics[width=0.45\textwidth]{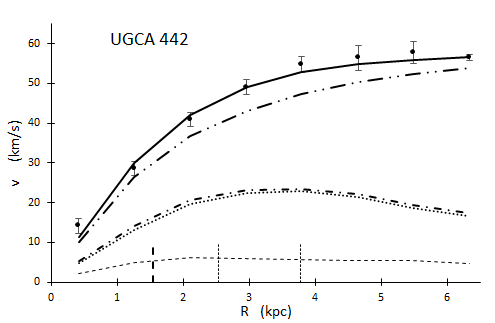}	
	\includegraphics[width=0.45\textwidth]{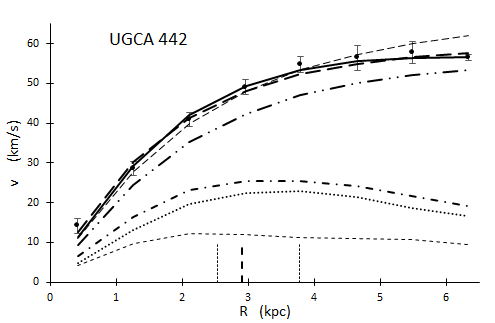}		
	\caption{\small Examples of decomposed rotation curves (RC):  points with error bars represent the observed RCs of UGC 128 (upper panels) and UGCA 442 (lower panels), applying the PITS halo model (left panels) and the Burkert halo model (right panels). UGC 128 is brighter (absolute magnitude $M_{[3.6]}=-22.0$ mag) than UGCA 442 ($M_{[3.6]}=-17.1$ mag) and correspondingly reaches a faster maximum circular velocity. The gaseous (thin dotted lines) and the stellar contribution (thin dashed lines) add up to form the baryonic component (dot-dashed lines) that together with the halo contribution (dot-dot-dashed lines) sum up to the total best-fit model RC (solid lines). Immediately above the $R$-axis three short vertical lines indicate $r_0$ (fat long-dashed) as well as 2.2 $R_d$ and 3.2 $R_d$ (thin short-dashed); with respect to the disk scale lengths $R_d$, the two galaxies have relatively small halo core sizes. UGC 128 is markedly better fit by means of a PITS halo ($\chi^2_n=2.88$, $M/L = 0.8$) than with a Burkert halo ($\chi^2_n=6.88$, $M/L=0.2$). Contrarily, UGCA 442 is slightly better fit with the Burkert halo ($\chi^2_n=0.46$, $M/L=0.8$) than with the PITS halo ($\chi^2_n=0.60$, $M/L=0.2$). In addition, in each of the right panels two universal rotation curves (URC) are superposed: that of \citet{Karukes17} (assuming $M/L=0.5$ and calculated by means of equation \ref{URCKS17}, thin dashed lines) as well as that proposed in this paper (according to equation \ref{URCv}, thick long-dashed lines). \normalsize} 
	\label{Fig2}
\end{figure*}
Stated differently, for a \emph{given size} (measured as a multiple of $R_d$), HSB galaxies are brighter in absolute magnitude than LSB galaxies, and vice versa. The lower left corner in Fig. \ref{Fig1} is populated by low-luminous galaxies with relatively smaller extensions, the so-called \emph{dwarf galaxies}. These may nevertheless be splitted into HSB or LSB galaxies according to the clear-cut criterium given above (and that is included in Table \ref{Table1}). Similar diagrams for spiral galaxies data in other bandpasses are shown ---with explicit visualization of this luminosity independent high-versus-low surface brightness dependence--- in, e.g., \citet{Bergvall99} ($<$$\mu_{0,B}$$>$$\approx 23.0\, \mathrm{mag\,arcsec^{-2}}$ with a focus on LSB galaxies), \citet{Zhong08} ($<$$\mu_{0,B}$$>$$\approx 22.0\, \mathrm{mag\,arcsec^{-2}}$, i.e., in the vicinity of the original Freeman law value 21.65), \citet{Janowiecki14} ($<$$\mu_{0,B}$$>$$\approx 20.7\, \mathrm{mag\,arcsec^{-2}}$ for BCD galaxies),  \citet{Fathi10} ($<$$\mu_{0,R}$$>$$\approx 20.2\, \mathrm{mag\,arcsec^{-2}}$), or \citet{Courteau07} and \citet{Dutton07} ($<$$\mu_{0,I}$$>$$\approx 19.0\, \mathrm{mag\,arcsec^{-2}}$, for dominantly brighter galaxies). At all bandpasses, there's a characteristic global mean extrapolated central SB irrespective of absolute magnitude;  and there seemingly is some trend for this mean surface brightness to become brighter towards longer wave lengths. 

Speaking of LSB and HSB galaxies as galaxies with scale lengths longer or shorter as compared to the mean scale length at a given luminosity (or at a given baryonic mass if some mass-to-light ratio is additionally given) motivates the definition of some relative-size or compactness parameter (e.g., recently, \citet{Karukes17} or \citet{DiPaolo18}) like  
\begin{equation}
C = \frac{<R_d>}{R_d}\approx\frac{10^{\log(R_d)_\mathrm{RL}}}{R_d}, \label{C}
\end{equation}
with $\log(R_d)_\mathrm{RL}$ as given by eq. (\ref{logRd}) being a function of absolute magnitude. An analogous but different definition with $\log(R_d)_\mathrm{RL}$ being a function of the luminous disk mass is applied by Salucci and coworkers. With our definition one actually has the identity $\log(C)$= $\log(R_d)_\mathrm{RL}-\log(R_d)= -\delta\log(R_d)$ (Fig. 1, top right). Values $C<1$ and $C>1$ correspond to LSB and HSB galaxies, respectively.  Hence, compactness is included in Table \ref{Table1}, too. Using equation (2), the values for the present sample range from 0.22 to 3.49 (Table B.1), and, consistent with the argument above, they (weakly) correlate with central surface brightness. 

Summing up, a crude clear-cut distinction between LSB and HSB galaxies (ignoring albeit some category of intermediate surface brightness galaxies) is accomplished either by directly comparing the extrapolated central surface brightness of a galaxy with the global mean value, i.e. $\mu_0/$$<$$\mu_0$$>$, or indirectly by refering to the compactness parameter $C=$$<$$R_d$$>$$/R_d$ (as an alternative to the residuals $\delta\log(R_d)$). Both these descriptions formally lead to theoretical loci of either constant central surface brightness or constant compactness. The relationsships introduced so far are included in Table \ref{Table1}, that gives a brief overview on some properties distinguishing (or not distinguishing) LSB and HSB galaxies from each other.

\subsection{Rotation curve decomposition} 

\subsubsection{Mass models}

For the RC decomposition of the observed velocity $v_\mathrm{obs}$, we assume at each radius the usual contributions,  i.e., some gaseous, stellar, and halo contribution, adding up to 
\begin{equation}
v_\mathrm{obs}^2=v_\mathrm{halo}^2 + \Upsilon_{[3.6]} v_\mathrm{stellar}^2 + v_\mathrm{gas}^2, \label{vobs}
\end{equation}
where $\Upsilon_{3.6}$ is the stellar mass-to-light ratio at 3.6 $\mu$m, assumed to be constant throughout a galaxy. Neither bulges nor bars are taken into account. The contributions due to gas and the stellar disk are part of the SPARC data: the stellar contribution was processed assuming a self-gravitating and rotationally supported exponential disk \citep{Freeman70}, calculated for a mass-to-light ratio of 1. Adopting a different mass-to-light ratio is therefore achieved by simple scaling. 
	
The dark matter halo is assumed to be spherically symmetric and is modelled twofold, by means of a pseudo-isothermal sphere (PITS, $\eta=0$) and by means of the Burkert (1995) halo ($\eta=1$), where $\eta$ is the parameter entering the hybrid Burkert-PITS (BP) density profile 
\begin{equation}
\rho_\mathrm{halo}(r) = \frac{\rho_0}{\left(1+\frac{r}{r_0}\right)^\eta\left(1+(\frac{r}{r_0})^2\right)}.\label{halodens} 
\end{equation}

We note that these models can be considered as special cases of a modified, cored \citet{Zhao96} profile with \emph{four} parameters in the form $\rho(r)=\rho_0 \,[(1+r/r_0)^{\gamma+\eta}(1+(r/r_0)^\alpha)^{(\beta-\gamma)/\alpha}]^{-1}$: the ($\alpha,\beta,\gamma,\eta$)-tuplet for the Burkert-PITS halo density parameters is (2,2,0,$\eta$), with according representations (2,2,0,0) for the PITS and (2,2,0,1) for the Burkert halo. The PITS and the Burkert halo models have mass distributions $M(<r)=4\pi\int_0^r r'^2\rho(r')dr'$ given by
\begin{equation}
M_\mathrm{halo}^\mathrm{PITS}(\le r) = 4\pi\rho_0r_0^3 \left[ \frac{r}{r_0} - \arctan\left(\frac{r}{r_0}\right) \right]\label{MhaloPITS}
\end{equation}
\begin{equation}
M_\mathrm{halo}^\mathrm{Burk}(\le r) = 2\pi\rho_0r_0^3 \left[ \ln\left( \left(1+\frac{r}{r_0}\right)\sqrt{1+\left(\frac{r}{r_0}\right)^2} \,\right) - \arctan\left(\frac{r}{r_0}\right) \right],\label{MhaloBurk}
\end{equation} 
respectively. The halo velocity contributions are calculated by means of $v_\mathrm{halo}^2(r)= G\,M_\mathrm{halo}(\le r)/r$. Because for large radii $r$$\gg$$r_0$ the $\arctan$-term is about constant, the former model has a constant asymptotic velocity $v_\infty^\mathrm{PITS}=(4\pi G\rho_0r_0^2)^{0.5}$, while the latter model reaches a peak and afterwards slowly decreases according to $v_{r \gg r_0}^\mathrm{Burk} \approx (4\pi G\rho_0r_0^3 \ln(r/r_0)/r)^{0.5}$. Choosing some typical value for $\rho_0$ and plotting the halo velocity contributions of the two models in the same $v(r)$-$r$-diagram reveals that for small galactocentric radii (out to about half a core size) the two curves look rather similar and hence must have similar velocity gradients (see Fig. \ref{Fig4}, bottom panels). At larger radii the PITS halo causes higher velocities than the Burkert halo. However, as illustrated in the figure and valid within a few core radii, rescaling the PITS model using successively smaller core radii creates intermittent rotation curves and finally provides an approximation for the RC of the Burkert halo; and vice versa. Moreover, scaling a PITS model (i.e., with $\eta =0$) by means of a decreasing core size $r_0$ obviously corresponds to increasing the value of $\eta >0$ in the hybrid profile provided by equation (\ref{halodens}). Whenever the PITS model is favored over the Burkert model, as discussed below and as can be inferred from Table B.2, its core size is indeed found to be smaller than that of the Burkert halo.  In conclusion, the model represented by equation (\ref{halodens}) seems to be more viable than the Burkert halo model or the PITS model alone. However, for values 0$<$$\eta$$<$1 the analytic calculation of the halo mass $M(r)$ is demanding and beyond the scope of this paper. For the sake of simple tractability, we restrain in the following to model values $\eta=0$ and $\eta=1$ and keep in mind the approximate scaleability as highlighted above. 

In order to discriminate a favored model for each galaxy we proceed as follows. For each rotation curve (RC) we numerically look for the best-fit with respect to the halo parameters $\rho_0$ and $r_0$ and as quantified by means of the (pseudo-)reduced $\chi^2_n$ value, i.e.,
\begin{equation}
\chi^2_n=\frac{1}{n}\sum_i \left( \frac{v_\mathrm{obs,i}-(v_\mathrm{halo,i}^2 +\Upsilon_{[3.6]} v_\mathrm{stellar,i}^2 +v_\mathrm{gas,i}^2 )^{0.5}}{\delta v_i}\right)^2,
\end{equation} 
where $n$ is the number of data points ($i=1,2,...,n$) and $\delta v_i$ are the attributed observational errors at discrete radii $r_i$ (as provied by the SPARC database). 
\begin{figure*}
	\centering
	\includegraphics[width=0.32\textwidth]{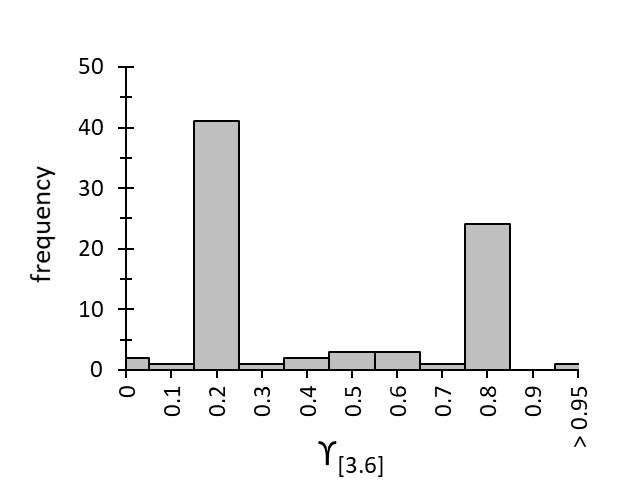}	\includegraphics[width=0.32\textwidth]{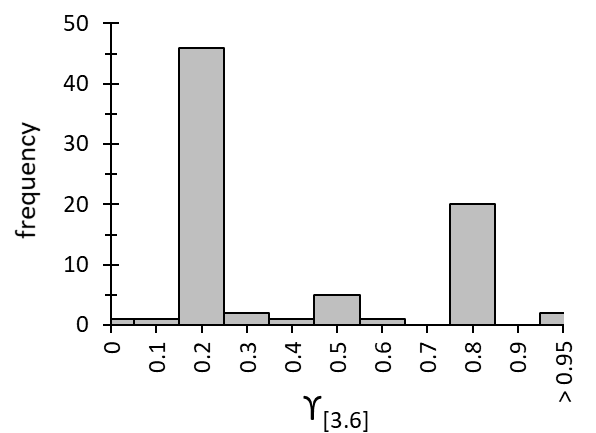}	
	\includegraphics[width=0.32\textwidth]{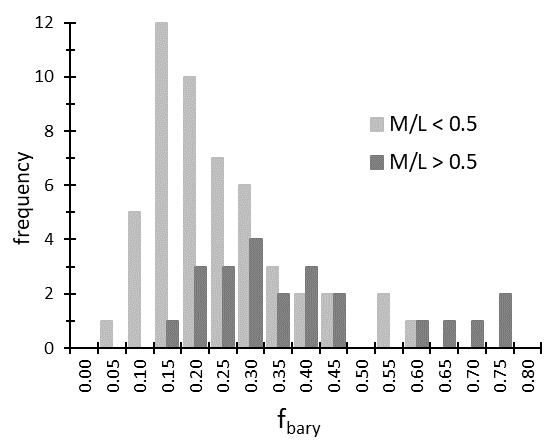}			
	\caption{\small Asymmetric bimodal distributions of the semi-free best-fit values for the stellar mass-to-light ratio $\Upsilon_{[3.6]}$, resulting from RC decompositions using either the PITS halo model (left-hand panel) or the Burkert halo model (middle panel). For the latter the frequency distribution according to the baryonic mass fraction $f_\mathrm{bary}$ measured out to 2.2 disk scale lengths (see Sect. 3.1) is shown too (right-hand panel), divided into subsamples with $\Upsilon_{[3.6]}<0.5$ (light gray bars) and $\Upsilon_{[3.6]}>0.5$ (dark gray bars). \vspace{1.3cm} \normalsize} 
	\label{Fig3}
\end{figure*}

\subsubsection{Mass-to-light ratio}

For the determination of the stellar mass-to-light ratio of a galaxy, assumed to be constant with galactocentric radius, a preliminary selection of three values was considered: $\Upsilon_{[3.6]}$= 0.2, 0.5, and 0.8 $M_\odot/L_\odot$, representing minimum, intermediate, and maximum disks, respectively. These values cover the range of typical values discussed in the literature (see below). Technically speaking, appropriately lowering or increasing $\Upsilon_{[3.6]}$ triggers a better match of the innermost measured velocity with the fitted model RC \citep{Kormendy16}. The RC-decomposition that yielded the lowest $\chi^2_n$ value, corresponding to the best overall goodness-of-fit, was finally chosen (Table B.2). In a few cases, we adopted some optimized values $\Upsilon_{[3.6]}$ slightly differing from the ones mentioned above, too. Varying the mass-to-light ratio from 0.2 to 0.8 substantially impacts the deduced values of the halo parameters: in extreme cases it may decrease $\rho_0$ by 90\% or increase $r_0$ by a factor of about 3, with changes in $\chi^2_n$ by more than 100\%. Hence, within our approach, $\Upsilon_{[3.6]}$ has a considerable impact on the finally adopted halo structure. 

For the whole sample of our irregulars (i.e., with morphological types $T\ge8$), the three semi-free best-fit stellar mass-to-light ratios exhibit an asymmetric bimodal distribution (Fig. \ref{Fig3}). Dominant are values around the median of 0.2 (for both the Burkert halo and the PITS halo case, and for about 60\% and 52\% of the galaxies, respectively), corresponding to submaximal disks and in accordance with the results of \citet{Martinsson13}, \citet{Swaters14}, and \citet{Angus16}. In the Burkert and in the PITS case, the sample means are $\Upsilon_{[3.6]}=0.39\pm0.04$  ($\sigma=0.35$) and $0.42\pm0.03$ ($\sigma=0.28$), respectively (excluding UGC11820 with an outlier value of 3). \citet{VanderKruit11} took already the view that disk galaxies in general are mostly submaximal. This was founded on the observation that their sample galaxies fullfill the velocity-ratio criterium for massive disk galaxies (maximum disk hypothesis) introduced by \citet{Sackett97}, i.e., $v_\mathrm{disk}(R)/v_\mathrm{obs} (R)\,\vline\,_{R=2.2R_d} =0.85\pm0.1$. Replacing $v_\mathrm{disk}$ by $v_\mathrm{bary}$ (as is appropriate for DM rich galaxies), our sample galaxies yield a mean modified ratio 
\begin{equation}
v_\mathrm{bary}(2.2R_d)/v_\mathrm{obs}(2.2R_d) =0.51\pm0.06\,\,\,\,\,\,\,(\sigma=0.16), \label{vrc}
\end{equation}
i.e., a value clearly below 0.85. Our concise value includes the considerable gaseous content of the galaxies and is thus provisionally proposed to be the velocity-ratio criterium for submaximality of disks within very late-type spirals and dwarf irregular galaxies. 

Recently, \citet{Ponomareva18} found the stellar mass-to-light ratios of individual galaxies as determined by several methods to cover a wide range of values, with, for example, a K-band median of $\Upsilon_{K}\approx 0.3$. This was determined either dynamically or by means of spectral energy density (SED) fitting, the latter with a number of outliers at values as high as 0.8 that are however not forming a dichotomy as seen as clear by us. Weighting their values and including thereby the value of 0.5 predominantly seen in approaches with constant mass-to-light ratios (see below), they finally extract a constant [3.6]-band mass-to-light ratio of $\Upsilon_{[3.6]}=0.35$ for their study. Apparently, the findings mentioned so far with median and mean values $\Upsilon_{[3.6]} < 0.5 $ all provide a homogeneous picture.
\begin{figure*} 
	\centering
	\includegraphics[width=0.45\textwidth]{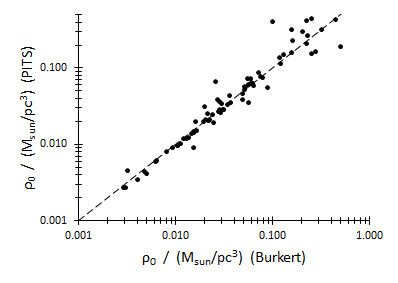}	\includegraphics[width=0.45\textwidth]{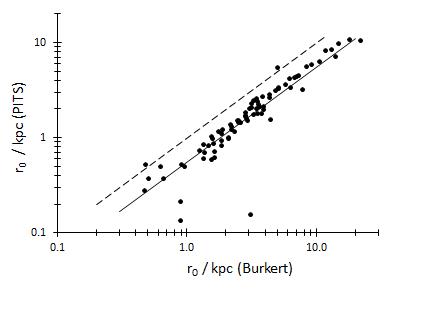}
	\includegraphics[width=0.45\textwidth]{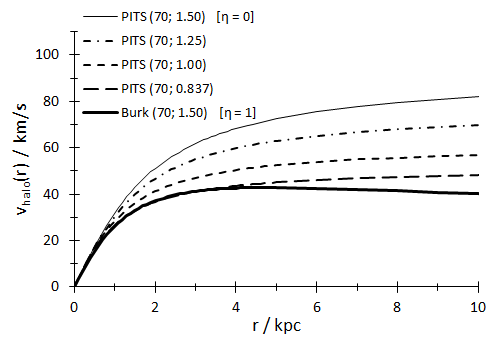}		
	\includegraphics[width=0.45\textwidth]{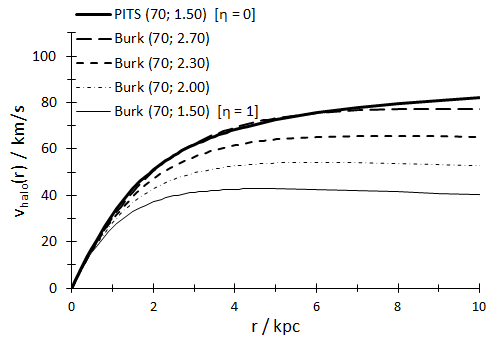}		
	\caption{\small \emph{Top:} Comparison of best-fit halo parameter values for RC decompositions with the PITS and the Burkert halo model. 1:1 lines are shown dashed. While the central densities grossly have similar values (left panel), the core sizes of the PITS halos are systematically lower than the Burkert halos by a factor of 0.558 (corresponding to a downward shift of $\log_{10}0.558=-0.253$ in the log-log-plot, cf. right panel), implying correspondingly smaller halo core masses. 
	\emph{Bottom:} The radial circular velocity components $v_\mathrm{halo}$ for different values $0\le\eta\le 1$ of the hybrid PITS-Burkert halo model are mimicked by means of selected pairs of parameter values $(\rho_0/( 10^6\, \mathrm{M_\odot}\mathrm{kpc}^{-3}); r_0/\mathrm{kpc})$ of the PITS ($\eta$=0) or the Burkert ($\eta$=1) model.    \normalsize} 
	\label{Fig4}
\end{figure*}
Our data processing produced few best-fit values $\Upsilon_{[3.6]}=0.5$. Based on stellar population synthesis models \citep{McGaugh14b, Meidt14}, this value is currently favored by the SPARC team despite of the frequent occurence of unrealistic ratios $v_\mathrm{bary}/v_\mathrm{obs}>1$ \citep[][Fig. 7]{Lelli16a}. The other peak of our bimodal distribution lays at 0.8 (in the Burkert halo case for about 25\% of the galaxies). The frequent occurence of this relatively high value, that implies maximum disks, is remarkable as its general application would lead to unphysically high disk circular-velocities for many high-mass spirals of morphological types $T\le7$ \citep{Lelli16a}. \citet{deBlok08} find the free-fit 3.6-micron $\Upsilon$ values for PITS fits typically to lay between 0.2 and 0.8 (in agreement with our findings) and to be peaked around a mean of 0.5 (see their Fig. 60, disagreeing with our finding). In addition, their ratios show some weak trends with absolute magnitude and with color (see, e.g., their Fig. 59). The former trend is neither seen in our data, and due to the lack of color data we can't say anything about the latter finding. The RC decompositions of \citet{Pace16} resulted in bimodal DC14 \citep{DiCintio14b} and PITS halo mass distributions, with a claimed similar degeneracy for the mass-to-light ratio. However, most of the larger masses were unrealistically high (comparable to galaxy group or cluster halo masses), thus mainly the lower mass solutions and the corresponding lower mass-to-light ratios were accepted for the best-fit decompositions of their galaxies. 

The striking mass-to-light ratio dichotomy seen with our data is not mirrored with other galaxy parameters (as far they are currently at disposal to us), with one unsharp exception. With the data at hand we found no correlation of $\Upsilon_{[3.6]}$ with other disk or halo structural or kinematical galaxy parameters (e.g., central surface brightness, central halo surface density, halo scale length; we did not check for galaxy color, type, or environment, however). The only variable that exhibits a recognizable kinship with the mass-to-light ratio is the baryonic mass fraction $f_\mathrm{bary}(2.2\,R_d)$ (at 2.2 disk scale lengths $R_d$, see Sect. 3.1). As seen in the right panel of Fig. \ref{Fig3}, the peak for the ($\Upsilon_{[3.6]}<0.5$)-subsample goes with a lower baryonic mass fraction than the peak for the ($\Upsilon_{[3.6]}>0.5$)-subsample. Despite of the distributions being heavily skewed, there seems to be a moderate tendency for minimum disks to exhibit lower inner-disk baryonic mass fractions than maximum disks. This is plausible and in line with the conclusion of \citet{Courteau99} that maximum disks preferably occur in galaxies that have higher rotational support at 2.2$\,R_d$. However, while they assign this to the more compact (HSB) galaxies, we do not see a correlation with $\mu_0$ or $C$. We additionally note that, again without an obvious relevance to our observed dichotomy, \citet{Kuzio08} sees a trend towards larger $r_0$ (PITS fits) with increasing stellar $\Upsilon$. All in all, there's still quite some ambiguity concerning the distribution of $\Upsilon$. Finally, but importantly, based on stellar population synthesis models several authors consistently find that low mass-to-light ratios are tightly correlated with bluer galaxy colors in various bands, and vice versa, and with lower metallicity \citep[e.g.,][]{Bell01, Meidt14}. A mass-to-light ratio dichotomy could then imply two populations of presumably younger and older galaxies. This seems improbable, but in a follow-up study we must check for a possible color bimodality. 

\subsubsection{Hybrid Burkert-PITS model}

Comparing the best-fit halo parameters of either decomposition approach (i.e., with the Burkert or the PITS halo adopted), the central densities $\rho_0$ have similar values ($\rho_{0,\mathrm{PITS}} = (1.05\pm0.5) \,\rho_{0,\mathrm{Burk}}+0.00\pm 0.01$, excluding four outliers), but the scale lengths (core radii) $r_0$ of the Burkert model are systematically higher by a factor of about 1.8 ($r_{0,\mathrm{PITS}} =(0.56\pm 0.02)\,r_{0,\mathrm{Burk}}+0.05\pm0.09$). This is illustrated in the two upper panels of Fig. \ref{Fig4}. Depending on the halo model used, some scaling relations based on these parameters will consequently be influenced. For example, the sample mean for the central halo surface density $\rho_0 r_0$ calculated with the Burkert halo parameters comes out higher by a factor of 1.8 than when applying PITS halos (Sect. 3.3). 

The radial circular velocity components $v_\mathrm{halo}$ for the hybrid Burkert-PITS halo model (as given by equation \ref{halodens}) are shown in Fig. \ref{Fig4} (lower panels). Illustrated are the cases for a fix central halo density $\rho_0 = 70\cdot 10^6\,\mathrm{M_\odot}/\mathrm{kpc}^3$ and varying cores sizes $r_0$ (in kpc). This mimics varying parameter values $0\le\eta\le1$ for the halo density profile (equation \ref{halodens}), with $\eta=0$ and $\eta=1$ representing the PITS and the Burkert model, respectively. In the left panel the core sizes of the PITS model halo are reduced from 1.5 to 0.837 kpc (=1.5$\cdot$0.558) approaching a Burkert halo with core size 1.5 kpc. Similarly, in the right panel the core sizes of the Burkert halo with initial core size 1.5 kpc are increased to 2.7 kpc ($\approx 1.5/0.558$) closely leading to the velocity contribution of a PITS halo with core size 1.5 kpc. Finally, at radii smaller than about four core sizes ($r<4\,r_0$) the halo components $v_\mathrm{halo}$ of the PITS and the Burkert model are approximately proportional. We conclude that the hybrid PITS-Burkert model may be a valuable candidate for RC mass model decompositions, with the real numbers $\eta \in [0;1]$ and $\Upsilon_{[3.6]} \in [0.2;0.8]$ possibly being interrelated. However, no simple analytic formalism is available and one must therefore calculate $v_\mathrm{halo}$ numerically. An application to real data seems nevertheless worthwhile, but is out of the scope of this paper.

\section{Scaling relations}

The various parameters involved in describing photometric and kinematic or structural and dynamical phenomena with respect to disk galaxies show some characteristic interrelationships; for a recent overview, see for example \citet{Kormendy16} (using PITS converted to isothermal halo models) or the series of papers by the SPARC team \citep[starting with][]{Lelli16b}. Based on our homogenous data set with its focus on the Burkert halo model interpretation, we present a non-comprehensive selection of such scaling relations. While most are well-known, we highlight and controversely discuss some details, partly provide new approaches in a consistent way, and introduce some new scaling relations not discussed so far in the literature. A list with acronyms used in the text can be found in Sect. 4, Table \ref{Table3}.

\subsection{Decreasing baryonic mass fraction from HSB to LSB galaxies}

\begin{figure*} 
	\centering
	\begin{minipage}{0.7\linewidth}
		\includegraphics[width=1.0\linewidth]{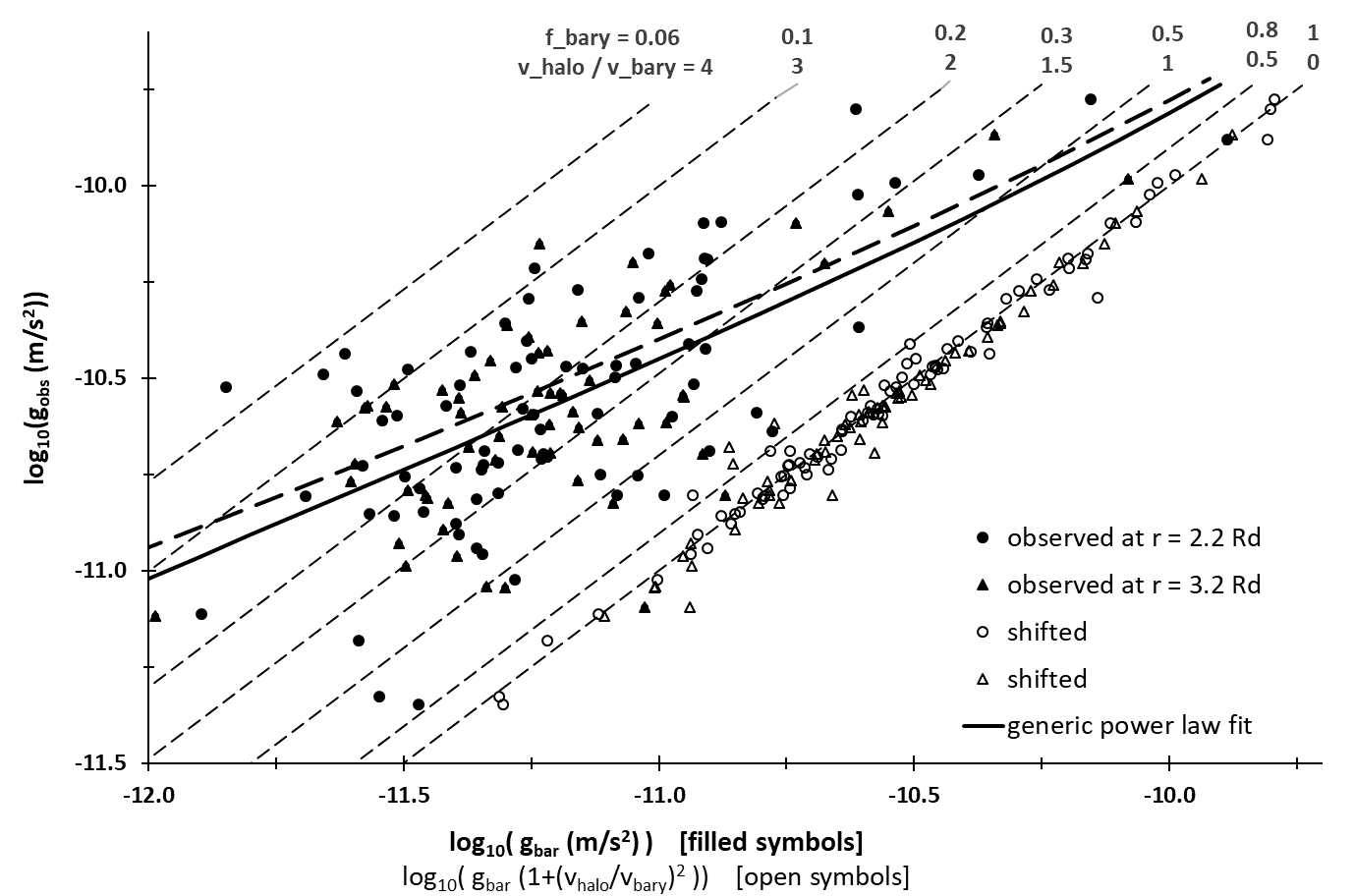}
	\end{minipage}\hfill
	\begin{minipage}{0.3\linewidth}
		\includegraphics[width=0.99\linewidth, height=4cm]{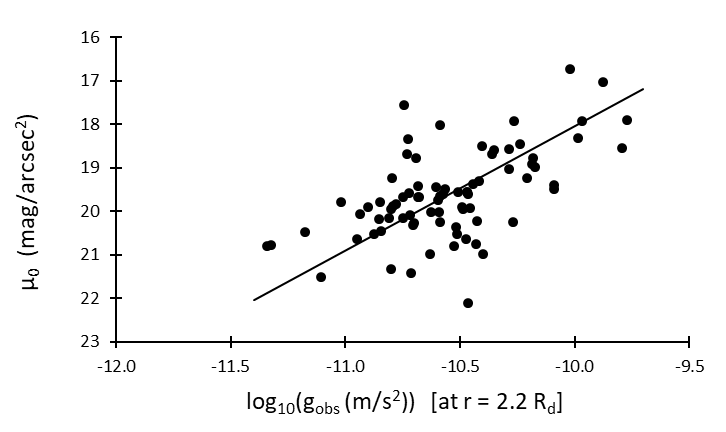}
		\includegraphics[width=0.99\linewidth, height=4cm]{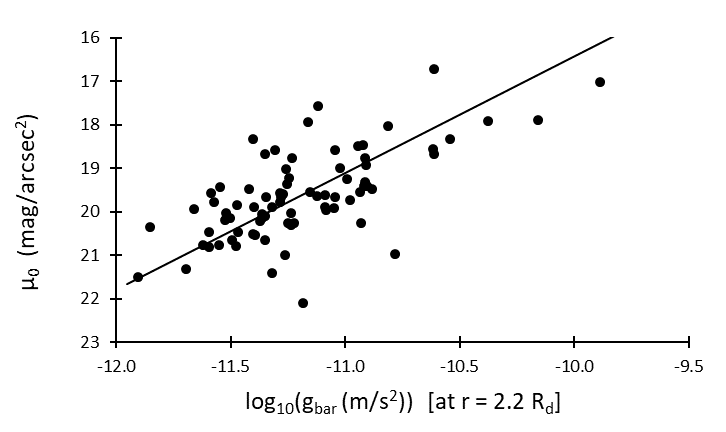}
	\end{minipage}	
	\caption{\small \emph{Left:} Radial acceleration relation (RAR) with data taken at radii 2.2 $R_d$ and 3.2 $R_d$ (filled circles and triangles, respectively). The scatter is largely due to the galaxies' individual baryonic mass fractions $f_\mathrm{bary}(2.2R_d)$ or $f_\mathrm{bary}(3.2R_d)$ made evident by plotting $g_\mathrm{obs}$ versus $g_{\mathrm{bary}}/ f_{\mathrm{bary}}$ in the same diagram (open symbols): the calculated points align (with a spread of about 0.1 dex) along the 1:1 line. Selected lines of constant baryonic mass fractions $f_\mathrm{bary}$ are shown as short-dashed lines with slope 1. There is an overall trend that with decreasing radial accelerations the baryonic mass fraction is decreasing as well, and vice versa. For the regression lines drawn see the text. 
	\emph{Right:} Observed and baryonic radial accelerations exhibit some correlation with extrapolated central surface brightness (here shown for the radius 2.2 $R_d$). Comparing the two stacked panels, we note the horizontal shift of each data point with a given $\mu_0$ due to $\log (f_\mathrm{bary})$. The left panel allows the interpretation that the baryonic mass fraction tends to be gently lower for LSB galaxies than for HSB galaxies, and vice versa. \normalsize} 
	\label{Fig5}
\end{figure*}

Dividing equation (\ref{vobs}) by $r$ in order to have $g_{\mathrm{obs}}= g_{\mathrm{gas}}+g_{\mathrm{stellar}}+g_{\mathrm{halo}}$ with radial accelerations $g=v^2(r)/r$ (and with $\Upsilon_{[3.6]}$ included within $g_{\mathrm{stellar}}$), unifying the gaseous and the stellar part by means of a baryonic amount $g_{\mathrm{bary}} \equiv g_{\mathrm{gas}}+ g_{\mathrm{stellar}}$, and rearranging and rewritting terms, one easily arrives at the radial acceleration relation (RAR)
\begin{equation}
g_\mathrm{obs}(r)=g_\mathrm{bary}(r)\left( 1+\left(\frac{v_\mathrm{halo}(r)}{v_\mathrm{bary}(r)}\right)^2 \right).\label{gobs}
\end{equation} 
The values on the left-hand side are obtained from the observed RCs, those of the right-hand side follow from model-based contributions for the rotating baryonic disk and for the Burkert halo. If the RC decompositions are accurate, this relation should produce data points aligned along on the 1:1-line. As can be seen by means of the open symbols in Fig. \ref{Fig5}, this is largely true for our decompositions at both radii considered, at 2.2$R_d$ and at 3.2$R_d$. Deviations reflect mismatches between observations and modelled results, incorporating systematic errors in both variables. 

The characteristic RAR-diagram logarithmically plots $g_\mathrm{obs}$ versus $g_\mathrm{bary}$ and is represented in Fig. \ref{Fig5} by means of the filled symbols. Each filled symbol at position  ($g_\mathrm{bary}$, $g_\mathrm{obs}$) corresponds to a horizontally shifted open symbol located at ($g_\mathrm{bary}/f_\mathrm{bary}$, $g_\mathrm{obs}$) (omitting the $\log$-notation and with $1/f_\mathrm{bary}$ being the \emph{mass-degeneracy factor} $D=M_{\mathrm{obs}}/M_{\mathrm{bary}}$ \citep{McGaugh14a} contained in equation \ref{gobs}). The location of the data points and their scatter in the diagram is basically the result of this mass-degeneracy factor and hence is explained by the individual relative content of DM within each of the sample galaxies out to the chosen radius. Galaxies with no DM would lay on the slope-1-line where $g_\mathrm{obs}=g_\mathrm{bary}$ (shown as the rightmost of the dashed lines). For any fixed ratio $v_{\mathrm{halo}}/v_{\mathrm{bary}}\ne 0$ the logarithmic representation in Fig. \ref{Fig5} will generate a line parallel to this 1:1 line; drawn are the straight lines for the ratios 0, 0.5, 1, 1.5, 2, 3, and 4. These ratios correspond to baryonic mass fractions 1, 0.8, 0.5, 0.3, 0.2, 0.1, and 0.06, respectively (indicated at the top of the figure). They are dynamically calculated within Newtonian mechanics, i.e., $f_{\mathrm{bary}}\equiv M_{\mathrm{bary}}/M_{\mathrm{obs}}=v^2_{\mathrm{bary}}/ v^2_{\mathrm{obs}}=g_\mathrm{bary}/g_\mathrm{obs}$ and hence with equation (\ref{gobs})
\begin{equation}
 f_\mathrm{bary}(r)= \frac{v^2_{\mathrm{bary}}(r)}{v^2_{\mathrm{obs}}(r)}=\left( 1+
 \left( \frac{v_\mathrm{halo}(r)}{v_\mathrm{bary}(r)} \right)^2 \right)^{-1}. \label{fbary}
\end{equation}

At the radius $r=2.2\,R_d$, for example, we have a mean $<$$f_\mathrm{bary}(2.2R_d)$$>$$\,\,=0.28\pm 0.02$ with a meaningful spread of $\sigma=0.18$ (Table \ref{Table1} on p.15). (The square-root of this corresponds to the value of the velocity-ratio for minimum disks discussed in Sect. 2.3.2). Comparing in the RAR diagram two data points on the regression line, say one with higher and the other with lower values for both the  $g_\mathrm{bary}$ and $g_\mathrm{obs}$ values, the latter has a tendency to have a fainter central surface brightness. This is due to the gentle correlation of both radial accelerations with extrapolated central surface brightness (see the stacked panels on the right-hand panel of Fig. \ref{Fig5}, shown for the radius 2.2 $R_d$). Thus the RAR diagram allows for the interpretation that the baryonic mass fraction $f_\mathrm{bary}$ tends to be lower for LSB galaxies than for HSB galaxies, and vice versa. However, as can be inferred from Table \ref{Table1}, the effect is too shallow to actually contribute to a clear-cut distinction of HSB and LSB galaxies. Reminding the luminosity-independent distinction of HSB and LSB galaxies discussed in Sect. 2.2, the above tendency is not to be confused with the analogous, well-known relation that more luminous galaxies tend to be less DM dominated than less luminous galaxies: the fainter the galaxies, the lower the mean baryon fractions, and vice versa \citep[e.g.,][]{Roberts94, deBlok08}; in the words of \citet{Kormendy16}: "Smaller dwarf galaxies form a sequence of decreasing baryon retention" that "is generated primarily by supernova-driven baryon loss or another process." An extrapolation to the extreme leads to very slowly rotating and faint, even dark, galaxies \citep{DiCintio17}. Within an individual galaxy, the radial dependence of $f_\mathrm{bary}(r)$ typically is slowly decaying and nearly vanishes at larger radii where DM strongly dominates (cf. equation \ref{fbary2} in Sect. 3.5 for a formal treatment and Fig. 58 of \citet{deBlok08} for an illustration with respect to low-mass galaxies). To guide the eye, a generic power law of the form 
\begin{equation}
g_\mathrm{obs}=g_\mathrm{bary}\left( 1+\frac{g_0}{g_\mathrm{bary}}\right)^{\gamma} \label{gobsgbary}
\end{equation}
\begin{figure*} 
	\centering
	\includegraphics[width=0.45\textwidth]{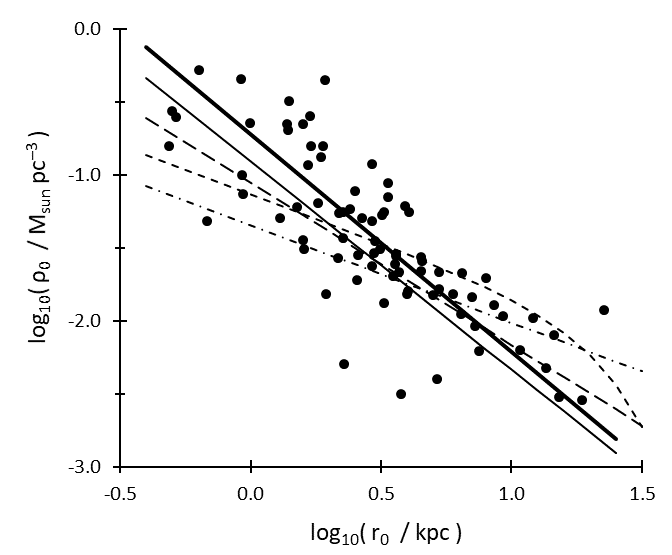}
	\includegraphics[width=0.45\textwidth]{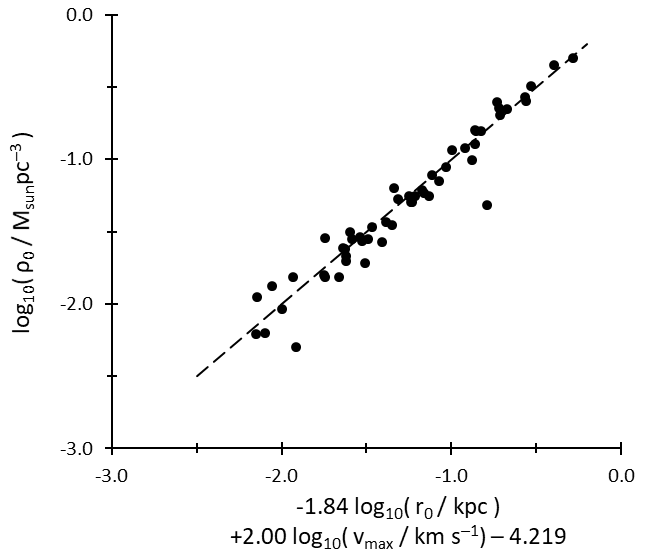}
	\caption{\small \emph{Left:} The two Burkert halo parameters correlate according to a power law with an approximate exponent -3/2 that corresponds to a slope of -3/2 in the $\log$-$\log$-plot. The fat and thin solid lines are OLSB fits to the present data set and to the smaller one in \citet{Karukes17} (for dwarf disk galaxies, slope $-1.4 \pm 0.1$), respectively. For comparison, the long-dashed and short-dashed lines are the OLS fit of \citet{Kormendy16} (for Sc-Im and dSph galaxies, slope $-1.11 \pm 0.07$, using nonsingular isothermal halos) and the eyeball fit of \citet{Salucci00} to spiral galaxies of all luminosities, respectively. Finally, the dot-dashed line shows the first guess of \citet{Burkert95} with a slope of -2/3. \emph{Right:} The central halo parameters are tightly connected to the maximum circular velocity. The 1:1 line is shown dashed. \normalsize} 
	\label{Fig6}
\end{figure*}
is fitted to the data in Fig. \ref{Fig5}, yielding best-fit parameter values $g_0=1.70\cdot 10^{-10}$ m$\,$s$^{-2}$ and $\gamma=0.44$, with the graph shown in Fig. \ref{Fig5} as fat solid line. This line goes with a spread of about 0.3 dex and asymptotically approaches the 1:1-line. For comparison, the correlation suggested by \citet{McGaugh16}, motivated by the Modified Newtonian Dynamics (MOND) scenario and given by $g_\mathrm{obs}=g_\mathrm{bary}(1- \exp(-\sqrt{g_\mathrm{bary}/ g_{0,\mathrm{MOND}}}\,))^{-1}$ (with $g_{0,\mathrm{MOND}}=1.2\cdot 10^{-10}$ m$\,$s$^{-2}$), is drawn as well (fat dashed line). Using nearly 2700 data points for 153 SPARC galaxies, their scattering statistics gives a total spread of amazingly low 0.12 dex, encouraging the authors to speak of this tight coupling between DM and LM as a "new natural law". Recently, now in addition allowing $\Upsilon_{[3.6]}$ to be a free parameter, \citet{Li18} even reduced the spread to well below 0.1 dex. Within our formalism from above, their divisor $1- \exp(-\sqrt{g_\mathrm{bary}/g_{0,\mathrm{MOND}}}\,)$ acts as an \emph{averaged} baryonic mass fraction function $<$$f_\mathrm{bary}(g_\mathrm{bary})$$>$: data points along the graph of the power-law function (\ref{gobsgbary}) (solid line) correspond to galaxies matching the mean-ratio $<v_\mathrm{halo}/v_\mathrm{bary}>=\sqrt{(1+g_0/g_\mathrm{bary})^\gamma -1}$, as a function of baryonic radial acceleration. 

In general, the width of the data band in Fig. \ref{Fig5} represents physical information, if the accelerations are measured within the luminous domain. The significance of deviations from this observed mass discrepancy$-$radial acceleration relation (MDAR) is discussed in, e.g., \citet{Navarro17} and \citet{Salucci18}. The width of the data band being substantial was emphasized by \citet{OBrien18} within a conformal gravity approach, too. Indeed, irrespective of the number of data points used, our comparably wide scatter is fundamentally considered not to be a statistical spread due to, e.g., observational uncertainties, but to be mainly intrinsic: it bears, as discussed above, relevant and quantifiable information on the individual galaxies' baryonic mass fractions at given inner or intermediate radii. For a given $g_\mathrm{bary}$ it makes perfectly sense to observe a whole range of galaxies with different $g_\mathrm{obs}$, and vice versa. 

Relation (\ref{gobsgbary}) or in the form promoted by \citet{McGaugh16} is nonlinear, but asymptotically approaches linearity for larger values of the radial accelerations. This feature bears an implication for the somewhat surprising square-root relation for the average halo contribution to the circular velocity as observed by \citet{McGaugh07} and confirmed by several teams \citep{deBlok08, Walker10, Castignani12}, namely $v_\mathrm{halo}(r) \propto r^{1/2}$ (typically evaluated at radii $r>1\,\mathrm{kpc}$ and claimed to be irrespective of luminosity). This relation implies $g_\mathrm{halo}=\mathrm{const}$ and hence a linear dependence $g_\mathrm{bary}\propto g_\mathrm{obs}$. The above asymtotics property imposes the restriction that the square-root relation only becomes valid for radial acceleration values larger than about $3\cdot10^{-10}$ m$\,$s$^{-2}$. 

\subsection{Halo core density vs. size relation}

As manifested in the left panel of Fig. \ref{Fig6}, the two Burkert halo parameters show some robust correlation, with considerable scatter, however. An OLSB fit yields
\begin{equation} 
\log\left( \frac{\rho_0}{\mathrm{M_\odot pc^{-3}}} \right) =(-1.489\pm0.108)\,\log\left(\frac{r_0}{\mathrm{kpc}}\right) -0.723,  \label{lgrho0lgr0}
\end{equation}
shown as fat solid line. For comparison, some similar relations given in the literature are plotted as well (see the figure caption for the references). The values for the slopes range from $-2/3$ to $-3/2$, with the steeper end  representing more recent findings in the context of Burkert halo profile decompositions. However, using the PITS model and taking $\Upsilon_{[3.6]}$ to be a free parameter, the 16 LSB galaxies of \citet{deBlok08} with 3.6 $\mu$m-data available exhibit an even steeper slope of -2.2$\pm$0.3 (OLSB-fit not shown here, but we refer to their Fig. 66 with data from their Table 5); with $\Upsilon_{[3.6]}$ fixed the slope becomes less steep again. Similarly, \citet{Kuzio08} find for LSB galaxies with PITS halos slopes that are the steeper the higher the adopted value of $\Upsilon$ is. It thus seems that somewhat steeper slopes are observed (i) with smaller-mass systems (mean slope around -3/2) as opposed to more heavy galaxies (mean slope around -1), and (ii) with decompositions that allow for some variability in the mass-to-light ratio. While these dependencies are neither new nor strict, they are in support of our comparably steep slope.

Our results are independent of the cored halo model chosen, i.e., whether the Burkert or the PITS halo is involved in the best-fit procedure. To be more concrete, despite of PITS halo decompositions systematically delivering smaller core radii as compared with the Burkert halos (we remind of Fig. \ref{Fig4}, upper right panel), a $\log\rho_0-\log r_0$-diagram with only the best-fit Burkert or PITS halo parameters plotted will still provide a similar slope and simply shift the regression line downwards. This holds if the $r_0$-values for best-fit PITS halos cover the full range of core densities. Indeed, a correponding OLSB fit yields $\log(\rho_0) =(-1.471\pm0.110) \,\log(r_0)-0.853$, thus with about the same slope as in equation (\ref{lgrho0lgr0}) but with a downward shift of the intercept by 0.13 dex (together with a slightly increased scatter, the coefficient of determination increases from $R^2=0.612$ to $R^2=0.671$). Using instead best-fit PITS halos only the shift amounts to 0.25. All of this is consistent.

From a particular cosmological perspective, one may expect an even somewhat steeper slope than -3/2 for the $\rho_0 - r_0$-relation. To elaborate on this in line with \citet[][Sect. 9.10]{Kormendy16}, we enter (i) a flat universe with (ii) hierarchically clustered structures (iii) originating in primordial density fluctuations subject to a power law in wavenumber $k$ or wavelength $\lambda$, i.e., $\vline\delta_k\vline\,^2\propto k^n \propto \lambda^{-n}$. Assuming furthermore (iv) present DM halos being bound and virialized objects, a scaling relation holds between the density $\rho$ and the size $R$ of such objects, with the exponent $n$ being called the slope of the power spectrum. According to \citet{Padmanabhan92}, the density-size-relation reads $\rho\propto R^\gamma$, where $\gamma = -3(3+n)/(5+n)$. Hence, the steeper (i.e., the more negative) the slope of the power spectrum the shallower the central density profiles of the resulting objects. One often translates $n=n_s-1$, with the scalar spectral index $n_s=0.968\pm0.006$ being known to high precision from observations of large scale structures (LSS), in particular, from the temperature anisotropies of the CMB radiation and related effects due to gravitational lensing with galaxy clusters \citep{Planck16}. Thus measured at large scales the power spectrum is close to (but significantly not) scale invariant, i.e., $n=-0.032$, hence larger primordial objects ($\lambda$ high) are slightly more abundant than smaller ones and the density distribution has a slope $\gamma = -1.792\pm0.003$. Assuming (v) these conditions to prevail to the present universe and galactic halos to be indeed virialized objects (thus accepting $\rho=\rho_0$ and $R=r_0$), our result $\gamma = -1.49\pm0.11$ ($n=-1.03$) provides a less steep central core density profile and the slope of the power spectrum is much lower. We note that this is determined for small scale structures (SSS) as probed by the halos of late-type spirals and dwarf irregular galaxies (at basically zero redshift). While our result disagrees with the CDM scenario that favours an exponent $-1$ ($n=-2$), \citet{Kormendy16} yield a scaling relation $\gamma =-1.11\pm0.07$ for their sample of Sc to Im  and dSph galaxies (their equation 51, corresponding to $n=-1.83\pm0.17$ and included in our Fig. \ref{Fig6}) and thus are still consistent with the CDM scenario. However, due to assumptions (i) to (v) stated above, we are urged to excercise caution with far reaching conclusions. For example, if halos are not virialized objects (but presumably still collapse or expand) the above cosmological density-size-relation cannot safely be applied.

If one accepts the assumptions, the following third variable dependency could get things staightened again: the scatter seen in Fig. \ref{Fig6} (left panel) around relation (\ref{lgrho0lgr0}) relates to observable kinematics, in particular to the maximum circular velocity. A multiple linear regression (MLR) to the 54 galaxies with data for $v_\mathrm{max}$ available yields 
\begin{eqnarray} 
\log\left( \frac{\rho_0}{\mathrm{M_\odot pc^{-3}}} \right) &=&(-1.839\pm0.066)\,\log\left(\frac{r_0}{\mathrm{kpc}}\right) \nonumber\\
& &+(1.996\pm0.113)\,\log\left(\frac{v_\mathrm{max}}{\mathrm{km\, s^{-1}}}\right) \nonumber\\  & & -(4.212\pm0.192),  \label{lgrho0lgr0lgvmax}
\end{eqnarray}
with a coefficient of determination $R^2=0.939$ and a standard deviation $\sigma = 0.124$. (The MLRs for the two other variable combinations go with much smaller coefficients of determination and are therefore ignored.) The corresponding plot is shown in Fig. \ref{Fig6} (right-hand panel), with the dashed line indicating 1:1 correspondence. If halos are virialized objects, the above slope of $-1.839\pm0.066$ would be consistent with the cosmological expectation $-1.792\pm0.003$ (still assuming the applicability of the $\rho$-$R$-relation and with the $v_\mathrm{max}$-term entering the normalization). 

Whether or not halos are virialized or the other assumptions mentioned above are valid, relation (\ref{lgrho0lgr0lgvmax}) is a crucial new actor on the scene. It relates the maximum circular velocity (typically measured beyond the optical radius) with the DM halo core region (typically smaller than the disk extension), as exemplified in Fig. \ref{Fig2} for two galaxies). Dynamically, LM in the disk seems to be of marginal influence for our sample galaxies. This feature will be encountered some more times in this study: the inner circular velocity gradient is linked to the maximum circular velocity (Sect. 3.4); the tight \emph{adjusted} baryonic Tully-Fisher relation discussed in Sect. 3.5 highlights the dominant role of DM; and in Sect. 3.6 we will argue that the above new relation is closely related to the central halo core column density. 
\begin{figure} 
	\includegraphics[width=0.48\textwidth]{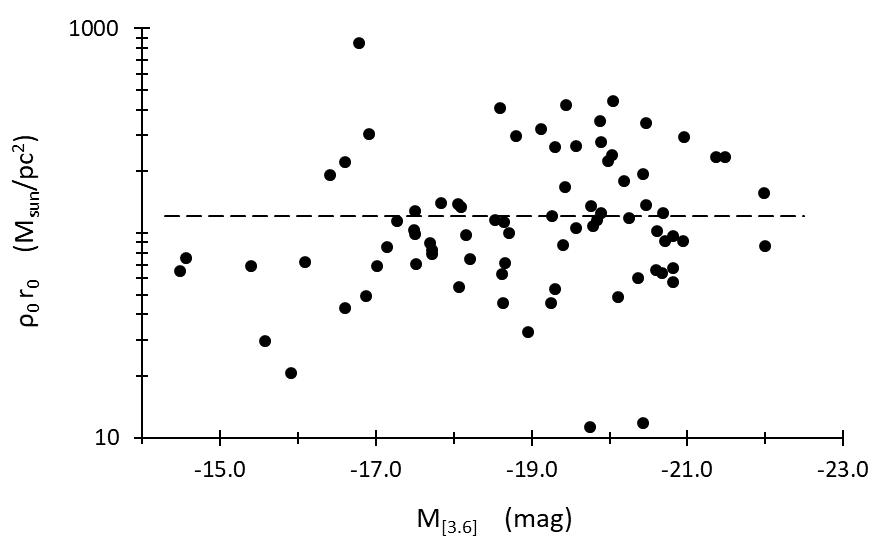}
	\caption{\small Central (Burkert) halo surface density $\rho_0r_0$ versus absolute magnitude $M_{[3.6]}$. The dashed line at 121 $M_\odot\mathrm{pc}^{-2}$ represents the median. \normalsize} 
	\label{Fig7}
\end{figure}

\subsection{Central halo surface density dependencies}

\subsubsection{Radial acceleration}

\begin{figure*}
	\centering
	\includegraphics[width=0.45\textwidth]{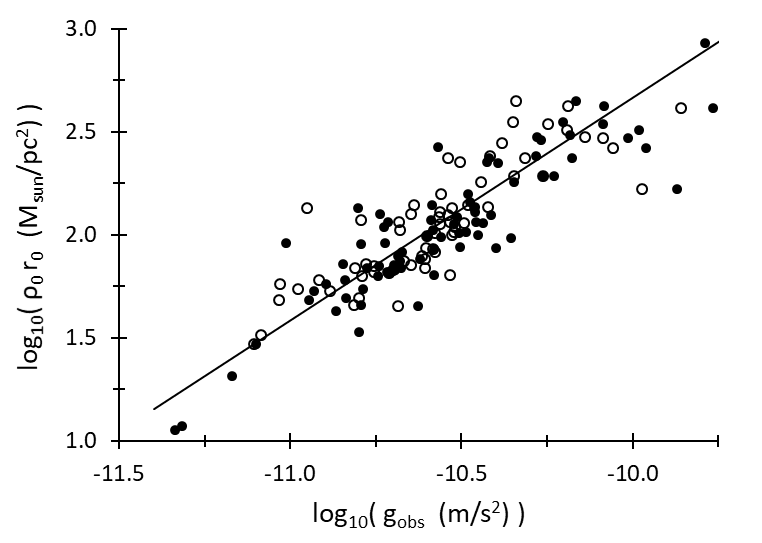}
	\includegraphics[width=0.45\textwidth]{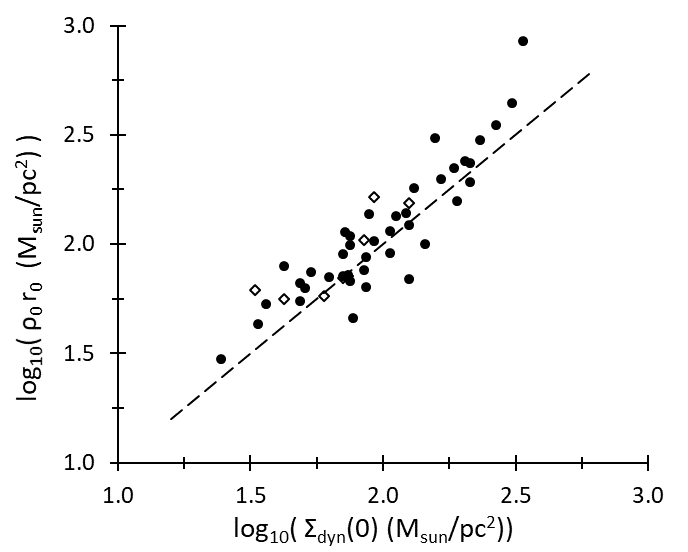}
	\includegraphics[width=0.45\textwidth]{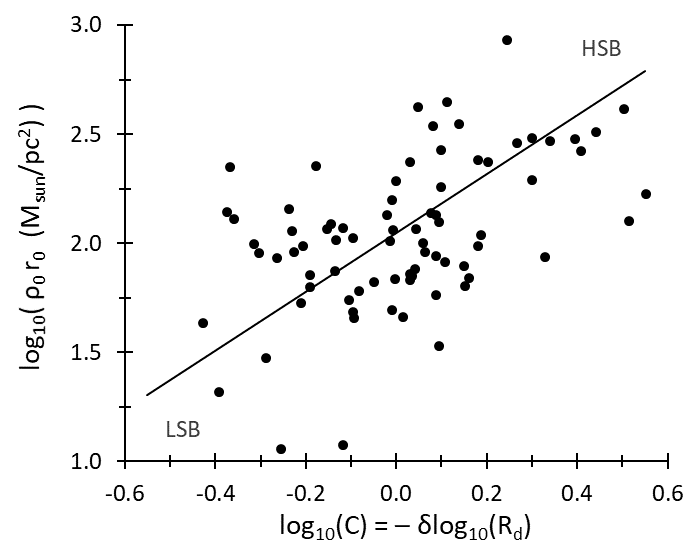}
	\includegraphics[width=0.45\textwidth]{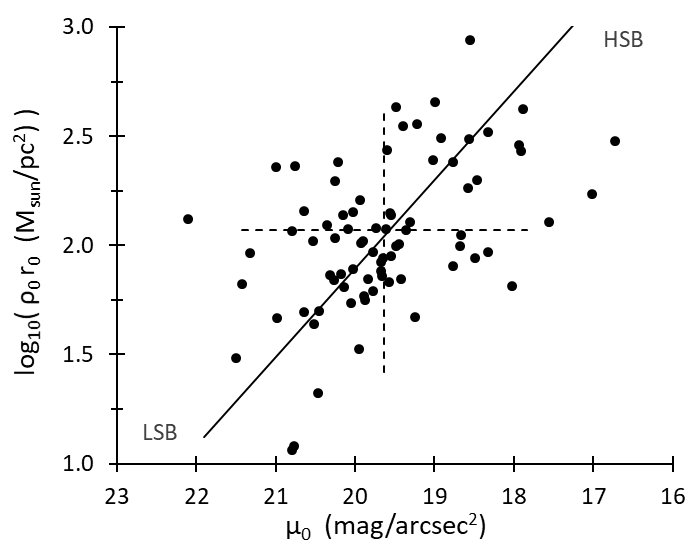}
	\caption{The scatter seen in the the previous figure is related to kinematic and photometric observables. \emph{Top left:} Radial acceleration $g_{\mathrm{obs}}=v^2_{\mathrm{obs}}(r)/r$ is strongly linked to central halo surface density $\rho_0\,r_0$. Shown are the values at the Freeman disk-peak radius $r=2.15\, R_d$ (filled symbols) and at the optical radius $r=3.2\, R_d$ (open symbols). The solid line is an OLSB fit to the combined sample. \emph{Top right:} DM halo central surface density correlates with the dynamically determined total central surface density. Open rhombic symbols are galaxies from \citet{Karukes17}. The dashed 1:1 line is to guide the eye. \emph{Bottom left:} There is some shallow trend for the less compact LSB galaxies to have smaller halo central surface densities than the more compact HSB galaxies. \emph{Bottom right:} Consistent with the lower left panel, DM central surface density weakly correlates with luminous matter central surface brightness. The short-dashed cross highlights the sample median $<$$\rho_0 r_0$$>$$\,\approx 121\, \mathrm{M_\odot/pc^2}$ and mean $<$$\mu_0$$>$$=19.63$ mag/arcsec$^2$.  \normalsize } 
	\label{Fig8}
\end{figure*}

Taken at face value, for the galaxies of the present sample Fig. \ref{Fig7} states the \emph{independency} of the central halo  surface density $\rho_0r_0$  with absolute magnitude $M_{[3.6]}$. The sample median for the Burkert halo parameters is $<$$\rho_0 r_0$$>$$ \approx 121\, \mathrm{M_\odot/pc^2}$, with a large sample spread of $\sigma = 112$. While the central surface density differs by about 2 orders of magnitude, the absolute magnitudes cover a range of about 8 magnitudes, and from Fig. \ref{Fig8} (lower right-hand side panel) one may infer that the values for the central surface brightness lay within an interval range of 6 mag/arcsec$^2$. For this reason (and decisively further strengthened by considering all types of spirals and dwarf spheroidal galaxies as well, not shown here) the halo central surface density of late-type spirals and dwarf irregular galaxies is sometimes considered to be of approximate \emph{constancy} \citep{Kormendy04,Kormendy16, Donato09, Gentile09, Walker10, Castignani12, Lelli14, Karukes17, DiPaolo18}. Given that the surface densities differ by two orders of magnitude, it would be more appropriate to speak of relative constancy. Precise constancy would imply true anti-correlation, i.e., inverse proportion requiring a slope of $-1$. The significantly different slope found in equation (\ref{lgrho0lgr0}) hints to hidden relations. Indeed, the previous section already revealed a very strong third variable dependency. The evidence for true constancy has been disputed on theoretical grounds by, e.g., \citet{Lin16} and \citet{delPopolo16} who both find a surface density versus halo mass relation with slope 0.18$\pm$0.05 and find in addition a weak dependence on absolute luminosity as well. A similar dependence was noted in \citet{Kormendy16} (statistically insignificant, however). Thus, compared with the multitude of massive galaxies, the near-constancy assumption is understandable for the subclass of late-type spirals and dwarf irregular galaxies with a relatively narrow range of masses. However, zooming in on the matter there are subleties to be recognized on observational grounds. Using equation (\ref{MhaloBurk}) and solving $g_\mathrm{obs}(r)$ $=g_\mathrm{halo}(r) +g_\mathrm{bary}(r)$ $=G M_\mathrm{halo}(r)/r^2 +g_\mathrm{bary}(r)$ for $\rho_0 r_0$ at a radius expressed as some real multiple of the core size, i.e., $r=k\,r_0$ ($k\in R^+$), gives
\begin{eqnarray}
\rho_0 r_0 &=& \frac{1}{2\pi G}\frac{k^2}{f(k)} \left( g_\mathrm{obs}(kr_0)-g_\mathrm{bary}(kr_0) \right)\nonumber \\
 & \approx& \frac{1}{2\pi G}\frac{k^2}{f(k)} g_\mathrm{obs}(kr_0).
 \label{rho0r0}
\end{eqnarray}
Herein, $f(k)\equiv \ln\left( (1+k)\sqrt{1+k^2}\,\right) - \arctan(k)$ is an auxillary function and for the approximation we neglected the baryonic term assuming values of $k$ that correspond to radii where dark matter is strongly dominating. For example, at $r=r_0$ ($k=1$) one has a median acceleration of $g_\mathrm{obs}(r_0)\approx g_\mathrm{halo}(r_0)=0.255\cdot2 \pi G$ $<$$\rho_0 r_0$$>=2.4\cdot 10^{-9}\,\mathrm{ms}^{-2}$, theoretically explaining and agreeing with the universal value $3.2^{+1.8}_{-1.2}\cdot 10^{-9}\,\mathrm{ms}^{-2}$ of \citet{Gentile09}. In DM dominated dwarf galaxies the scaling law given by relation (\ref{rho0r0}) should hold for other radii than $r_0$, too. Indeed, in the upper left panel of Fig. \ref{Fig8} the central halo surface density is seen to be rather tightly related to the observed radial acceleration at the two selected radii, 2.2 $R_d$ (filled symbols) and 3.2 $R_d$ (open symbols). Corresponding OLSB fits give $\log(\rho_0\,r_0) = (1.072\pm0.057)\,\log(g_{\mathrm{obs}}) +13.361 $ and $\log(\rho_0\,r_0) = (1.252\pm0.086)\,\log(g_{\mathrm{obs}}) +15.252$, equivalent to
\begin{eqnarray} 
\rho_0 r_0 &=& 2.293\cdot 10^{13} g_{\mathrm{obs}}^{1.072\pm 0.057}\,\,\,\,\,[\mathrm{at}\,\,2.2\,R_d]\label{rho0r0vsgobs1}\\
\rho_0 r_0 &=& 1.792\cdot 10^{15} g_{\mathrm{obs}}^{1.252\pm 0.086}\,\,\,\,\,[\mathrm{at}\,\,3.2\,R_d]\label{rho0r0vsgobs2}, 
\end{eqnarray}
with units [$\rho_0 r_0$] = $\mathrm{M_\odot pc}^{-2}$ and [$g_{\mathrm{obs}}$] = $\mathrm{ms}^{-2}$. This implies that $g_\mathrm{obs}$ is smaller at 3.2 $R_d$ than at 2.2 $R_d$, and indeed, on average the factor is about 0.73. The solid line in Fig. \ref{Fig8} (upper left panel) is for the combined sample, with a slope of 1.077 and only shown to guide the eye. In principle, given two kinematic observations for the circular velocities at two radii one may solve for the two halo parameters; however, as the scatter is moderate but considerable this won't work in practice. Nevertheless, the clear relationship implies that it is the inner halo surface density that is mainly responsible for the observed kinematics at most radii, in particular at outer regions. As will be discussed in Sect. 3.6, the central halo column density turns out to be an even more relevant ingredient for a structure-kinematics scaling relation. 

For the sake of completeness, we add the following note: for the 54 galaxies with RC data available out to the flat part, we similarly get $\log(\rho_0\,r_0) = 1.153\,\log(g_\mathrm{obs}) +14.355 \label{lgrho0r0vslggobs3}$, or equivalently, $\rho_0 r_0 = 2.265\cdot 10^{14}\, g^{1.153}_\mathrm{obs}$ at the radius $R_\mathrm{max}$, with units as above. We note that we have a median $r_0 \approx 0.64 R_\mathrm{opt} \approx 2 R_d$ and that on average $R_\mathrm{max}\approx 1.22 R_\mathrm{opt} \approx 3.9 R_d$. 

\subsubsection{Surface brightness and baryonic surface density}

The above fits do not significantly improve when one performs multilinear regressions  with additional quantities like $\mu_0$ or $\log C$ beside $g_\mathrm{obs}$ or if one takes $f_\mathrm{bary}$ as the independent variable. This is consistent with the only shallow relations shown in the lower panels of Fig. \ref{Fig8}: one perceives some gentle relationships with luminous matter central surface brightness and, correspondingly (Sect. 2.2), with the compactness parameter. OLSB fits yield
\begin{equation} 
	\log(\rho_0 \,r_0\,\,\mathrm{M_{\sun} pc^{-2}}) =-(0.406\pm 0.005)\, \mu_0+10.011,\,\,\,\,\,\,(\sigma=0.30) \label{SDSB}
\end{equation}
and $\log (\rho_0 \,r_0 \,\,\mathrm{M_{\sun} pc^{-2}}) =1.348 \log_{10}C+2.047$, $\sigma=0.30$. A similar statement holds for the case with $f_\mathrm{bary}$.

However, there are some hidden subleties involving the mass-to-light ratio and in particular the baryonic mass fraction. The halo central surface density dominates the central mass surface density, i.e., $\Sigma_\mathrm{halo}(r_0) \equiv \rho_0r_0 \approx \Sigma_\mathrm{dyn}(0)$, the stellar mass central surface density $\Sigma_\mathrm{stellar}(0)$ seems to be of no or of minor importance. This is evidenced in Fig. \ref{Fig8} (upper right panel) where the abscissa represents the central surface density estimates by \citet{Lelli16c} who applied the RC data provided by the SPARC database to a discretized version of Toomre's (1963) formula for a self-gravitating disk,
\begin{equation}
\Sigma_{\mathrm{obs}}(0) = \frac{1+q_0}{2\pi G}\int_0^\infty \frac{v_\mathrm{obs}^2(r)}{r^2}dr. \label{Toomre}
\end{equation}
\begin{figure*} 
	\centering
	\includegraphics[width=0.32\textwidth]{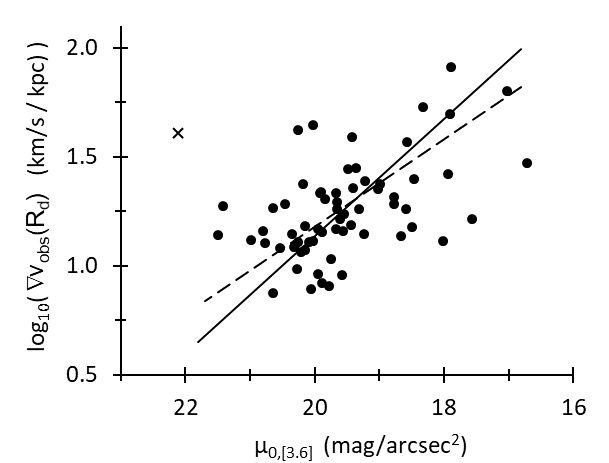}	
	\includegraphics[width=0.32\textwidth]{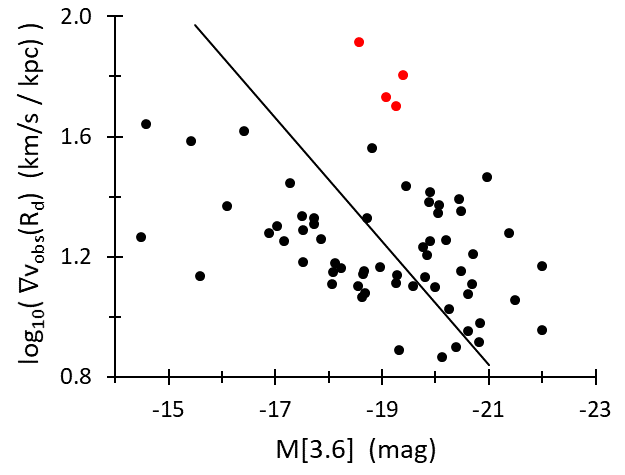}	
	\includegraphics[width=0.32\textwidth]{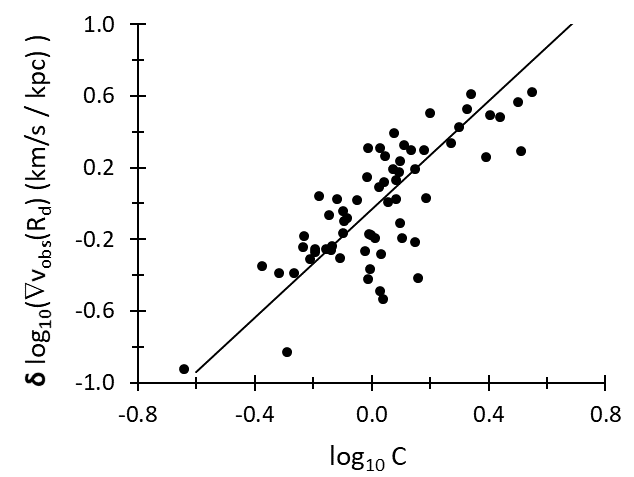}
	\caption{\small \emph{Left:} The circular velocity gradient at the galactocentric distance $R_d$ weakly correlates with extrapolated central surface brightness. Solid lines are in all panels OLSB fits. The cross marks UGCA281 that has been arbitrarily excluded for the fit. For comparison, the dashed line has slope -0.2 corresponding to idealized theoretical expectations in the limit $r\rightarrow 0$ (Lelli et al. 2013). \emph{Middle:} The velocity gradient versus absolute magnitude diagram shows that, roughly, brighter galaxies have smaller gradients. \emph{Right:} The vertical deviations from the mean line shown in the middle panel correlate with the compactness parameter C: at a given luminosity (and consistent with the left diagram), more compact (HSB) galaxies go with steeper inner circular velocity gradients than less compact (LSB) galaxies.\normalsize} 
	\label{Fig9}
\end{figure*}
Herein $q_0 \in [0.15, 0.40]$ is some axial ratio that is a function of the stellar mass and that accounts for finite disk thickness. The data for $\Sigma_{\mathrm{obs}}(0)$ (=$\Sigma_{\mathrm{dyn}}(0)$) are available at the SPARC homepage. Obviously, the points group along the 1:1 line (shown as dashed line), hence the two very different derivations for the central surface density nevertheless coincide. The central surface density for an artefactual disk tightly correlating with our spherical halo central surface density is owed to the fact that our sample galaxies are DM dominated at all galactocentric radii with the rotation curve crucially shaped by the halo velocity component. Hence the central density as determined from a distribution extending from inner to outer radii (equation \ref{Toomre}) is rather insensitive to baryonic mass, and the equivalence $\log(\Sigma_\mathrm{obs}(0))\approx \log(\rho_0r_0)$ must not come as a surprise.

The baryonic mass fraction not only links the observed with the baryonic radial acceleration (Sect. 3.1), the observed and the baryonic surface density show a similar kinship: for the mean surface density holds $\Sigma\propto M/R^2$, providing a ratio $\Sigma_\mathrm{bary}/\Sigma_\mathrm{obs}= (M_\mathrm{bary}/R^2)/ (M_\mathrm{obs}/R^2)= M_\mathrm{bary} /M_\mathrm{obs}= f_\mathrm{bary}$. Analogous to equation (\ref{gobsgbary}) we thus expect a two-surface densities relation (SDR)
\begin{equation}
\Sigma_\mathrm{obs}(0)=\Sigma_\mathrm{bary}(0)\left( 1+\frac{\Sigma_0}{\Sigma_\mathrm{bary}(0)}\right)^{\delta} \label{SigmaobsSigmabary}
\end{equation}
at $r=0$. Indeed, rewriting the relationship found by \citet{Lelli16c} between dynamical and stellar central surface density (their equation 10) with fixed parameter values $\alpha = 1$ and $\Sigma_\mathrm{crit}\approx \Sigma_0$ (as given by their equation 11) in the form of relation (\ref{SigmaobsSigmabary}), we may identify $\delta=0.38\pm0.04$ and $\Sigma_0 \approx 1500$ M$_{\sun}\,$pc$^{-2}$. We note that the two bracketed, single-power law factors in equations  (\ref{gobsgbary}) and  (\ref{SigmaobsSigmabary}) both are equal to the inverse baryonic mass fraction (or, equivalently, to the mass deficiency) and hence should be in principle equal to each other. 

Linking surface density with surface brightness according to the substitution $\Sigma_\mathrm{bary}(0) \propto \Upsilon I_0 = \Upsilon 10^{-0.4\mu_0}$ \citep[e.g.,][]{Bakos08, Swaters14} and given the equivalence $\log(\Sigma_\mathrm{obs}(0))\approx \log(\rho_0r_0)$, equation (\ref{SigmaobsSigmabary}) reads in logarithmic terms 
\begin{equation} 
\log(\rho_0\,r_0)=-0.4\,\mu_0 + \log\left( 1+\frac{\Sigma_0}{\Sigma_\mathrm{bary}(0)}\right)^{\delta} +\log\Upsilon + \mathrm{const}, \label{SDSB2}
\end{equation}
The $-0.4\mu_0$-dependence seen in equation (\ref{SDSB}) is recovered. The bracketed term in equation (\ref{SDSB2}) ---equivalent to the inverse baryonic mass fraction--- must be responsible for the trend observed in the $\mu-\Sigma$-diagram by \citet{Swaters14} and \citet{Lelli16c}, namely, that fainter galaxies with on average lower baryonic content lay above the one-to-one correspondence line. 

The eight galaxies of \citet{Karukes17} that are in common with ours (i.e., contained within the SPARC database) nicely fit the above relation as well (open rhombic symbols in Fig. \ref{Fig8}, upper right hand panel). This is somewhat surprising, because in their approach they apply averaged Burkert halo core sizes that are directly reused to calculate central halo density. Their Burkert halo core sizes $r_0$ are all constrained to lay on the line defined by $r_0=2.95\,R_d^{1.38}$ (equ. \ref{r0KS17} below), corresponding to the short dashed line in the right-hand panel of Fig. \ref{Fig12}; this is in sharp contrast to our fitted parameter values that exhibit a disperse distribution without any correlation (Fig. \ref{Fig12}). Hence their approach generates similar values for the central surface density despite adopting different halo parameter values. Remarkably, using their own disk mass-dependent compactness parameter, adopting the Burkert halo for the RC decompositions, and making in addition various  model assumptions, \citet{Karukes17} and in their footsteps \citet{DiPaolo18} manage to find a nearly perfect correlation among the three (logarithmized) variables $\rho_0$, $r_0$, and $C$. We cannot reproduce such a desirable result with our different treatment of $r_0$ and $C$ (the discussion below in Sect. 3.7.1 will take up this point again). Our most promising candidates for a third variable are $g_\mathrm{obs}$ or, more convincingly, $v_\mathrm{max}$ (Sect. 3.2). 

\subsection{Inner circular velocity gradient}
\subsubsection{Definition}

If two galaxies are located at the same spot in the TF-diagram (thus having about the same absolute magnitude and hence about the same maximum circular velocity $v_\mathrm{flat}$) they may nevertheless have different disk scale lengths and hence differ in size. For a smaller, i.e., for a HSB galaxy this implies a faster increase of the RC at small radii, while a relatively more extended LSB galaxy is expected to have a more moderate increase. Stated differently, at small galactocentric distances we expect the velocity gradient to depend on the compactness parameter or, correlating with it as discussed in Sect. 2.2, with the central surface brightness. We will investigate this and other issues in the next subsection. For that purpose we construct the following estimator for the one-dimensional circular velocity gradient at one disk-scale length: on the RC diagram with points described by the coordinates ($r$;\,$v(r)$) we interpolate the three selected points (0;0), (2.15 $R_d$; $v_\mathrm{obs}(2.15 R_d)$), and (3.2 $R_d$; $v_\mathrm{obs}(3.2R_d)$) (as listed in Table \ref{TableB3} the Appendix) by means of a parabola $v(r)=a_2r^2+a_1r+a_0$ (where $a_0=0$ due to the first point) with slope function $dv(r)/dr=2a_2r+a_1$. Solving the implied two-dimensional system of equations for $a_1$ and $a_2$ and inserting the radial values, one has at the selected radius $r=R_d$ the velocity gradient $\nabla v(R_d)\equiv \frac{dv(r)}{dr}\vline_{\,r=R_d}$ 
\begin{equation}
\nabla v_\mathrm{obs}(R_d) \approx \frac{0.532\,v_\mathrm{obs}(2.15 R_d)-0.045\,v_\mathrm{obs}(3.2 R_d)}{R_d}.\label{velograd}
\end{equation}

Applying this measure to the 65 galaxies with RC data available out to at least the optical radius $3.2\,R_d$ (see Table \ref{TableB3}) we find a median value for the gradient at one $R_d$ of about $17$ km$\,$s$^{-1}$$\,$kpc$^{-1}$ ($\sigma = 13$). As a consistency check, \citet{Swaters09} calculate the mean logarithmic slope between 2 and 3 scale lengths for 48 galaxies by means of $S_{(2,3)}=\log_{10}(v(3R_d)/v(2R_d))/\log_{10}(3/2)=0.32\pm0.26$ which fully agrees with the value obtained similarly for 65 of our galaxies, $S_{(2.15,3.2)}=0.35\pm0.25$.

\subsubsection{Dependencies with photometric quantities}

The inner circular velocity gradient only shows shallow linear trends with photometric quantities (disk scale length, central surface brightness, absolute magnitude) or with halo variables (core size, core density, and central surface density) if inquired pairwise. For example, the inner velocity gradient at $R_d$ weakly scales with the extrapolated central surface brightness of the luminous matter (Fig. \ref{Fig9}, left panel). An OLSB fit (shown as solid line) yields
\begin{equation}
\log\nabla v_\mathrm{obs}(R_d) =(-0.269\pm 0.034)\,\, \mu_{0,[3.6]}+6.519.
\label{nablav1}
\end{equation} 

This observational evidence for the late-type spirals in the SPARC sample adds to similar findings of several research teams \citep{Tully77, deBlok96, Garrido05, Swaters09, Lelli13}. In particular, the latter apply a velocity gradient measure valid for $r\rightarrow 0$ to a selection of the full SPARC sample, obtaining an OLS fit given by $\log \nabla v(0) = (-0.22\pm0.02) \mu_{0,C}+(6.28\pm 0.40)$ (with $\mu_{0,C}$ being not the extra\-polated but the true central surface brightness in the R-band). An identical slope is found by \citet{Lelli13} for dwarf galaxies, $\log(v(R_d)/R_d)= (-0.22\pm0.03) \mu_{0,R}+(6.4\pm 0.6)$  (with $\mu_{0,R}$ being the inclination corrected extrapolated central surface brightness in the R-band). Within the errors our observed off-center gradient at one $R_d$ (equation \ref{nablav1}) is consistent with this (at the steep end for the slope, however). According to equation (8) in \citet{Lelli13}, the scatter in the diagram may be attributed to differences in the mass-to-light ratios, baryonic mass fractions, disk thicknesses, and geometrical mass distributions for the individual galaxies. Irrespective of the detailed interrelationsships among these quantities, together with the results of Sect. 2.2 concerning compactness and the clear-cut distinction of LSB and HSB galaxies equation (\ref{nablav1}) reminds for the moment of the following well-known interpretation: comparing two galaxies with about the same absolute magnitude (and, according to the Tully-Fisher relation, with about the same maximum circular velocity $v_\mathrm{max}$), the galaxy with the brighter central surface brightness (and, correspondingly, with higher compactness) exhibits a steeper inner rotation curve than the less compact galaxy with the fainter central surface brightness  (Fig. \ref{Fig9}, left panel). With caution one may state that on average LSB galaxies have smaller inner circular velocity gradients than HSB galaxies (Table \ref{Table1}). We will add refined evidence to this statement immediately below and again after having introduced a novel RC parameter in Sect. 3.7.2.

As a second example, we consider the circular velocity gradient versus absolute magnitude diagram (Fig. \ref{Fig9}, middle panel). Formally, $\log \nabla v_\mathrm{obs}(R_d)=0.206\,M_{[3.6]}+5.161$. The spread is large, but the general trend is, maybe counterintuitively, that fainter galaxies have steeper velocity gradients. But consistent with the discussion above, the  vertical deviations from this OLSB-fit line correlate with the luminous compactness of the galaxies (Fig. \ref{Fig9}, right-hand panel); formally, $\delta \log \nabla v_\mathrm{obs}(R_d)=1.515\,\log(C)-0.034$. Hence at a given luminosity, more compact (i.e., HSB) galaxies do have steeper-than-mean circular velocity gradients than less compact (i.e., LSB) galaxies (Table \ref{Table1}). Luminous matter obviously is linked to some degree with the observed kinematic behaviour at the chosen inner radius.

However, a similar statement can be asserted about the influence of the dark matter. But as already mentioned above, while there are some shallow trends neither the halo core size or core density nor the central surface density shows any thrilling trend with the velocity gradient. For example, we have $\log \nabla v_\mathrm{obs}(R_d) = (0.81\pm 0.07)\,\log (\rho_0r_0)-0.41$. And the inner circular velocity gradient (at $R_d$) is completely independent of the baryonic mass fraction $f_\mathrm{bary}$ or of the baryon-mass-to-halo-mass ratio $f_\mathrm{bary}/(1-f_\mathrm{bary})$ (both measured at 2.2$\,R_d$). One would however expect to see such a dependence, based one the prediction of high-resolution simulations for feedback-driven galaxy evolution \citep{DiCintio14a}, if the baryonic mass fraction is measured at the virial radius (not tackled here). 

\subsubsection{Rotation curve geometry}
\begin{figure} 
	\centering
	\includegraphics[width=0.45\textwidth]{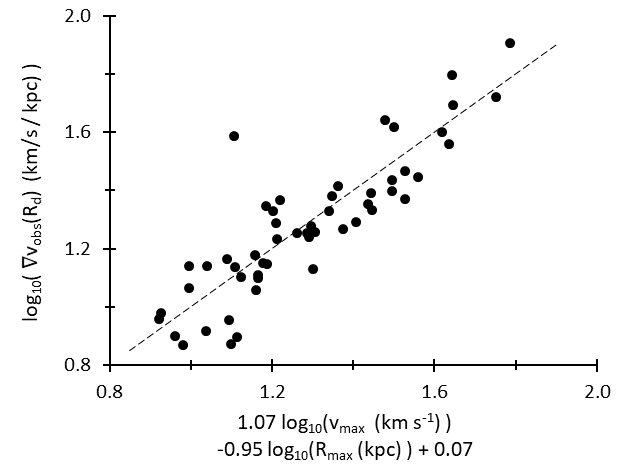}
	\caption{\small The circular velocity gradient at $R_d$ versus a combination of variables at $R_\mathrm{max}$. The dashed solid line shows the 1:1 correspondence. \normalsize} 
	\label{Fig10}
\end{figure}

Instead, there is a conspicuous connection of the inner circular velocity gradient with a couple of outer radii variables. In particular, with the help of a multiple-linear regression we retrieve the circular velocity gradient versus maximum velocity (VGMV) relation
\begin{equation}
\log\nabla v_\mathrm{obs}(R_d) =\log\left( \frac{v_\mathrm{max}^{1.07\pm 0.12}} {R_\mathrm{max}^{0.95\pm 0.07}} \right) + 0.07\pm0.19, \,\,\,\,(\sigma = 0.12)
\label{nablav2}
\end{equation}
with the gradient evaluated at $R_d$ (depicted in Fig. \ref{Fig10}). This relation links the inner part of an average RC with its outer part. Within the errors we simply have $\nabla v_\mathrm{obs}(R_d) \approx v_\mathrm{max}/R_\mathrm{max}$. Actually, \citet{Navarro96} (their Fig. 10) and especially \citet{Kravtsov98} already noted a correlation of $v_\mathrm{max}$ and $R_\mathrm{max}$, the latter particularly for LSB galaxies and dwarfs. Here we state this law more precisely by identifying the galaxy dependent factor of proportionality by means of another kinematic variable, i.e., $v_\mathrm{max}\approx \nabla v_\mathrm{obs}(R_d) \, R_\mathrm{max}$. Unfortunately, the spread of $\sigma = 0.12$ in the logarithmic relation (\ref{nablav2}) translates into a factor 1.32 for the spread in $\nabla v_\mathrm{obs}(R_d)$, hence for practical purposes relation (\ref{nablav2}) is of limited predictive power. A difficulty in practice contributing to the uncertainty is to adequately decipher the value of $R_\mathrm{max}$. Simulations may nevertheless use the VGMV-relation as a validity check, because from a theoretial viewpoint it essentially constitutes a formal statement of statistical value about the universal geometry of a typical RC.

\subsection{A baryonic TF relation for varying radii}
\begin{table}\centering
	\small
	\begin{tabular}{lcc}
		\hline
		& LSB		&	HSB  \\
		\hline
		disk: 	& & \\
		$\mu_{0,[3.6]}$	& $>19.6$ 		& $<19.6$\\
		$R_d\,\, / <R_d> \,\,\vline_{M_{[3.6]}} = 1/C$ & $>1$ & $<1$\\				
		$f_\mathrm{gas}$ & $>0.5$ & $<0.5$ \\
		$<f_\mathrm{bary}>\,\vline \,_{r=2.2\,R_d}$ & $0.28\pm0.03$  & $0.28\pm0.03$ \\		
		$v_{\mathrm{disk}}/v_\mathrm{obs}\,\,\vline \,_{r=2.2\,R_d}$ &
		$0.32\pm0.02$ & $0.40\pm0.02$ \\		\hline	
		halo: 	& & \\ 
		$\rho_0 r_0\,\,/ \,<\rho_0 r_0>$	 & $<1$  & $>1$ \\ 
		$v_{\mathrm{halo}}/v_\mathrm{obs}\,\,\vline \,_{r=2.2\,R_d}$ & $0.83\pm0.02$ & $0.84\pm0.03$ \\
		\hline
		$\nabla v_\mathrm{obs}\,\,/\,<\nabla v_\mathrm{obs}> \,\,\vline \,_{r=\,R_d}$ &$<1$ & $>1$\\
		$\Upsilon_{[3.6]}$	& $0.2\,(-0.8)$ & $0.2\,(-0.8)$\\
		\hline \hline
	\end{tabular}
	\caption{\small Distinction of LSB and HSB galaxies (morphological types $T\ge 8$) with respect to selected \emph{average} structural and kinematical features. An intermediate group of ISB galaxies is omitted here for brevity.  The mean extrapolated central surface brightness $<$$\mu_{0,[3.6]}$$>$$ =19.6\pm0.1$ $\mathrm{mag}\,\mathrm{arcsec}^{-2}$ is a universal value that holds irrespective of absolute magnitude (given $T\ge 8$). The entry for the baryonic gas fraction $f_\mathrm{gas}=M_\mathrm{gas}/M_\mathrm{bary}$ is according to \citet{Schombert11} and \citet[][their Fig. 7]{Sorce16}, in agreement with the mean values $0.57\pm0.04$ ($\sigma=0.24$) and $0.37\pm0.04$ ($\sigma=0.24$) for our LSB and HSB subsamples, respectively. The median central surface brightness is $<\rho_0 r_0>=121$ M$_\odot\,\mathrm{pc}^{-2}$ and the mean values for the LSB and HSB subsamples are $94\pm8$ ($\sigma=56$) and $216\pm27$ ($\sigma=162$), respectively.  The median circular velocity gradient at $R_d$ is $<\nabla v_\mathrm{obs}(R_d)> =17\pm3$ km$\,$s$^{-1}$$\,$kpc$^{-1}$, and the mean values for the LSB and HSB subsamples are $16\pm1$ ($\sigma=8$) and $27\pm3$ ($\sigma=17$), respectively. Neither the other kinematic ratios given here nor the baryonic mass fractions or the mass-to-light ratios are helpfull for a LSB-HSB-distinction. No colour distinction in terms of LSB galaxies being bluer than HSB galaxies is included.}
	\label{Table1}
\end{table}
\begin{figure*} 
	\centering
	\includegraphics[width=0.45\textwidth]{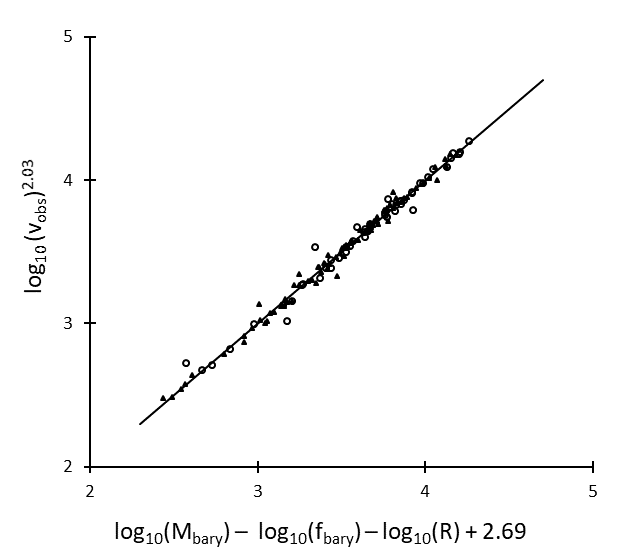}
	\includegraphics[width=0.45\textwidth]{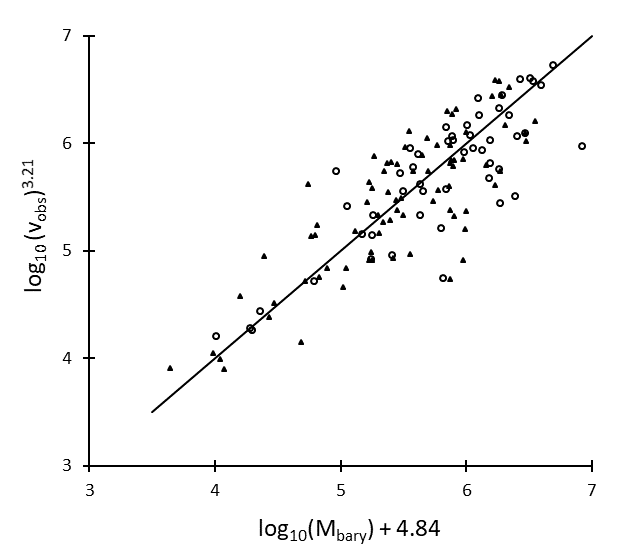}
	\caption{\small \emph{Left:} DM adjusted baryonic Tully-Fisher relation: observed versus calculated circular velocity at galactocentric radii $R=2.15\,R_d$ (filled symbols) and $R=3.2\,R_d$ (open symbols). According to an OLSB fit, the observed velocity is taken to the power of 2.03$\pm$0.03. \emph{Right:} Baryonic Tully-Fisher relation without adjustment (i.e., omitting  variables other than $\log(M_\mathrm{bary})$) for the same data as shown in the left panel. An OLSB-fit provides the velocity exponent 3.21$\pm0.23$. In both panels solid lines are 1:1 lines. \normalsize} 
	\label{Fig11}
\end{figure*}

Writing for any radius $v_\mathrm{obs}^2\,$ = $\,v_\mathrm{bary}^2\,$ + $\, v_\mathrm{halo}^2\,$ = $\,v_{\mathrm{bary}}^2
  ( 1$$+$$(\frac{v_\mathrm{halo}}{v_\mathrm{bary}})^2 ) $ = $\,v_\mathrm{bary}^2/f_\mathrm{bary}$ and inserting  $v_\mathrm{bary}^2$ = $G\,M_\mathrm{bary}(R)/R$, 
one readily gets the theoretical relation for a DM-adjusted baryonic Tully-Fisher relation (aBTFR)
\begin{equation}
 \log v_\mathrm{obs}^2(R)=\log M_\mathrm{bary}(\le R)-\log f_\mathrm{bary}(R) -\log(R) + \mathrm{const}.\label{TF}
\end{equation}
It is the second term on the right-hand side of this equation, i.e., the mass-degeneracy term, that regulates the explicit and non-neglectable interplay of the baryonic matter with the dark matter dominated total circular velocity. The baryonic mass is calculated dynamically by means of $M_\mathrm{bary}(\le R)=(\Upsilon_{[3.6]}v_\mathrm{stellar}^2(R)+ v_\mathrm{gas}^2(R))R/G$ and the mass-fraction $f_\mathrm{bary}(R)$ stems from equation (\ref{fbary}). As shown below, adopting instead in equation (\ref{TF}) a \emph{mean} mass-fraction or even omit it would result in a different slope together with a considerably increased scatter. Relation (\ref{TF}) allows for the choice of a particular galactocentric position. \citet{Sancisi04} noted such a radial or local applicability of the TFR, nowadays sometimes called "Renzo's rule" \citep[as in, e.g.,][]{Lelli16a}, and an optical radial TF relation was exploited in detail by \citet{Yegorova07} finding increasing slopes with increasing radii. The DM contribution being relevant at intermediate radii was shown by \citet{McGaugh07}, too. Leaving in equation (\ref{TF}) both the exponent of the velocity and the unspecified constant on the right-hand side as free fitting parameters, an OLSB-fit yields for the amalgamated data shown in Fig. \ref{Fig13} (left)
 \begin{equation}
\log (v_\mathrm{obs}(R)^{2.03\pm 0.03})=\log M_\mathrm{bary}(\le R)-\log f_\mathrm{bary}(R) - \log(R) + \mathrm{2.69}.\label{TF2}
\end{equation}   
We note that $R \in \{2.2\,R_d,\,3.2\,R_d\}$. The consistency of our approach is seen by means of the small scatter. The implied slope of $2.03\pm 0.03$ does not reconcile with the theoretical slopes of 3 to 4 given in the literature. Obviously, those approaches usually ignore the adjusting terms and directly plot the $\log (v_\mathrm{obs}) - \log (M_\mathrm{bary})$-diagram. Thus, omitting these terms for the same data as above results in 
 \begin{equation}
\log (v_\mathrm{obs}(R)^{3.21\pm 0.23})=\log M_\mathrm{bary}(\le R)+\mathrm{4.84}\label{TF3}
\end{equation}   
and in the scattered pattern of data points shown in Fig. \ref{Fig11} (right). The slopes for the 2.2$R_d$- and for the 3.2$R_d$-subsample are 3.12 and 3.26, respectively. Within the errors given in \citet{Yegorova07} our [3.6]-band values partly agree with their I-band optical TFR values at the correspoding radii, namely $2.93\pm0.14$ and $3.28\pm0.03$. Accordingly and for I-band data as well, \citet{Dutton07} mainly use $v_\mathrm{max}$ and get for a large sample of galaxies a slope of 3.33. Using $M_\mathrm{star}$ instead of $M_\mathrm{bary}$ and adopting the velocity measure at 2.2$R_d$, \citet{Dutton11} report a stellar TFR for mostly massive (and hence gas-poor) galaxies with a steeper slope of 3.86$\pm$0.16. Recently, \citet{Ponomareva18} adopted $\Upsilon_{[3.6]}=0.35$ to determine stellar masses from luminosity according to $M_\mathrm{star}= \Upsilon_{[3.6]}\,L_{\odot,[3.6]}$ and obtained a slope for the BTFR of $3.0\pm0.2$. This is rather similar to our finding. In quest for a third variable they surmise similar to \citet{Papastergis16} that the gaseous content particularly in gas-rich galaxies may be responsible for smaller slopes. In general, one finds systematically lower values for the BTFR as compared to the TFR \citep{Bell01, Gurovich10}: the latter, for example, weight their results with the Cepheid distances-based values of \citet{Sakai00} and use bivariate fits to get a BTFR-slope of $3.2\pm0.1$. So far, we are in good company with our low value for the slope of the BTFR.

Subtracting equation (\ref{TF2}) from equation (\ref{TF3}) we learn that 
\begin{equation}
f_\mathrm{bary}(R) \approx 7.08\cdot 10^{-2}\,\frac{v_\mathrm{obs}(R)^{1.18}}{R},
\label{fbary2}
\end{equation}
\begin{figure*} 
	\centering
	\includegraphics[width=0.48\textwidth]{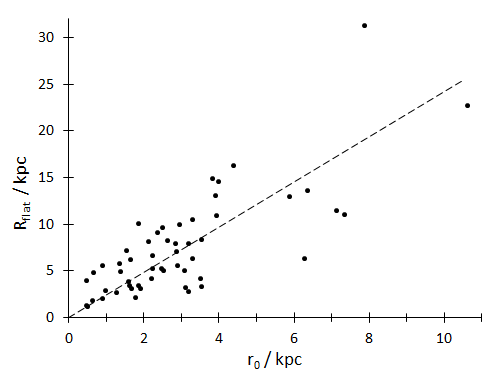}
	\includegraphics[width=0.48\textwidth]{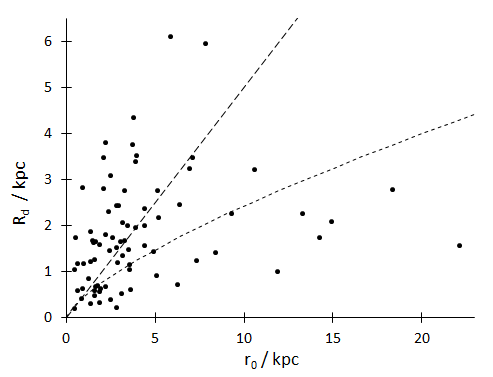}
	\caption{\small \emph{Left:} The radius $R_\mathrm{flat}$ where the flat (or moderately declining) regime of the RC starts is weakly related to the core size. The slope of the dashed line is equal to the sample median ratio $R_\mathrm{flat}/r_0$ = 2.4 (54 galaxies). \emph{Right:} Exponential disk scale length of the luminous matter versus DM halo scale length. The slope of the long-dashed line is equal to the median ratio $R_\mathrm{d}/r_0$ = 0.65 for the full sample (79 galaxies). For comparison, the pairs of constrained values used by \citet{Karukes17} within their URC approach would all lay along the short-dashed line.  
		\normalsize} 
	\label{Fig12}
\end{figure*}		
at $R \in \{2.2\,R_d,\,3.2\,R_d\}$ and where the values entering the right-hand side can be read off the RC of a galaxy. This formula implies a decrease of the baryonic mass fraction towards outer galaxy radii \citep{deBlok08}. If the exponent in this relation would be 2 (instead of 1.18) we would precisely recover the  baryonic mass fraction versus radial acceleration relation discussed in \citet{Lelli16b}: the theoretical relation (\ref{TF}) can be written with the usual Ansatz $M_\mathrm{bary}=A\,v_\mathrm{obs}^4$ (where $A$ is some constant of normalization), if one identifies $f_\mathrm{bary}(R)= v_\mathrm{obs}(R)^2/R$ (up to a constant). Comparison with our empirical relation (\ref{fbary2}) reveals that our data do not recover this idealized expectation, because empirically $f_\mathrm{bary}(R)\ne g_\mathrm{bary}(R)$. However, as mentioned above, at \emph{larger} radii (like $R_\mathrm{max}$, where we have $<$$R_\mathrm{max}$$>$$\,\approx 5.3 R_d$ for our sample) one may expect steeper slopes for the nonadjusted BTFR. This indeed would be in accordance with recent results for observations at 3.6 $\mu$m:  \citet{Lelli16b} found slopes between 3.7 and 4.0 for the BTFR of selected SPARC galaxies, and in a multi-wavelength TF relation study with 32 spiral galaxies of all Hubble types, and \citet{Ponomareva17} found a slope of $3.8\pm0.1$. Both these authors adopt $v_\mathrm{flat}\equiv v(R_\mathrm{max})$ for the velocity measure (according to \citet{Lelli16b}, this choice minimizes the scatter), use mass-to-light ratios around 0.5, and consider errors in both directions for the fitting procedure. In order to have a more consistent picture, the generalization of relation (\ref{fbary2}) would imply an exponent that itself depends on $R$, i.e., $f_\mathrm{bary}(R) \propto v_\mathrm{obs}(R)^{f(R)}/R$, for example a linear function $f(R)=[(2-1.18)/(R_\mathrm{max}-2.7R_d)] \,(R-R_\mathrm{max})+2$. However, instead of entering constructions like this one may question the accuracy and unambiguity of the data and the processing procedure. The slope of the optical TF relation is known since its discovery \citep{Tully77} to increase with longer wavelengths (this effect nearly diminishes for gas-rich galaxies, where the stellar mass or luminosity plays a secondary role), to depend on Hubble type (less so in the mid-IR bandpass), and to depend on the choice of the velocity measure. For example, in agreement with earlier findings of \citet{Geha06} on the BTFR, \citet{Swaters09} emphasize that "late-type dwarf galaxies do not appear to obey the [optical] TFR as derived for brighter spiral galaxies", implying shallower slopes. Our low value found for our sample of late-type spirals and dwarf galaxies falls into this category.  The choice of the velocity measure was shown by several teams to indeed significantly influence the resulting slope and hence to produce rather different baryonic TF relations \citep[e.g., ][]{Trachternach09, McGaugh12, Brook16, Bradford16}. The same holds true if different mass-to-light ratios are allowed for \citep{Bell01, McGaugh05, Ponomareva18}. We therefore think that our values $v(R) < v_\mathrm{max}\equiv v(R_\mathrm{max})$ and our granted for semi-free best-fit mass-to-light ratios $\Upsilon_{[3.6]}$ do have a decisive impact on the slope as observed. In addition, the fitting procedure being a crucial means in the determination of the BTFR slope was already emphasized by others \citep{McGaugh12, Lelli16b}. Moreover, our crude selection criteria for the inclusion of a galaxy into our sample (in particular with respect to inclination) may have a more pronounced effect than expected \citep{McGaugh12}. 

After these considerations, it is rewarding that all of these disturbing implications seem to become unimportant as soon as the third variable, i.e., the $\log(f_\mathrm{bary})$-term, is no more ignored (equation \ref{TF2}). (The fourth variable, $R$, actually only contributes scatter-increasing shifts if different radii are involved, as in our amalgamated data set, otherwise it only contributes to the constant). Incorporating this quantity (that is calculated by means of photometric and kinematical information to obtain $v_\mathrm{bary}$ and $v_\mathrm{halo}$, respectively) restitutes a nonambiguous aBTFR with rather small scatter. It is to emphasize that this same adjusted BTFR is observed for HSB and LSB galaxies (with HSB and LSB as defined in Sect. 2.2), in accordance with earlier findings for the BTFR \citep[e.g.,][]{Zwaan95, Sprayberry95, McGaugh98}. Our formulation circumvents the introduction of surface mass density or surface brightness (and the mass-to-light ratio), that could serve as a third (and fourth) variable and that can be shown to lead to a theoretical slope of 4, too (supposing constancy). The expenses of our approach are the reasonable but uncommon slope of 2 and, relying on a DM-related third variable, the restitution of the pure baryonic TFR as a hybrid TFR.

\subsection{Diversity of rotation curves}

\begin{figure*} 
	\centering
	\includegraphics[width=0.4\textwidth]{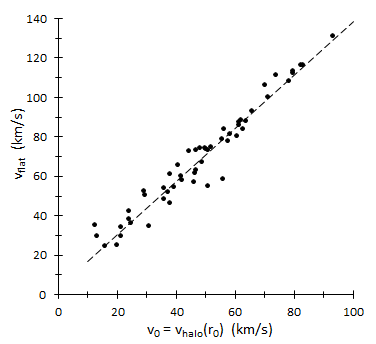}
	\includegraphics[width=0.55\textwidth]{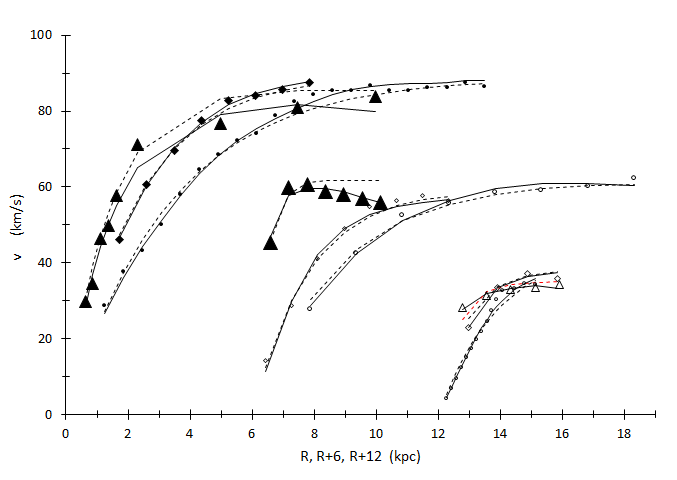}
	\caption{\small \emph{Left:} Maximum velocity $v_\mathrm{max}$ of the observed RC at the beginning of the flat (or decreasing) regime versus calculated halo contribution $v_0 =  v_\mathrm{halo}(r_0) \propto\sqrt{\rho_0r_0^2}$. The dashed line corresponds to the OLSB-fit given by equation (\ref{vmax}).
	\emph{Right:} Diversity of RCs, illustrated for selected galaxies with RCs grouped in triplets according to similar values of $v_\mathrm{max}$ = 86$\pm$2, 59$\pm$1.5, and 35$\pm$1  km$\,$s$^{-1}$. To avoid superposed RCs for the different groups, galactocentric radii are shifted by 6 and 12 kpc for the second and the third group, respectively. Within each triplet the halo core sizes $r_0$ (and correspondingly $R_\mathrm{max}$) of the galaxies increase from left to right providing RCs with successively more extended curvatures (CRRC shape relation). Shown are observed data points (symbols) and the mass modelled RCs as explained in Sect. 2 (solid lines) as well as the heuristic URCs that will be introduced in Sect. 3.6.2 (dashed lines). Symbol sizes are about proportional to the compactness parameter $C$ (in order to illustrate the CRC relation), and filled (open) symbols correspond to HSB (LSB) galaxies (following the distinction already adopted in Fig. \ref{Fig1} and presented in Table \ref{Table1}). The galaxies (and in brackets their Burkert halo core sizes $r_0$/kpc) are: UGC 191 (1.9), UGC 6399 (3.2), and NGC 55 (5.9) for the first group; UGC 7690 (0.64), UGCA 442 (2.9), and DDO 17 (3.9) for the second group; UGC 9992 (0.49), D512-2 (1.6), and KK98-251 (3.2) for the third group. \normalsize} 
	\label{Fig13}
\end{figure*}
The RCs of galaxies with about the same absolute magnitude or maximum circular velocity may look rather different. In particular and as discussed in Sect. 3.4, they may exhibit RCs with different inner velocity gradients and correspondingly with different $r_\mathrm{max}$ \citep{deBlok08, Oman15}. \citet{Verheijen99} attributed these different RC shapes to either LSB or HSB galaxies and spoke of "a kinematic dichotomy between LSB and HSB galaxies within similar halos". It is, however, the different structure of the halo that may explain this observation, as discussed now. Equation (\ref{rho0r0}) reads  $\rho_0 r_0 = (2\pi Gf(1))^{-1}\, g_\mathrm{halo}(r_0)$ at the radius $r=r_0$ ($k$=1), and  inserting $g_\mathrm{halo}(r_0)=v_0^2/r_0^{}$ leads to the relation 
\begin{equation}
v_\mathrm{0} \,\,(\mathrm{km\,s}^{-1}) = (2\pi G f(1))^{0.5} \rho_0^{0.5} r_0^{}.
\end{equation} 
Coincidentaly, the maximum velocity $v_\mathrm{max}$ representing the flat (or, in a few cases, the highest) part of the observed RC tightly correlates with this halo velocity contribution $v_0\equiv v_\mathrm{halo}(r_0)$, as shown in Fig. \ref{Fig11} (left). In particular, the dashed OLSB-fit-line obeys
\begin{equation}
v_\mathrm{max} = (1.352\pm 0.037)\,v_0+3.371. \,\,\,\,(\mathrm{in\,\,km\,s^{-1}})  \label{vmax}
\end{equation}
This relation, holding for late-type spirals and dwarf disk galaxies, is rather tight. Irrespective of the baryonic matter the value of the observed constant circular velocity is dictated by the velocity contribution due to the DM contained in the core of the halo. This is somewhat surprising because of the core radius $r_0$ typically being on average a factor of 2.8 smaller than the starting radius $R_\mathrm{max}$ of the flat regime of the RC; more precisely, $<$$R_\mathrm{max}/ r_0$$>$ = 2.8$\pm$1.5, with a median at 2.4 (see Fig. \ref{Fig12}, left). The inner halo structure determines about the outer kinematics. The minor influence of the baryonic mass is reflected in the \emph{adjusted} TFR where the inclusion of the typiclly low-valued baryonic mass fraction, or equivalently, of the high ratio $v_\mathrm{halo}/v_\mathrm{bary}$ is crucial, as discussed in the previous section.

As a consequence of the previous couple of empirical relations, two galaxies with different halos, i.e., with $\rho_{0,1} \ne \rho_{0,2}$ and $r_{0,1} \ne r_{0,2}$, will have approximately the same maximum circular velocity if the halo parameters correspond to the same central halo column density. We thus have a central halo column density versus maximum circular velocity (CDMV) relation
\begin{equation}
	v_\mathrm{max}^2\propto\rho_{0,1}^{} r_{0,1}^{2-\epsilon} \approx \rho_{0,2}^{} r_{0,2}^{2-\epsilon}.\label{CDMV}
\end{equation}
The quantity $\epsilon$ is to possibly adapt the value of the exponent to the empirical result obtained in equation (\ref{lgrho0lgr0lgvmax}), i.e., $\epsilon=0.161$; for the following discussion we set $\epsilon=0$. Before proceeding, we note that on theoretical grounds a relation alike to this one is expected for any spherical density distribution according to Gauss's law of gravity (or, equivalently, Poisson's equation), stating in differential form $\nabla g(r) =4\pi G\,\rho(r)$, hence, $v^2(r)/r = 4\pi G \int_0^{r} \rho(r')\,dr'$. Assuming here for the sake of simplicity a mean density $\rho(r) = \rho_0$, one gets $v^2(r)\propto \rho_0r^2$. If indeed $v_\mathrm{max}\propto v_0=v(r_0)$, as indicated by equation (\ref{vmax}) that informs on the kinematic dominance of the halo, we arrive at the CDMV relation.

According to relation (\ref{CDMV}), the galaxy with the larger core will have a correspondingly lower core density $\rho_0 \propto v_\mathrm{max}^2/r_0^2$. A larger core radius implies a more extended halo mass distribution,too, thus $v_\mathrm{max}$ is reached at a larger radius $R_\mathrm{max}$. The RC profile correspondingly has a smaller slope at small and intermediate radii, compatible with the findings in Sect. 3.4 concerning the inner RC gradient. In Fig. \ref{Fig13} (right-hand panel) a selection of RCs with comparable values for $v_\mathrm{max}$ are shown and indeed exhibit curvatures that are the more extended the larger the halo core radii are. For a verbal reference, we call this qualitative dependence the core radius versus rotation curve shape (CRRC) relation. (In Sect. 3.6.2 we will formulate a similar, but quantitative CRRC relation). In addition, for the majority of RCs shown the compactness of the luminous matter (as discussed in Sect. 2.2 and quantified by the parameter $C$) is decreasing with increasing halo core size, too (we note the symbol sizes to vary accordingly). DM halo core radius (and according to Sect. 3.3, core surface density as well) and LM compactness are (loosely) related, too. A (tight) correlation between these two quantities, which may be called the core radius versus compactness (CRC) relation, was already noted and quantified by others \citep{Lelli16c,Karukes17}.

The question arises whether different baryonic mass fractions may lead to different RC shapes. It turns out that $f_\mathrm{bary}$ (evaluated for galaxies with similar rotational velocities at $r=2.15 R_d$) is not correlated with $r_0$ and hence is not (or not directly) responsible for the varying RC shapes. Thus while the tightness of the aBTFR leaves no room for correlations with galaxy luminous structural parameters (i.e., scale length and central surface brightness) we cannot exclude yet some correlation with the halo structure as given by the column density.

The main results of this section, i.e., (i) the strong correlation between the observed (squared) maximum circular velocity and the central halo column density (CDMV relation), and (ii) the influence of the halo core radius on the RC shape (CRRC) shed some more light on the "unexpected diversity of dwarf galaxy rotation curves" \citep{Oman15}. Their Fig. 5 illustrates the diversity of RC shapes in a similar way as does our Fig. \ref{Fig11}. The involved "unexpectedness" stems from the observation that simulations based on stongly cuspy halo density profiles (like the gNFW profile advocated within the $\Lambda$CDM scenario) only may match the inner part of the RCs of low-mass galaxies if the corresponding decompositions with cored DM halos exhibit small core radii $r_0$. For larger cores with lower central core densities there's an inner mass deficit \citep{Oman15}. Maybe the ongoing core-vs-cusp debate finds some convergeing turn as soon as nonphysical singular cusps are omitted and only models with nonsingular cusps (in the sense of steep central core density profiles) are allowed for. We come back to this issue in Sect. 3.7.2. 
 
\subsection{Universal rotation curve}
\subsubsection{Conventional URC}

In order to obtain a "universal rotation curve" (URC) for irregualr spiral \emph{and} dwarf galaxies \citet{Karukes17} and \citet{DiPaolo18} use a Burkert halo for the DM contribution and normalize the galactocentric radius $r$ and the total velocity $v(r)$ by means of the optical radius $R_\mathrm{opt}=3.2\,R_d$ and the corresponding velocity $v(R_\mathrm{opt})$, respectively. A peculiarity of their approach is (i) the definition of the value of the halo core radius by means of a power-law of the exponential disk scale length that is (ii) reused to calculate the central halo density as follows:
\begin{eqnarray}
r_0^{KS}\,\,(\mathrm{kpc}) &=& 2.951\, (R_d\,\, (\mathrm{kpc}))^{1.38} \label{r0KS17}\\
\rho_0^{KS}\,\,(\mathrm{g/cm^3})&=& 7.24\cdot 10^{-24}(r_0^{KS}\,\, (\mathrm{kpc}))^{-0.97}\,C^{2.18}_{KS}.\label{rho0KS17}
\end{eqnarray}
Citing \citet{Salucci07} the first equation is said to hold on average for normal spirals with much higher masses and is seemingly adopted for dwarf disks under the assumption of universal applicability; hence $r_0$ is not a free parameter but determined by a luminous-disk parameter. The parameter $C_{KS}$ in the second equation is some compactness parameter defined analogously to our equation (\ref{C}) but depending both on $R_d$ and on the total disk mass. Because the latter is expected to be proportional to $10^{-0.4\mu_0}R_d^2$ for an exponential disk, both Burkert halo parameters strongly depend on $R_d$, with some scatter added due to the additional presence of $\mu_0$. In other words, in this approach the structure of the disk is claimed to fully determine the structure of the halo. Within this setting, \citet{Karukes17} present their universal rotation curve (URC) by means of 14 data points that are destilled from radially binned and averaged empirical data points and that are listed in their Table 2. We interpolate their discrete points by means of a 4$^\mathrm{th}$-order polynomial 
\begin{equation}
v_n(r_n) \approx -0.07\,r_n^4 + 0.50\,r_n^3 -1.39\,r_n^2 +1.93\,r_n, \label{URCKS17pts}
\end{equation}  
with normalized coordinates $r_n\equiv r/R_\mathrm{opt}$ and $v_n\equiv v/V_\mathrm{opt}$, where $R_\mathrm{opt}\equiv 3.2R_d$ and $V_\mathrm{opt}\equiv v(R_\mathrm{opt})$. In order to derive this regression, we additionaly assumed a 0th point at $(r_n;v_n)$=(0;0), corrected the 14th point to (1.88;1.19) (due to their Fig. 2), and added an extra\-polated 15th point at (2.5;1.234); the uncertainty measure of the fit (unweighted for errors in $v$) amounts to $R^2=0.9989$ and thus legitimates our approach. For radii larger than $r=2.5\,R_\mathrm{opt}=8\,R_d$ flatness at the level $v = 1.234\,V_\mathrm{opt}$ is assumed. Thus, given $R_\mathrm{opt}$ and $V_\mathrm{opt}$ of an individual galaxy, its continuous RC is presumed to be calculable as a scaled URC according to  
\begin{equation}
v_\mathrm{obs}(r) = v_n(r/R_\mathrm{opt})\,v(R_\mathrm{opt}). \label{URCKS17}
\end{equation}  
No individual halo parameters enter anymore this final description. However, for our sample $r_0$ and $R_d$ (or $R_\mathrm{opt}=3.2R_d$) actually do not correlate (see Fig. \ref{Fig12}, right-hand panel), contrary to the  assumption met in equation(\ref{r0KS17}). It is thus not surprising that superposing averaged RCs as described above onto the empirical data do not lead to very good individual matches in general; this can be seen in Fig. \ref{Fig2} (right-hand panels) for a couple of example RCs with superposed  URCs that were calculated according to the above prescription. Only if the value $r_0$ of our best-fit RCs and the one constrained according to equation (\ref{r0KS17}) accidentally coincide there may be some quantitative agreement, otherwise there are significant deviations. This undermines the conclusion of \citet{Karukes17} who claim that "the universality of this curve can be inferred from its very small rms values". The main advantage of the conventional URC with its double normalized axes lies, beside its claim of being applicable to all types of disk galaxies, in its decomposition that explicitely distinguishes between a luminous and a dark contribution. However, there are, in addition to the strict dependence on the luminous structure parameter $R_d$, three more uncertainties or ambivalencies, as \citet{DiPaolo18} concede: First, galaxies with RCs that do not match with this prescription are considered to be "outliers" that "are expected to be taken care by few other parameters". Second, their approach ignores the gas content of galaxies assuming that this contribution "is usually a minor component to the circular velocities" within the inner regions ($r<6\,R_d$) of galaxies. This assumption discriminates all low-luminous, but gas-rich galaxies like, e.g., D631-7 that will be an example galaxy of the next section. Third, the conventional URC does not unambiguously account for the different velocity gradients observed with individual galaxies of similar maximum circular velocities (their Figs. 2 to 4 in the inner region $r<3\,R_d$). However, as we reconfirmed by different means the observation of \citet{Lelli13} in Sect. 3.4 this is an observational constraint that needs to be incorporated in any successful model URC. We conclude that at intermediate radii the luminous and the non-luminous matter are not as tightly related  as claimed by \citet{Karukes17} and others \citep[e.g.,][]{Trachternach09, McGaugh16}.

\subsubsection{Alternative URC}	

Instead of solely refering to the luminous parameter $R_d$, it seems promising to look for an URC that either explicitely involves some DM halo parameter or that uses other structural or kinematic observables like $v_\mathrm{max}$ and $R_{max}$. I suggest to work with the following non-singular, spherically symmetric \emph{total matter} density function 
\begin{eqnarray}
\rho(r)&=&\frac{v_{max}^2}{4\pi G}\,\frac{1}{r^2}\left[ 1-\left(1-\frac{r}{r_c}\right)\,e^{-r/r_c}  \right]^2 
\label{URCrho1}\\
&=&\frac{v_{max}^2}{4\pi G}\left( \frac{1-e^{-r/r_c}}{r} + \frac{e^{-r/r_c}}{r_c} \right)^2
\label{URCrho2}
\end{eqnarray} 
with finite central density $\rho(0) = \lim \limits_{r\rightarrow 0}\rho(r)=v_\mathrm{max}^2 / (\pi G r_c^2)$, central density slope $d\rho(0)/dr$ $= \lim \limits_{r\rightarrow 0}(d\rho(r)/dr)  = -3v_\mathrm{max}^2 / (2\pi G r_c^3)$, and $\rho \propto r^{-2}$ at large radii (no inbuilt truncation yet). Herein $v_\mathrm{max}\equiv v(R_\mathrm{max})$ is the maximum circular velocity as simply inferred from the observed RC at the radius $r=R_\mathrm{max}$ (where the flat or declining part begins), and $r_c$ is some total matter scaling parameter governing the curvature of the RC and that is determined by means of fitting the RC. 

This total matter density profile scales the radius for an isothermal profile ($\propto r^{-2}$) by means of the squared-brackets factor; it thus becomes finite at the center and further out has a shallower slope as compared with the isothermal profile. The smaller $r_c$, the higher the central density; in the limit $r_c \rightarrow 0$ one recovers the isothermal profile. The slope function for the inner total matter density profile becomes 
\begin{equation}
\alpha_\mathrm{total}(x) \equiv 
\frac{d\,\log \rho(r)}{d\,\log \,r} \bigg\vert_{x \ll 1} \approx -\frac{3(1-3x/4)}{2(1-3x/2)}\,x, 
\end{equation}
with $x \equiv r/r_c$. For example, at radii $r/r_c = $ 0, 0.1, 0.15, and 0.2 the estimates are $\alpha_\mathrm{total} \approx $ 0, -0.16, -0.26, and -0.36, respectively; we note that on average $r_c\approx 0.9\cdot(2.15\,R_d)\approx 1.94\,R_d$. Such values are typical for the \emph{halos} of observed and simulated dwarf galaxies \citep{Kravtsov98, Oh11a,Oh11b}. The gravitational potential corresponding to the density given above will be composed of two additive contributions, one stemming from the factor proportional to $r^{-2}$ (isothermal sphere potential $\propto \ln\,r\,$) and a more complicated one related to the squared-brackets factor (involving among others a Yukawa-like potential and the lower incomplete gamma function). Here we go without it. The radial mass distribution $M(\le r)=4\pi \int_0^r r'^2 \rho(r') dr'$ can be written as 
\begin{equation}
M(\le r)= \frac{v_\mathrm{max}^2r_c}{G} \left( x\left(1-e^{-x}\right)^2
-\frac{x}{2}\left(1+x\right)e^{-2x}
+\frac{1}{4}\left(1-e^{-2x}\right)\right),
\label{URCM}
\end{equation}
with $M \propto r$ at large radii, and the circular velocity profile $v(r)=(GM(\le r)/r)^{1/2}$ becomes
\begin{equation}
v(x) = v_\mathrm{max}\sqrt{ \left( 1-e^{-x} \right)^2 + \frac{1}{2} \left( \frac{1-e^{-2x}}{2x}-(1+x)
e^{-2x} \right) },
\label{URCv}
\end{equation}
where $x\equiv r/r_c$. For (very) small radii $r$$\ll$$r_c$ one has $v(r)\propto r^{1/2}$ (implying a central logarithmic slope $d\,\log(v(r))/d\, \log(r)=0.5$) and for large radii $r$$\gg$$r_c$ one approaches a constant value $v(r)\approx v_\mathrm{max}$. No truncation function for larger radii is considered here. In the appendix we show how a couple of possible decompositions of this velocity curve into some DM halo and baryonic disk components fairly well agrees with observed kinematic data.

\begin{table}\centering
	\small
	\begin{tabular}{lcccc}
		\hline
		galaxy&$v_\mathrm{max}$&$r_c$&$M_{[3.6]}$&$\nabla v(R_d)$\\
		     & (km$\,$s$^{-1}$)& (kpc)& (mag)& (km$\,$s$^{-1}\,$kpc$^{-1}$)\\
		\hline
		UGC 191	&	83.9	&	1.6	& -20.01 & 22.6\\
		UGC 6399&	87.6	&	2.6	& -20.16 & 18.3\\
		NGC 55	&	86.8	&	3.8	& -20.92 &  7.6\\
		\hline
		UGC 7690&	60.7	&	0.5	& -19.09 & 53.6\\
		UGCA 442&	57.8	&	2.0	& -17.12 & 18.2\\
		DDO 170	&	60.0	&	3.1	& -18.59 & 11.8\\
		\hline
		UGC 9992&	34.3	&	0.7	& -18.07 & 15.3\\
		D512-2	&	35.9	&	1.0	& -18.04 & 14.3\\
		KK98-251&	34.6	&	2.3	& -16.58 & 13.7\\
		\hline
		D631-7	&	58.5	&	2.8	& -17.49 & 19.7\\
		F583-1	&	84.7	&	3.7	& -19.24 & 14.0\\
		UGC 128	&	131.0	&	6.0	& -21.96 &  9.2\\
		\hline \hline
	\end{tabular}
	\caption{\small URC parameter values $v_\mathrm{max}$ and $r_c$ for selected triplets of galaxies with similar $v_\mathrm{max}$. In addition, the absolute magnitudes $M_{[3.6]}$ of the galaxies (related to $v_\mathrm{max}$ according to the optical Tully-Fisher relation) and the inner velocity gradients $\nabla v(R_d)$ (anticorrelated with $r_c$ that acts as a measure for the extension of a RC curvature and that correlates with $r_0$) are reproduced for comparison. Corresponding RCs are shown in Figs. 2, 11 (right), and A.1. \normalsize}
	\label{Table2}
\end{table}

\begin{figure*} 
	\centering
	\includegraphics[width=0.32\textwidth]{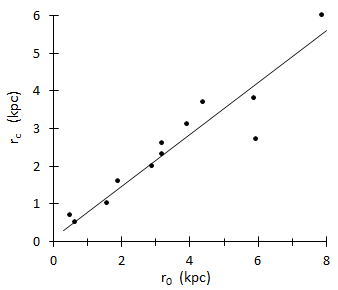}		\includegraphics[width=0.32\textwidth]{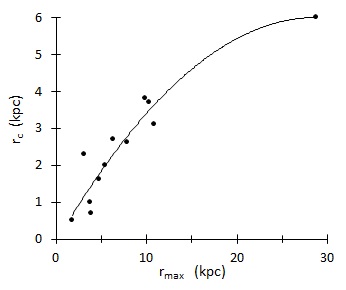}	\includegraphics[width=0.32\textwidth]{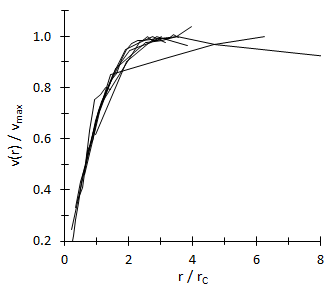}
	\caption{\small Universal RC for selected galaxies. \emph{Left:} The scaling parameter $r_c$ for the one-parameter heuristic RCs (some shown in Fig. \ref{Fig13}, right-hand panel, as dashed lines) versus the Burkert halo scaling parameter $r_0$ (corresponding to the RCs shown in Fig. \ref{Fig13}, right-hand panel, as solid lines). The solid line obeys equation (\ref{rcvsr0}). \emph{Middle:} The scaling parameter $r_c$ versus $R_{max}$, where the latter can be directly inferred from the RC. \emph{Right:} Fiducial URC: co-added doubly-normalized RCs shown in Fig. \ref{Fig13} (right-hand panel), i.e., normalized in both axes according to equation (\ref{vURC}). \normalsize} 
	\label{Fig14}
\end{figure*}

Considering only the leading term under the square-root of equation (\ref{URCv}) would give $v(r) = v_\mathrm{max}(1-e^{-r/r_c})$. This is similar to the simplyfied \emph{polyex} model of \citet{Giovanelli02} (with their slope parameter $\alpha =0$ for the outer RC). \citet{Bell02} used this approximate form in his study on viscous evolution in LSB galaxies, and \citet{Feng11} used this idealized form as a starting point and compared it to other and modified models. The mass density suggested here is different from theirs but provides a more realistic match with the data, too. In particular, for late-type spirals and dwarf irregulars the simple exponential limited-growth model often does not allow for a satisfactory RC fit at intermediate radii (i.e. within the range $r_\mathrm{max}/2<r<r_\mathrm{max}$) where it predicts too low RC values. This motivates investigating modified models, as the one proposed here. The URCs as calculated by means of equation (\ref{URCv}), using the parameter values given in Table \ref{Table2}, are superposed in Figs. \ref{Fig2} (right-hand panel, thick dashed lines), \ref{Fig13} (right-hand panel, thin dashed lines), and \ref{UGCA442_figs} (red lines) for each of the galaxies shown. Given only two parameters, i.e., $v_\mathrm{max}$ (as directly inferred from the observerd RC) and $r_c$ (as a model fit parameter), and going without any explicit reference to stellar and gaseous disk data, the URC represents the observed data remarkably well. In Fig. \ref{Fig13} (right-hand panel) three triplets of galaxies each with similar maximum circular velocity are plotted in order to illustrate that for a given maximum circular velocity different values of $r_c$ scale the RC in the $r$-direction: the larger this scaling parameter's value the larger is $r_\mathrm{max}$ and, correspondingly, the lower is the gradient of the velocity curve at intermediate radii like, e.g., $r=r_c$ or $r=r_\mathrm{max}/2$ (see Table \ref{Table3}).  
Underpinning the simplicity of our approach, $r_c$ is related to $r_\mathrm{max}$, as is evidenced in Fig. \ref{Fig14} (middle panel). The quadratic regression shown as solid line obeys
\begin{equation}
r_c = -0.0763 + 0.4214\,r_\mathrm{max}-0.0073\,r_\mathrm{max}^2. \label{rcvsr0}
\end{equation}  
Unfortunately, the correlation is loose (uncertainty measure  $R^2=0.794$), largely dependent on the lonesome upper data point and thus cannot safely replace the fitting procedure to obtain $r_c$; otherwise a simple inspection of the observed RC going along with a determination of the single point ($r_\mathrm{max}$, $v_\mathrm{max}$) would suffice to get the total URC. 

With more precision, the values of $r_c$ correlate with the core sizes $r_0$ of the Burkert halo; the OLSB-fit shown as solid line in Fig. \ref{Fig14} (left-hand panel) obeys
\begin{equation}
r_c = (0.644\pm 0.044)\,r_0+0.213. \label{rcvsr0}
\end{equation}  
Curiously enough, on average $r_c$ is about equal to $r_{0,\mathrm{PITS}}$ that itself is about equal to $0.6\,r_{0,\mathrm{Burk}}$ (Sect. 2.3). In principle, the set of equations (\ref{lgrho0lgr0lgvmax}) and (\ref{rcvsr0}) describes the formal conversion between the Burkert halo parameters and the alternative URC parameters for any individual galaxy, with no luminous matter variables being involved. Thus fitting $v_\mathrm{max}$ and $r_c$ to a RC with a flat part allows to immediately estimate the core size and together with relation (\ref{lgrho0lgr0lgvmax}) the core density as well. Hence, without relying on photometry (in particular, on $R_d$) the RC immediately delivers a description of the underlying DM halo structure.

Intriguingly, the doubly normalized observed RCs shown in Fig. \ref{Fig14} (right-hand panel), with the axes normalized by means of $v_\mathrm{max}$ and $r_c$ for each galaxy, i.e., $v_\mathrm{obs}(r)/v_\mathrm{max}$ versus $r/r_c$,  provide rather convergent curves and hence motivate our normalized fiducial URC to be simply expressed as
\begin{equation}
v_\mathrm{URC}(r/r_c)\equiv v(r)/v_\mathrm{max}, \label{vURC}
\end{equation}
with $v(r)=v(r/r_c)$ given by equation (\ref{URCv}). The exception that strongly deviates from similarity is UGC 191 with a peculiar transition from the steep to the flat part of the doubly-normalized RC. Decreasing RCs (like with UGC 7690) are not properly mapped as well.  We note that the individual RCs stem from truely observed RCs and are not constructed by means of any binned or averaged data points. In general, the co-added doubly normalized RC shown cover a rather narrow region in the rising part that may be considered as the statistical interval of uncertainty for our URC. Finally, even judged by visual inspection only, we mean to observe in Fig. \ref{Fig13} that the synthetic RCs calculated with equation (\ref{URCv}) (dashed lines) provide fits to the data points as good as or even better than the mass modelled RCs (solid lines). Given all these promising preliminary results, our novel URC candidate function, applied so far only to a handfull of galaxies, deserves further investigation. In an upcoming study we will have to take into account as many of the galaxies of the SPARC sample as possible and to include a comparison with doubly normalized RCs based on Burkert halo model decompositions (or, as a next step, by adopting the hybrid BP halo model) and an examination of the hidden relationships between luminous and dark matter. 
 
Different to the approaches of \citet{Karukes17}, \citet{DiPaolo18}, and  \citet[][providing precision templates for non-dwarf spirals of Hubble types Sb to Sm]{Lapi18} who basically rely on the photometric quantity $R_d$ (and related disk mass dependent quantities), we rely on the kinematic quantity $v_\mathrm{max}$ and additionally on the formal parameter $r_c$ (related to $R_\mathrm{max}$). Comparing equations (\ref{rho0KS17}) and (\ref{lgrho0lgr0lgvmax}), if both theirs and our results are correct there must be a tight connection between $C_{KS}$ and $v_\mathrm{max}$. 

The RC-shaping scaling parameter $r_c$ anticorrelating with the inner circular velocity gradient (at a given $v_\mathrm{max}$, see Table \ref{Table2}) is plausible, but its physical meaning and its relation to other physical quantities remains --beside its kinship with the Burkert halo core size $r_0$ (equation \ref{rcvsr0})-- to be found. The underlying total matter density profile is simple and effective: formally it has the structure of the density of an isothermal sphere, with a nonlinearly scaled radius, however: for small radii the singular factor $\propto r^{-2}$ is counterbalanced by the squared-brackets factor such that the central density is finite. A novelty of our approach is that the particular decompositions proposed in the Appendix do not match with the observed luminous contribution for the circular velocity (usually modelled by a Freeman disk) but with that of the total disk (sometimes modelled by the sum of two Freeman disks as in \citet{Tonini06} and \citet{Karukes17}). Other ways of decompositions may in addition hide within the formalism proposed. Non-constant mass-to-light ratio profiles, being however unpopular but not prohibitively counter-factual, would even further increase the variety of possible decompositions.

\section{Summary \& conclusions}

In this study, we confirmed, refined, or controversely discussed some details concerning well-known scaling relations or laws for late-type spiral and dwarf irregualar galaxies, linking the structure and the dynamics of these rotationally supported systems. We additionally contributed some new scaling relations that were not yet (to the best of our knowledge) discussed in the literature. The following results were obtained. 

\begin{itemize}
\item Rotation curve (RC) mass model decompositions for 79 late-type spiral and dwarf irregular galaxies of the SPARC sample are provided, assuming spherically-symmetric dark matter halos of Burkert and of pseudo-isothermal (PITS) type. \item Disk mass-to-light ratios are restricted to have constant, but  semi-free best-fit values 0.2, 0.5, or 0.8 $M_{\sun}/L_{\sun}$ at [3.6]. They exhibit an asymmetric bimodal distribution with the dominant peak located at the median value of 0.2 (minimum disks) and the subdominant peak at 0.8 (maximum disks). It remains to be checked, whether this is artefactual or really related to some physical parameters like baryonic mass fraction or galaxy color and metallicity. 
\item As compared to decompositions with PITS halos, those with Burkert halos provide better fits for about two thirds of all galaxies. Comparing the best-fit halo parameters for either approach, the central densities $\rho_0$ have similar values, but the scale lengths $r_0$ of the Burkert model are systematically higher by a factor of 1.80$\pm$0.05. This is of relevance for the calculation of, e.g., mean central surface densities for a given sample of galaxies. Within galactocentric radii smaller than a few halo scale lengths, $r_0$-scaled versions of the Burkert and the PITS model represent the generalized Burkert-PITS halo density profile $\rho (r)=\rho_0 (1+r/r_0)^{-\eta}$ $(1+(r/r_0)^2)^{-1}$ with 0$\le$$\eta$$\le 1$.
\item Focussing on the Burkert halo, we use the two-radial accelerations relation (RAR) diagram for three purposes: based on the mass model relation $g_{\mathrm{obs}}=g_{\mathrm{bar}} (1+(v_{\mathrm{halo}}/ v_{\mathrm{bar}})^2)$, with the factor within the outer brackets being the inverse baryonic mass fraction that quantifies the mass deficiency, we (i) illustrate the numerical accuracy of our RC decompositions; (ii) we emphasize the information content in the spreaded distribution of the data points; and (iii) we confirm some decrease of the baryonic mass fraction when going from HSB to LSB galaxies. 
\item The latter distinction is anchored on the observation that for any given absolute magnitude (i) the central surface brightnesses have a uniform mean value $<$$\mu_{0,[3.6]}$$>$$= 19.63\pm0.11\, \mathrm{mag\,arcsec^{{-2}}}$, and (ii) larger-than-mean disk scale lengths (lower compactness) correspond to LSB galaxies, while smaller-than-average scale lengths (higher compactness) correspond to HSB galaxies. Consistently, given the baryonic Tully-Fisher relation (BTFR), some expected increase of the inner velocity-gradient (inferred here at $R_d$) with brighter central surface brightness is indeed observed on average. At a given luminosity, more compact (i.e., HSB) galaxies do have steeper-than-mean circular velocity gradients than less compact (i.e., LSB) galaxies. 
\item The average inner circular velocity gradient at one disk scale length roughly follows $\nabla v(R_d) \approx v_\mathrm{max}/R_\mathrm{max}$ (equation \ref{nablav2}). This relation may be considered as a constraint concerning the universal shape of the RC of rotationally supported low-mass galaxies. It links the kinematics within the luminous inner part of a galaxy with variables taken from the outer part, where DM dictates the motion.
\item We formulate some minimum disk hypothesis as follows: DM-rich disk galaxies with low mass-to-light ratios fullfill the velocity-ratio criterium $v_\mathrm{bary}(R)/v_\mathrm{obs}(R)\,\rvert_{R=2.2R_d} =0.5\pm0.1$ (equation \ref{vrc}). 
\item The luminous and the dark matter content at inner radii are both closely linked with the observed kinematic behaviour. We report on an adjusted baryonic Tully-Fisher relation (aBTFR), that however incorporates the baryonic mass fraction as an essentially non-baryonic ingredient. Taking the DM content explicitely into account, the spread accompaning the usual BTFR relation is highly reduced.
\begin{table}\centering
	\small
	\begin{tabular}{lll}
		\hline
		scaling rel.: & & \\
		aBTFR	& adjusted baryonic TF relation & 3.5	\\
		CDC 	&  core density vs. compactness & 3.3	\\
		CDMV	&  column density vs. maximum velocity	& 3.2 \\
		CRC 	&  core radius vs. compactness	& 3.6	\\
		CRRC	&  core radius vs. rotation curve shape & 3.6\\
		MDAR	&  mass discrepancy - radial acceleration rel.& 3.1\\
		RAR		&  two-radial accelerations relation & 3.1 \\
		SDR		&  two-surface densities relation & 3.3.2 \\
		VGMV	&  velocity gradient vs. maximum velocity & 3.4.3 \\
		VRC		&  velocity-ratio criterium	& 2.3.2\\
		\hline
		others: & & \\
		DM / LM	&  dark / luminous matter	& \\
		LSB / HSB & low / high surface brightness & 2.2	\\
		OLSB	& ordinary-least squares bisector & \\
		RC		& individual rotation curve	 & \\
		URC 	& universal rotation curve & 3.7\\
		\hline \hline
	\end{tabular}
	\caption{\small Overview on acronyms for some scaling relations and other terms. The columns give the acronym, its meaning, and the number of the section in the text where the corresponding scaling relation or phenomenon is introduced or discussed in more depth. \normalsize}
	\label{Table3}
\end{table}

\item The baryonic mass fraction within individual galaxies decreases with growing galactocentric distance according to the statistical relation $f_\mathrm{bary}\propto v_\mathrm{obs}^{1.18}(R)/R$ (equation \ref{fbary2}).  
\item The Burkert halo parameters roughly follow $\rho_0$$\,\propto\,$$r_0^{-1.5}$. More subtle, allowing for the maximum circular velocity as a third variable, the tight relation $\rho_0$$\,\propto\,$$r_0^{-1.84\pm0.07} v_\mathrm{max}^{2.00\pm 0.11}$ with very small scatter emerges (equation \ref{lgrho0lgr0lgvmax}). As with the velocity gradient, this links the inner region of a galaxy (actually, the halo structure) with a kinematic variable taken from the outer part, where DM dictates the motion.
\item The halo central surface density $\rho_0 r_0$, with a sample median $<$$\rho_0 r_0$$>$$\,\approx 121\, \mathrm{M_\odot pc^{-2}}$, weakly correlates with the disk central surface brightness $\mu_0$ and strongly correlates with the observed radial acceleration $g_{\mathrm{obs}}=v^2_{\mathrm{obs}}(r)/r$ at different galactocentric radii. Consistently, and even more pronounced, the maximum velocity $v_\mathrm{max}$, that typically represents the flat part of the RC, is tightly proportional to the halo component of circular velocity at $r_0$, i.e. $v_\mathrm{max} \propto  v_\mathrm{halo}(r_0)  \propto \rho_0^{1/2} r_0$. This is equivalent to a maximum circular velocity versus central halo column density (CDMV) relation $v_\mathrm{max}^2 \propto \rho_0 r_0^2$ (equation \ref{CDMV}). Given some $v_\mathrm{max}$, larger halo cores not only correspond to smaller central halo densites but on average go with fainter central surface brightness or, equivalently, less compact disks, too. 
\item Overall, halo cores are small (we have a median $r_0 \approx 2 R_d \approx 0.64 R_\mathrm{opt}$, hence close to a Freeman disk peak radius or good half an optical radius, to be compared with the average $R_\mathrm{max}\approx 3.9 R_d \approx 1.22 R_\mathrm{opt}$), but their size does actually not correlate with the optical radius. We did not yet look at the transition of low mass galaxies to very luminous galaxies that usually exhibit high values of $v_\mathrm{max}$ and sometimes a decreasing outer RC. Whether or not the scaling relations discussed in this paper will change due to Hubble type was not investigated here.
\item Certainly, some of the scatter in our figures and the spreads in our relations could be reduced if we would restrain to a more restrictive selection of galaxies. This potentially may alter some of our fitted scaling relations, too. Whether or not the frequent occurence of a difference between observed central surface brightness and extrapolated central surface brightness plays a role in the interplay between LM and DM remains an open question. The simple ordinary least-squares bisector (OLSB) fitting procedure adopted throughout the paper allows for rudimentary inclusion of errors in both variables and seems to be a working alternative to more sophisticated routines that consider individual errors for all data points. However, the question on how to deal best with errors has no simple answer, despite its enormous influence on the produced relations. The overall consistency of the results presented above makes us nevertheless feel confident with the findings of our study. In order to attempt to further reduce the amount of uncertainty in the context of halo related relations, it seems worthwhile to adopt the hybrid Burkert-PITS (BP) model given by equation (\ref{halodens}) for the decompositions, with the fitting parameter $\eta$ possibly being somehow related to observable quantities.
\item Some of our best-fit model RCs only accidentally match with the conventional universal rotation curve (URC). A better conversion is provided by the non-singular total matter density profile $\rho_\mathrm{total}(r)=$ $(v_\mathrm{max}^2/4\pi G r^2)$ $( 1 - (1-r/r_c)\exp(-r/r_c))^2$, with the scaling parameter $r_c$ linearly scaling with $r_0$ of the Burkert model or being roughly estimable via a quadratic regression of $r_\mathrm{max}$, and with $v_\mathrm{max}$ being replaceable by means of the CDMV relation. Consistently, for galaxies with comparable luminosity this RC-shape parameter $r_c$ anticorrelates with the inner circular velocity gradient. We provide the corresponding formulae for the total radial mass distribution and the total velocity profile. The co-added doubly-normalized velocity profiles of a selection of nine galaxies exhibit a high degree of similarity. This is strongly encouraging us to continue the quest for a synthetic URC along this path. It is strengthened by the successful application of analytic RC decompositions of this synthetic URC into a baryonic disk and a halo component to a couple of example galaxies (Appendix A). An inquiry of this novel density profile by testing it against many more galaxies than done here is a postboned endeavour.
\item A plethora of acronyms is used for brief reference to the various scaling relations and phenomena (Table \ref{Table2}). 
\end{itemize}

If the usefullness of the alternative URC model density will be confirmed, a new playground would be provided. For example, if $r_c$ and eventually $v_\mathrm{max}$ or the parameters related to the decompositions are \emph{time-dependent}, evolutionary aspects could be addressed within our simple parameterization approach. Figs. \ref{Fig13} (right-hand panel) and \ref{UGCA442_figs} (lower right-hand panel) invite to be interpreted in terms of such a toy scenario. Of course, the astrophysical reason for the proposed total matter density profile would have to be addressed in any case, as well as its deeper interrelationship with the scaling relations or laws picked up and investigated in this study. Irrespective of any URC model these genuine scaling relations among structural and dynamical parameters stand alone and help to shape our understanding of late-type spiral and dwarf irregular galaxies. Explaining their origin will lead to refined and maybe surprising insights concerning theories of galaxy formation and on the nature of DM.

\section*{Acknowledgements}
The empirical starting point for this study was constituted by the database of the SPARC team, publicly available at astroweb.cwru.edu/SPARC/.

\bibliographystyle{mnras}
\bibliography{Parodi2018_StructureDynamicsRelations}


	
\setcounter{subsection}{0}
\renewcommand{\thesubsection}{A.\arabic{subsection}}
\setcounter{equation}{0}
\renewcommand{\theequation}{A.\arabic{equation}}
\setcounter{figure}{0}
\renewcommand{\thefigure}{A.\arabic{figure}}

\begin{figure*}
	\centering 
	\includegraphics[width=0.48\textwidth]{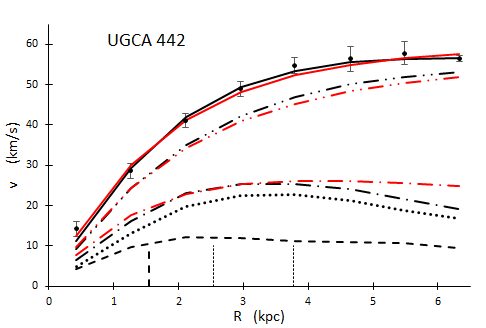}
	\includegraphics[width=0.48\textwidth]{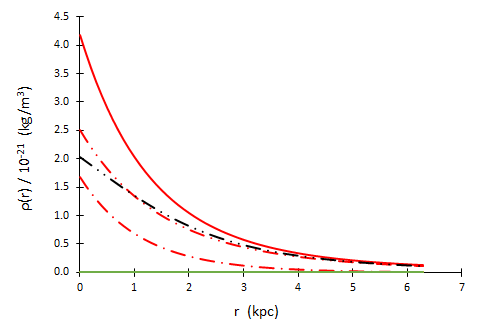}
	\includegraphics[width=0.48\textwidth]{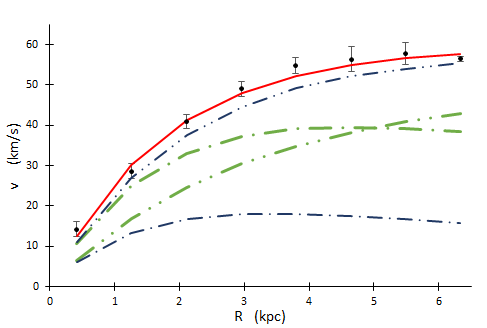}
	\includegraphics[width=0.48\textwidth]{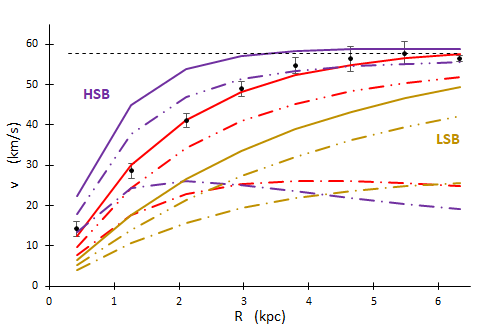}
	\caption{\small Illustration of the $\alpha\beta\gamma$-decomposition by means of the RC of the dI galaxy UGCA 442 (circles with error bars). \emph{Top left:} The black lines correspond to the mass model decomposition using a Burkert halo; they are already shown in Fig. 2. Overlaid in red is the URC according to equation (\ref{URCv}) with parameter values $v_\mathrm{max}=57.8$ km$\,$s$^{-1}$ and $r_c$ = 2.0 kpc (red solid line) and an $\alpha\beta\gamma$-decomposition with ($\alpha$,$\beta$,$\gamma$) = (1.0,0.7,0.0), with halo contribution $v_I(r)$ (red dot-dot-dashed line) and baryonic disk contribution $v_{II}$ (red dot-dashed line) according to equations (\ref{URCvI}) and (\ref{URCvII}), respectively. \emph{Top right:} Total matter density $\rho(r)$ and its contributions $\rho_I(r)$ and $\rho_{II}(r)$ corresponding to the circular velocities shown in red in the figure to the left and  calculated by means of equations (\ref{URC_rhoI}) and (\ref{URC_rhoII}). The Burkert halo density is shown, too (black dot-dot-dashedline). \emph{Bottom left:} In order to illustrate the broad range of  ($\alpha$,$\beta$,$\gamma$)-decompositions that are theoretically possible, the URC shown in the panel above (red line) is now $\alpha\beta\gamma$-decomposed using other parameter values: the green lines use (1,0,0) and represent a fictitious galaxy with a high baryonic mass fraction at inner radii while the blue lines use (1,1,0) and represent a galaxy that is strongly DM dominated at all radii, similar to UGCA 442. \emph{Bottom right:} Keeping the values for the maximum circular velocity and for the $\alpha\beta\gamma$-decomposition parameters constant and equal to those of UGCA 442 (upper left panel), only the scaling parameter $r_c$ is varied: $r_c=1$ kpc (lilac lines, representing an HSB galaxy), $r_c=2.8$ kpc (red lines, UGCA 442, with a central surface brightness at about the global mean value of 19.63 mag$\,$arcsec$^{-2}$), and $r_c=4$ kpc (ochre lines, representing a LSB or even an ultra-diffuse galaxy (UDG)).  \normalsize }
	\label{UGCA442_figs}
\end{figure*} 
\begin{figure} 
	\includegraphics[width=0.48\textwidth]{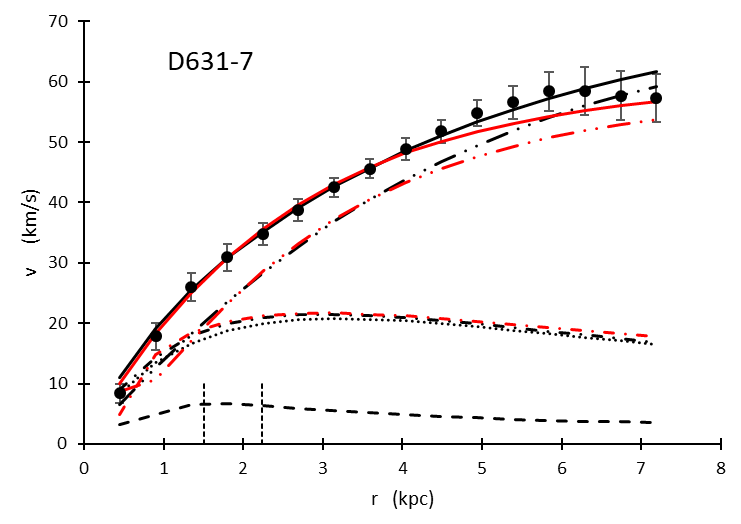}
	\caption{\small The RC of dI galaxy D631-7 (data points with error bars) with two decompositions overlaid: at one hand using a Burkert halo (black lines, line types as in Fig. 2), at the other hand a $\delta$-decomposition with parameter values $v_\mathrm{max}=58.5$ km$\,$s$^{-1}$, $r_c$ = 2.8 kpc, and $\delta=0.16$ (red lines).  \normalsize}
	\label{D631_7_fig}
\end{figure}

\section*{Appendix A: Preliminary URC decompositions}

The URC given by equation (\ref{URCv}) formally allows for various decompositions into a halo and a disk part in a top-down approach. Usually, mass modelling follows a bottom-up approach that frequently is based on Freeman disks for the stellar matter and sometimes for the gaseous matter too (with, for example, about three times as large a scale length) and on an adopted halo model \citep[e.g.,][]{Tonini06, Karukes17}. Here we instead start with the observed RC and assume some total baryonic disk contribution that is not necessarily following a single exponential disk or that is the sum of two exponential disks. A separate halo contribution is always added. The decompositions proposed below involve one to three parameters that allow for optimized fits to the observed baryonic mass model data (as provided here by the SPARC database).

\subsection{$\alpha \beta \gamma$-decomposition}
  
Using three real parameters $0\le\alpha, \beta, \gamma\le1$ and applying the binomial theorem we arbitrarily partition the total matter density (\ref{URCrho2}) as follows:
\begin{eqnarray}
\rho(r)&=&\rho_I(r)+\rho_{II}(r) \\
\rho_I(r)&=&\rho_I(r;\,\alpha,\beta,\gamma)=\frac{v_{max}^2}{4\pi G} \left( \,\,
	\alpha\left( \frac{1-e^{-r/r_c}}{r} \right)^2  \right. \nonumber \label{URC_rhoI}\\
 & &+\,\beta \left. \frac{2}{r\,r_c} e^{-r/r_c} (1-e^{-r/r_c}) +
	\gamma \left( \frac{e^{-r/r_c}}{r_c} \right)^2 \,\,\right) \\
\rho_{II}(r)&=&\rho_{II}(r;\,\alpha,\beta,\gamma)\nonumber \\
&=&\rho_I(r;\,1-\alpha, 1-\beta,1-\gamma) \label{URC_rhoII}
\end{eqnarray} 
The corresponding velocity contributions $v_i(r)=(GM_i(\le r) /r)^{0.5}$ ($i=I, II$), where $M_i(\le r)= 4\pi\int_0^rr'^2 \rho_i(r')dr'$, become
\begin{eqnarray}
v_I(r)&=& v_I(r;\,\alpha,\beta,\gamma)\nonumber\\
&=& v_\mathrm{max} \left[ \alpha - \beta +\beta\left( 1-e^{-r/r_c} \right)^2 -2(\alpha - \beta)\frac{1-e^{-r/r_c}}{r/r_c}  \right. \nonumber\\
& &\left. +\left(\alpha - \beta +\frac{\gamma}{2} \right) \frac{1-e^{-2r/r_c}}{2r/r_c}
-\frac{\gamma}{2}\left( 1+\frac{r}{r_c} \right)e^{-2r/r_c}  \right]^{0.5} \label{URCvI}\\
v_{II}(r)&=& v_{II}(r;\,\alpha,\beta,\gamma)\nonumber \\
&=&v_I(r;\,1-\alpha, 1-\beta,1-\gamma)\label{URCvII}
\end{eqnarray}
We note that $\sqrt{ v_I^2(r) + v_{II}^2(r)}=v(r)$, with $v(r)$ as given by equation (\ref{URCv}) for the URC, holds by construction, irrespective of the three parameter values chosen.

This decomposition is refered to as $\alpha \beta \gamma$-decomposition. In Fig. \ref{UGCA442_figs} its application to the dI galaxy UGCA 442 is illustrated. In the upper left-hand panel the decomposition including a Burkert DM halo is shown in black (with line types as in Fig. \ref{Fig2}) while the URC (with parameter values $v_\mathrm{max}=57.8$ km$\,$s$^{-1}$ and $r_c$ = 2.0 kpc) and its $\alpha\beta \gamma$-decomposed contributions (with parameter values ($\alpha$, $\beta$, $\gamma$) = (1.0, 0.7, 0.0)) are shown in red. Out to the optical radius at about 5.7 kpc the mutual match is acceptable for both, the total RC and the baryonic contribution. Further out, however, there's a mismatch between the observed and the baryonic matter contributions (dot-dashed lines). Out there, either our model decomposition fails (but not the URC itself) or the observed quantities of gas are actually underestimated. We note that the non-neglectable gaseous contribution dominates over the stellar contribution at all radii. As can be inferred from the upper right-hand panel, speaking of \emph{cored} Burkert halos in the sense of flat or at least non-steep core profiles is potentially misleading: for the present case, the initial halo density $\rho_0$ has already dropped by 75\% at  $r=r_0= 1.6$ kpc, hence dwarf galaxies with typically small core radii do actually exhibit nearly constant densities only at very small radii. In the lower left-hand panel we artificially play with the parameter values to get a feeling for the generality of the model decompositions: the green lines use ($\alpha$,$\beta$,$\gamma$)=(1,0,0) and represent a fictitous galaxy with a high baryonic mass fraction at inner radii while the blue lines use (1,1,0) and represent a galaxy that is strongly DM dominated at all radii, similar to UGCA 442. Finally, in the lower right-hand panel only the scale lengths $r_c$ are changed from high (ochre lines) to intermediate (red lines, corresponding to UGCA 442), to small (lilac lines), with all other parameters kept at constant values according to the situation in the upper left-hand panel. The peak of the baryonic contribution is located at $r_\mathrm{peak}\approx 2r_c$. Given the fixed maximum circular velocity that about corresponds to a given absolute magnitude (see Table \ref{Table2}), a small value for the scaling parameter $r_c$ hypothetically represents an HSB galaxy (corresponding to a higher inner velocity gradient) while a larger value represents a LSB galaxy (corresponding to a shallower inner velocity gradient) or even to an ultra-diffuse galaxy \citep[UDG, see][]{DiCintio17}. Consistently, the dwarf galaxy UGCA 442 between them has a central surface brightness $\mu_0$=19.65 mag$\,$arcsec$^{-2}$, i.e., very close to the global mean value identified in Sect. 2.2. Hence, despite $r_c$ being not proportional to the optical disk scale length $r_d$ (probably due to the non-neglectable effect of the gaseous matter content) this last Fig. \ref{UGCA442_figs} (lower right panel) and the very first figure in this paper (Fig. \ref{Fig1}, upper left panel) seem to be intimately related. Furthermore, it is tempting to think of the parameters $r_c$ and ($\alpha$,$\beta$,$\gamma$) as being time-dependent, for example, $r_c$ decreasing with time. The dynamical change of a time-dependent decomposition would then manifest different evolutionary stages of a galaxy. Perfoming accordingly numerical simulations of galaxy evolution in order to test this idea is wishfully postponed to a future work.

\subsection{$\delta$-decomposition}

For some galaxies the $\alpha\beta \gamma$-decomposition does not satisfactorily fit the data. A different decomposition is therefore proposed along the following lines. A two-dimensional exponential disk with scale length $r_d$ and surface density $\Sigma_D(r)=\Sigma_0 \exp(-r/r_d) $ has a mass distribution $ M_D(\le r)= 2\pi\int_0^r r'\Sigma_D(r')dr'= M_D\,(1-(1+r/r_d) \exp(-r/r_d))$, where $M_D=2\pi\Sigma_0 r_d^2$ is the total disk mass. Formally, the corresponding circular velocity profile is $v_D(r)=(G M_D(\le r)/r)^{1/2} = (G M_D)^{1/2} [(1-e^{-r/r_d})/r -e^{-r/r_d}/r_d]^{1/2}$, peaking at about $r\approx 2\,r_d$. The formal structure of this relation for $v_D(r)$ will be recognized in the following decomposition of the URC (equation \ref{URCv}). Rearranging terms in equation (\ref{URCv}) and thus rewriting the URC as 
\begin{eqnarray}
v(r) &=& \sqrt{ v_I^2(r) + v_{II}^2(r) } \\
v_I(r) &=& v_\mathrm{max} \sqrt{ \left( 1-e^{-r/r_c} \right)^2 - \frac{r}{2r_c}e^{-2r/r_c} }
\end{eqnarray}
\begin{equation}
v_{II}(r) = v_\mathrm{max} \sqrt{ \frac{r_c}{4} \left( \frac{1-e^{-2r/r_c}}{r}- 2\frac{e^{-2r/r_c}}{r_c} \right) }
\end{equation} 
allows to identify $v_{II}(r)$ as an exponential disk component with scale length $r_c/2$ and to attribute the remaining terms to the halo component $v_I(r)$. The component $v_{II}(r)$ peaks at about $r\approx r_c$. However, firstly, the scale length $r_c$ does not correlate with the photometric scale length $r_d$, as noted in Sect. 3.7.2. The disks of galaxies, secondly, do have some vertical extension, and they consist, thirdly, of combined stellar and gaseous disks with different scale lengths. To somehow account for these expected deviations from a pure thin exponential disk we introduce a baryonic disk parameter $\delta$ and modi\-fy the circular velocity contributions --without changing the total circular velocity $v(r)$-- as follows:
\begin{eqnarray}
v_I(r) &=& v_\mathrm{max} \sqrt{ \left( 1-e^{-r/r_c} \right)^2 - \frac{1}{2} \left( \frac{r}{r_c}-\delta \right) e^{-2r/r_c} }\\
v_{II}(r) &=& v_\mathrm{max} \sqrt{ \frac{r_c}{4} \left( \frac{1-e^{-2r/r_c}}{r}- 2 (1+\delta)\frac{e^{-2r/r_c}}{r_c} \right) }
\end{eqnarray}
We stress that the total velocity $v(r)$ is independent of $\delta$ and is equal to the URC given in equation (\ref{URCv}). The purpose of this parameter is to introduce another degree of freedom that allows for a better match of the decomposition term $v_{II}$ with the observed circular velocity due to the baryonic matter distribution (i.e. stellar and gaseous). We apply the $\delta$-decomposition to the dwarf irregular galaxy D631-7 and show the result in Fig. \ref{D631_7_fig}  The total baryonic contribution is adequately traced for most radii with $\delta=0.16$ (red dot-dashed line). At intermediate radii the halo contribution (red dot-dot-dashed line) closely follows the contribution given by the Burkert halo (black dot-dot-dashed line). At inner radii $r<r_c/2$ some mismatch can be recognized. 
  
Adapting $\alpha\beta\gamma$- or $\delta$-decompositions for all galaxies of our sample will hopefully be addressed in a future survey where we will give the explicit formulae for the particular $\delta$-dependent density contributions $\rho_I(r)= (4\pi G r^2)^{-1}\frac{d}{dr}(rv_I^2(r))$ and $\rho_{II}(r)= (4\pi G r^2)^{-1}\frac{d}{dr}(rv_{II}^2(r))$, too.

\section*{Appendix B: Structural and kinematic data}	
\setcounter{table}{0}
\renewcommand{\thetable}{B.\arabic{table}}

The following three tables contain the data declared in Sect. 2.1 and extensively used throughout this paper.

\clearpage
\onecolumn
\begin{longtable}{llccccc}
\caption{Selected photometry and luminous structure parameters}\\
\hline \hline 
$\,\,\,$Name		&		&	$D$		& $M\,[3.6]$&$\mu_0\,[3.6]$&$R_d\,[3.6]$&	$C$	\\
			&		&	Mpc		&	mag		& mag$\,$arcsec$^{-2}$&	kpc		&		\\
$\,\,\,\,\,\,$(1)			&	$\,\,\,\,\,\,$(2)	&	(3)		&	(4)		&	(5)	 	   &	(6)		&	(7)	\\
\hline
\endfirsthead
\caption{continued.}\\
\hline \hline
$\,\,\,$Name		&		&	$D$		& $M\,[3.6]$&$\mu_0\,[3.6]$	&$R_d\,[3.6]$&	$C$	\\
			&		&	Mpc		&	mag		& mag/arcsec$^2$  	&	kpc		 &		\\
$\,\,\,\,\,\,$(1)			&	$\,\,\,\,\,\,$(2)	&	(3)		&	(4)		&	(5)			&	(6)		 &	(7)	\\
\hline
\endhead
\hline
\endfoot
D512-2		&	-		&	15.20	&	-18.04	&	19.88	&	1.24	&	0.77	\\
D564-8		&	-		&	8.79	&	-15.55	&	21.50	&	0.61	&	0.50	\\
D631-7		&	-		&	7.72	&	-17.49	&	19.66	&	0.70	&	1.06	\\
DDO 64		&	UGC 5272&	6.80	&	-17.25	&	19.36	&	0.69	&	0.97	\\
DDO 154		&	UGC 8024&	4.04	&	-16.07	&	20.18	&	0.37	&	1.05	\\
DDO 161		&	UGCA 320&	7.50	&	-18.60	&	19.24	&	1.22	&	1.01	\\
DDO 168		&	UGC 8320&	4.25	&	-17.46	&	19.90	&	1.02	&	0.72	\\
DDO 170		&	UGC 8333&	15.40	&	-18.59	&	20.14	&	1.95	&	0.63	\\
ESO 444-84	&  PGC 48111&	4.83	&	-16.39	&	20.25	&	0.46	&	0.98	\\
F561-1		&	-		&	66.40	&	-20.78	&	19.89	&	2.79	&	1.20	\\
F563-1		&	-		&	48.90	&	-19.96	&	20.76	&	3.52	&	0.65	\\
F563-V1		&	-		&	54.00	&	-19.73	&	20.79	&	3.79	&	0.54	\\
F563-V2		&	-		&	59.70	&	-20.44	&	19.40	&	2.43	&	1.18	\\
F565-V2		&	-		&	51.80	&	-18.63	&	20.80	&	2.17	&	0.58	\\
F567-2		&	-		&	79.00	&	-20.08	&	20.64	&	3.08	&	0.79	\\
F574-2		&	-		&	89.10	&	-20.40	&	20.77	&	3.76	&	0.75	\\
F583-1		&	-		&	35.40	&	-19.24	&	20.35	&	2.36	&	0.70	\\
IC 2574		&	DDO 81	&	3.91	&	-19.27	&	20.05	&	2.78	&	0.60	\\
KK98-251	& PGC 166192&	6.80	&	-16.58	&	20.52	&	1.34	&	0.37	\\
NGC 55		& ESO 293-50&	2.11	&	-20.92	&	18.33	&	6.11	&	0.58	\\
NGC 1705	& ESO 158-13&	5.73	&	-18.57	&	17.89	&	0.39	&	3.13	\\
NGC 2366	& DDO 42	&	3.27	&	-17.69	&	19.67	&	0.65	&	1.25	\\
NGC 2915	& ESO 37-3	&	4.06	&	-18.77	&	18.57	&	0.55	&	2.43	\\
NGC 3109	& DDO 236	&	1.33	&	-17.48	&	19.44	&	1.56	&	0.47	\\
NGC 3741	& UGC 6572	&	3.21	&	-15.37	&	19.42	&	0.20	&	1.42	\\
NGC 4068	& UGC 7047	&	4.37	&	-17.69	&	18.77	&	0.59	&	1.38	\\
NGC 4214	& UGC 7278	&	2.87	&	-19.40	&	17.02	&	0.51	&	3.49	\\
NGC 6789	& UGC 11425	&	3.52	&	-16.76	&	18.55	&	0.31	&	1.72	\\
UGC 128		& -			&	64.50	&	-21.96	&	19.94	&	5.95	&	0.96	\\
UGC 191		& DDO 2		&	17.10	&	-20.01	&	19.02	&	1.58	&	1.49	\\
UGC 634		& DDO 7		&	30.90	&	-20.45	&	19.56	&	2.45	&	1.17	\\
UGC 731		& DDO 9		&	12.50	&	-18.03	&	20.02	&	2.30	&	0.41	\\
UGC 891		& DDO 10	&	10.20	&	-18.19	&	19.67	&	1.43	&	0.72	\\
UGC 1230	& -			&	53.70	&	-21.46	&	20.21	&	4.34	&	1.05	\\
UGC 1281	& -			&	5.27	&	-18.13	&	19.48	&	1.63	&	0.61	\\
UGC 2023	& DDO 25	&	10.40	&	-19.55	&	19.60	&	1.55	&	1.23	\\
UGC 2259	& -			&	10.50	&	-19.85	&	19.22	&	1.62	&	1.35	\\
UGC 2455	& NGC 1156	&	6.92	&	-20.66	&	17.56	&	0.99	&	3.20	\\
UGC 4305	& DDO 50	&	3.45	&	-18.92	&	19.95	&	1.16	&	1.23	\\
UGC 4325	& NGC 2552	&	9.60	&	-20.02	&	18.99	&	1.86	&	1.27	\\
UGC 4483	& -			&	3.34	&	-14.54	&	20.02	&	0.18	&	1.08	\\
UGC	4499	& -			&	12.50	&	-19.73	&	19.55	&	1.73	&	1.20	\\
UGC 5005	& -			&	53.70	&	-20.79	&	20.27	&	3.20	&	1.05	\\
UGC 5414	& NGC 3104	&	9.40	&	-19.38	&	19.55	&	1.47	&	1.20	\\
UGC 5716	& -			&	21.30	&	-18.68	&	19.92	&	1.14	&	1.12	\\
UGC 5750	& -			&	58.70	&	-20.56	&	19.57	&	3.46	&	0.87	\\
UGC 5764	& DDO 83	&	7.47	&	-16.58	&	20.99	&	1.17	&	0.42	\\
UGC 5829	& DDO 84	&	8.64	&	-18.64	&	20.31	&	1.99	&	0.63	\\
UGC 5918	& DDO 87	&	7.66	&	-17.68	&	21.32	&	1.66	&	0.49	\\
UGC 5986	& NGC 3432	&	8.63	&	-20.94	&	16.72	&	1.67	&	2.15	\\
UGC 5999	& DDO 90	&	47.70	&	-20.58	&	20.53	&	3.22	&	0.95	\\
UGC 6399	& -			&	18.00	&	-20.16	&	18.58	&	2.05	&	1.23	\\
UGC 6628	& -			&	15.10	&	-20.69	&	19.78	&	2.82	&	1.14	\\
UGC 6818	& -			&	18.00	&	-19.76	&	18.67	&	1.39	&	1.51	\\
UGC 6917	& -			&	18.00	&	-21.34	&	18.77	&	2.76	&	1.56	\\
UGC 6923	& -			&	18.00	&	-20.41	&	18.46	&	1.44	&	1.96	\\
UGC 7089	& -			&	18.00	&	-20.64	&	18.02	&	2.26	&	1.39	\\
UGC 7125	& -			&	19.80	&	-20.34	&	19.78	&	3.38	&	0.81	\\
UGC 7232	& NGC 4190	&	2.83	&	-16.89	&	18.92	&	0.29	&	1.95	\\
UGC 7261	& NGC 4204	&	13.10	&	-19.87	&	17.93	&	1.20	&	1.83	\\
UGC 7323	& NGC 4242	&	8.00	&	-20.79	&	18.68	&	2.26	&	1.48	\\
UGC 7399	& NGC 4288	&	8.43	&	-19.41	&	19.48	&	1.64	&	1.09	\\
UGC 7524	& NGC 4395	&	4.74	&	-20.22	&	19.74	&	3.46	&	0.75	\\
UGC 7559	& DDO 126	&	4.97	&	-16.85	&	20.46	&	0.58	&	0.96	\\
UGC 7577	& DDO 125	&	2.59	&	-15.89	&	20.47	&	0.90	&	0.40	\\
UGC 7608	& DDO 129	&	8.21	&	-17.81	&	20.64	&	1.50	&	0.57	\\
UGC 7690	& -			&	8.11	&	-19.09	&	18.32	&	0.57	&	2.71	\\
UGC 7866	& DDO 141	&	4.57	&	-16.99	&	19.84	&	0.61	&	0.97	\\
UGC 8490	& NGC 5204	&	4.65	&	-19.28	&	17.91	&	0.67	&	2.51	\\
UGC 8837	& DDO 185	&	7.21	&	-18.51	&	20.09	&	1.72	&	0.69	\\
UGC 9992	& -			&	10.70	&	-18.07	&	20.15	&	1.04	&	0.93	\\
UGC 10310	& DDO 204	&	15.20	&	-19.86	&	19.31	&	1.80	&	1.22	\\
UGC 11557	& -			&	24.20	&	-21.96	&	18.49	&	2.75	&	2.08	\\
UGC 11820	& -			&	18.10	&	-19.22	&	20.98	&	2.08	&	0.79	\\
UGC 12632	& DDO 217	&	9.77	&	-19.54	&	20.25	&	2.42	&	0.78	\\
UGC 12732	& -			&	13.20	&	-19.81	&	19.61	&	1.98	&	1.08	\\
UGCA 281	& -			&	5.68	&	-17.48	&	22.11	&	1.72	&	0.43	\\
UGCA 442	& ESO 471-6	&	4.35	&	-17.12	&	19.65	&	1.18	&	0.53	\\
UGCA 444	& DDO 221; WLM&	0.98	&	-14.45	&	21.42	&	0.83	&	0.22	\\
\label{TableB1}	
\end{longtable}


\begin{longtable}{lcccccccc}
\caption{Best-fit parameter values using Burkert and PITS halos}\\
\hline\hline
$\,\,\,$Name	& Burk: $\,\rho_0$& $r_0$	& $\Upsilon_{[3.6]}$& $\chi^2_n$&	PITS: $\,\rho_0$&	$r_0$& $\Upsilon_{[3.6]}$& $\chi^2_n$\\
		& $M_\odot$ pc$^{-3}$	&	kpc		& $M_\odot/L_\odot$	& 		&
		  $M_\odot$ pc$^{-3}$  &	kpc		& $M_\odot/L_\odot$	&			\\
$\,\,\,\,\,\,$(1)		&	(2)		&	(3)		&	(4)	&	(5)		&	(6)		&	(7)		&	(8)	&	(9)		\\
\hline
\endfirsthead
\caption{continued.}\\
\hline\hline
$\,\,\,$Name	& Burk: $\,\rho_0$& $r_0$	& $\Upsilon_{[3.6]}$	& $\chi^2_n$&	PITS: $\,\rho_0$&	$r_0$& $\Upsilon_{[3.6]}$	& $\chi^2_n$\\
		& $M_\odot$pc$^{-3}$		&	kpc	&	$M_\odot/L_\odot$	& 			& $M_\odot$pc$^{-3}$	  &	kpc	 &	$M_\odot/L_\odot$	&			\\
$\,\,\,\,\,\,$(1)		&	(2)		&	(3)		&	(4)	&	(5)		&	(6)			  &	(7)	 &	(8)	&	(9)		\\
\hline
\endhead
\hline
\endfoot
D512-2	&	0.0352	&	1.59	&	0.8	&	0.027	&	0.0322	&	0.95	&	0.8	&	0.046	\\
D564-8	&	0.0158	&	1.92	&	0.2	&	0.027	&	0.0139	&	1.20	&	0.2	&	0.028	\\
D631-7	&	0.0115	&	6.29	&	0.2	&	0.462	&	0.0099	&	4.04	&	0.2	&	0.427	\\
DDO 064	&	0.0657	&	1.78	&	0.2	&	0.286	&	0.0566	&	1.14	&	0.2	&	0.305	\\
DDO 154	&	0.0291	&	2.53	&	0.2	&	0.859	&	0.0276	&	1.48	&	0.2	&	1.686	\\
DDO 161	&	0.0064	&	7.36	&	0.5	&	0.248	&	0.0058	&	4.42	&	0.5	&	0.313	\\
DDO 168	&	0.0295	&	3.56	&	0.2	&	3.170	&	0.0251	&	2.35	&	0.2	&	3.095	\\
DDO 170	&	0.0164	&	3.94	&	0.2	&	1.245	&	0.0194	&	1.94	&	0.2	&	0.960	\\
ESO 444-G084	& 0.1204&	1.63	&	0.8	&	0.969	&	0.1316	&	0.85	&	0.8	&	0.725	\\
F561-1	&	0.0276	&	2.13	&	0.2	&	0.109	&	0.0372	&	0.95	&	0.2	&	0.223	\\
F563-1	&	0.0578	&	3.99	&	0.2	&	0.479	&	0.0582	&	2.07	&	0.2	&	0.489	\\
F563-V1	&	0.0052	&	2.24	&	0.5	&	0.063	&	0.0040	&	1.27	&	0.6	&	0.073	\\
F563-V2	&	0.1229	&	2.86	&	0.8	&	0.132	&	0.1105	&	1.67	&	0.8	&	0.210	\\
F565-V2	&	0.0223	&	5.19	&	0.2	&	0.035	&	0.0196	&	3.22	&	0.2	&	0.039	\\
F567-2	&	0.0196	&	2.52	&	0.8	&	0.224	&	0.0191	&	1.41	&	0.8	&	0.295	\\
F574-2	&	0.0032	&	3.72	&	0.2	&	0.010	&	0.0044	&	2.08	&	0.05&	0.009	\\
F583-1	&	0.0282	&	4.41	&	0.2	&	0.160	&	0.0261	&	2.57	&	0.2	&	0.316	\\
IC 2574	&	0.0030	&	18.38	&	0.8	&	2.046	&	0.0027	&	10.44	&	0.8	&	2.169	\\
KK98-251&	0.0137	&	3.19	&	0.2	&	0.253	&	0.0120	&	2.03	&	0.2	&	0.242	\\
NGC 55	&	0.0158	&	5.89	&	0.2	&	0.163	&	0.0145	&	3.54	&	0.2	&	0.269	\\
NGC 1705	&	0.4657	&	0.9		&	0.8	&	0.411	&	5.3576	&	0.13	&	0.4	&	0.051	\\
NGC 2366	&	0.0378	&	2.22	&	0.2	&	0.531	&	0.0338	&	1.35	&	0.2	&	0.757	\\
NGC 2915	&	0.1635	&	1.87	&	0.1	&	0.309	&	0.2248	&	0.80	&	0.1	&	0.448	\\
NGC 3109	&	0.0227	&	4.44	&	0.2	&	0.159	&	0.0202	&	2.75	&	0.2	&	0.184	\\
NGC 3741	&	0.0245	&	2.87	&	0.8	&	0.722	&	0.0239	&	1.63	&	0.8	&	0.643	\\
NGC 4068	&	0.0222	&	3.64	&	0.2	&	0.077	&	0.0198	&	2.19	&	0.2	&	0.068	\\
NGC 4214	&	0.0547	&	3.13	&	1.25&	0.639	&	4.7180	&	0.15	&	0.2	&	0.534	\\
NGC 6789	&	0.4597	&	1.89	&	0.8	&	0.111	&	0.4179	&	1.07	&	0.8	&	0.123	\\
UGC 128&	0.0204	&	7.88	&	0.2	&	6.884	&	0.0301	&	3.12	&	0.8	&	2.879	\\
UGC 191&	0.1315	&	1.87	&	0.2	&	2.864	&	0.1455	&	0.90	&	0.2	&	1.025	\\
UGC 634&	0.0220	&	6.37	&	0.2	&	0.122	&	0.0248	&	3.27	&	0.2	&	0.721	\\
UGC 731&	0.0600	&	2.37	&	0.8	&	0.466	&	0.0715	&	1.14	&	0.8	&	0.124	\\
UGC 891&	0.0155	&	4.93	&	0.2	&	0.066	&	0.0139	&	3.05	&	0.2	&	0.069	\\
UGC 1230&	0.0629	&	3.83	&	0.2	&	0.200	&	0.0604	&	1.75	&	0.8	&	0.566	\\
UGC 1281&	0.0322	&	3.09	&	0.2	&	0.180	&	0.0279	&	1.96	&	0.2	&	0.183	\\
UGC 2023&	0.0123	&	22.18	&	0.2	&	0.010	&	0.0115	&	10.24	&	0.2	&	0.011	\\
UGC 2259&	0.2301	&	1.56	&	0.2	&	1.483	&	0.4037	&	0.58	&	0.4	&	0.356	\\
UGC 2455&	0.0108	&	11.92	&	0.04&	0.701	&	0.0095	&	8.13	&	0.04&	0.729	\\
UGC 4305&	0.0499	&	0.67	&	0.5	&	1.519	&	0.0380	&	0.36	&	0.5	&	1.761	\\
UGC 4325&	0.3288	&	1.38	&	0.2	&	0.292	&	0.3109	&	0.68	&	0.8	&	1.002	\\
UGC 4483&	0.1616	&	0.48	&	0.2	&	0.170	&	0.1567	&	0.27	&	0.2	&	0.194	\\
UGC 4499&	0.0524	&	2.63	&	0.2	&	0.116	&	0.0550	&	1.40	&	0.2	&	0.145	\\
UGC 5005&	0.0065	&	10.64	&	0.5	&	0.011	&	0.0059	&	6.22	&	0.5	&	0.026	\\
UGC 5414&	0.0253	&	3.52	&	0.6	&	0.051	&	0.0185	&	2.51	&	0.7	&	0.051	\\
UGC 5716&	0.0288	&	3.56	&	0.8	&	2.755	&	0.0349	&	1.73	&	0.8	&	2.044	\\
UGC 5750&	0.0095	&	7.15	&	0.2	&	0.093	&	0.0087	&	4.25	&	0.2	&	0.156	\\
UGC 5764&	0.2325	&	0.98	&	0.2	&	1.989	&	0.2585	&	0.49	&	0.8	&	3.561	\\
UGC 5829&	0.0210	&	3.47	&	0.8	&	0.176	&	0.0204	&	1.98	&	0.8	&	0.142	\\
UGC 5918&	0.0620	&	1.48	&	0.2	&	0.030	&	0.0629	&	0.80	&	0.2	&	0.011	\\
UGC 5986&	0.0908	&	3.30	&	0.5	&	0.755	&	0.0536	&	2.42	&	0.8	&	1.161	\\
UGC 5999&	0.0151	&	6.93	&	0.2	&	0.420	&	0.0136	&	4.17	&	0.2	&	0.595	\\
UGC 6399&	0.0575	&	3.20	&	0.2	&	0.025	&	0.0341	&	2.22	&	0.8	&	0.049	\\
UGC 6628&	0.1025	&	0.91	&	0.2	&	0.101	&	0.3938	&	0.21	&	0.2	&	0.196	\\
UGC 6818&	0.0131	&	8.44	&	0.2	&	0.650	&	0.0114	&	5.41	&	0.2	&	0.641	\\
UGC 6917&	0.0730	&	3.30	&	0.3	&	0.146	&	0.0851	&	1.70	&	0.2	&	0.194	\\
UGC 6923&	0.0799	&	2.48	&	0.2	&	0.237	&	0.0739	&	1.47	&	0.2	&	0.289	\\
UGC 7089&	0.0049	&	13.32	&	0.8	&	0.095	&	0.0043	&	8.22	&	0.8	&	0.096	\\
UGC 7125&	0.0158	&	3.91	&	0.2	&	0.207	&	0.0087	&	2.66	&	0.6	&	0.386	\\
UGC 7232&	0.2288	&	1.36	&	0.2	&	0.084	&	0.2043	&	0.82	&	0.2	&	0.102	\\
UGC 7261&	0.2084	&	1.37	&	0.2	&	0.121	&	0.2950	&	0.59	&	0.2	&	0.013	\\
UGC 7323&	0.0106	&	9.31	&	0.8	&	0.165	&	0.0095	&	5.71	&	0.8	&	0.160	\\
UGC 7399&	0.2601	&	1.66	&	0.8	&	0.575	&	0.4362	&	0.7		&	0.2	&	0.163	\\
UGC 7524&	0.0563	&	2.14	&	0.2	&	0.601	&	0.0700	&	0.99	&	0.2	&	0.954	\\
UGC 7559&	0.0318	&	1.58	&	0.2	&	0.051	&	0.0275	&	1.00	&	0.2	&	0.060	\\
UGC 7577&	0.0041	&	5.11	&	0.3	&	0.0950	&	0.0034	&	5.28	&	0.3	&	0.098	\\
UGC 7608&	0.0501	&	2.86	&	0.2	&	0.054	&	0.0443	&	1.78	&	0.2	&	0.047	\\
UGC 7690&	0.5143	&	0.64	&	0.4	&	0.078	&	0.1846	&	0.48	&	0.8	&	0.187	\\
UGC 7866&	0.0764	&	0.92	&	0.2	&	0.045	&	0.0753	&	0.51	&	0.2	&	0.026	\\
UGC 8490&	0.1613	&	1.67	&	0.8	&	0.425	&	0.3056	&	0.60	&	0.8	&	0.161	\\
UGC 8837&	0.0083	&	14.30	&	0.2	&	0.133	&	0.0077	&	7.01	&	0.2	&	0.143	\\
UGC 9992&	0.2799	&	0.49	&	0.2	&	0.002	&	0.1600	&	0.51	&	0.8	&	0.012	\\
UGC 10310&	0.0573	&	2.23	&	0.8	&	0.053	&	0.0581	&	1.20	&	0.8	&	0.130	\\
UGC 11557&	0.0170	&	5.17	&	0.2	&	0.425	&	0.0146	&	3.26	&	0.2	&	0.442	\\
UGC 11820&	0.0031	&	14.95	&	3.0	&	0.535	&	0.0027	&	9.60	&	3.0	&	0.527	\\
UGC 12632&	0.0365	&	2.96	&	0.8	&	0.056	&	0.0422	&	1.47	&	0.8	&	0.095	\\
UGC 12732&	0.0266	&	4.45	&	0.8	&	1.057	&	0.0640	&	1.50	&	0.2	&	0.347	\\
UGCA 281	&	0.2576	&	0.51	&	0.2	&	0.077	&	0.1521	&	0.36	&	0.6	&	0.104	\\
UGCA 442	&	0.0301	&	2.91	&	0.8	&	0.464	&	0.0332	&	1.54	&	0.2	&	0.601	\\
UGCA 444	&	0.0523	&	1.27	&	0.8	&	0.199	&	0.0513	&	0.72	&	0.8	&	0.182	\\
\label{TableB2}	
\end{longtable}


\begin{longtable}{lccccccccc}
\caption{Best-fit Burkert halo: kinematic data at selected radii}\\
\hline\hline
		&			&		\multicolumn{3}{c}{2.15 $R_d$:} 
					&		\multicolumn{3}{c}{3.2 $R_d$:}  & & 				\\
$\,\,\,$Name	&$\nabla v(R_d)$	&	$v_\mathrm{g}$, $v_{\mathrm{s}}^\Upsilon$, $v_\mathrm{h}$	&	$v_\mathrm{total}$	&	$v_\mathrm{obs}$ &	$v_\mathrm{g}$, $v_{\mathrm{s}}^\Upsilon$, $v_\mathrm{h}$	&	$v_\mathrm{total}$ & 	$v_\mathrm{obs}$ & $R_\mathrm{flat}$ & $v_\mathrm{flat}$		\\
	& {km\,s$^{-1}$kpc$^{-1}$}& km\,s$^{-1}$& km\,s$^{-1}$& km\,s$^{-1}$& km\,s$^{-1}$& km\,s$^{-1}$& km\,s$^{-1}$ & kpc& km\,s$^{-1}$\\
$\,\,\,\,\,\,$(1)		&	(2)		&		(3)		&	(4)		&	(5)		&	(6)		&	(7)		&	(8)	& (9) &(10)\\
		\hline	
		\endfirsthead
		\caption{continued.}\\
		\hline\hline
		&			&		\multicolumn{3}{c}{2.15 $R_d$:}  
					&		\multicolumn{3}{c}{3.2 $R_d$:}  & & 				\\
$\,\,\,$Name	&$\nabla v(R_d)$	&	$v_\mathrm{g}$, $v_{\mathrm{s}}^\Upsilon$, $v_\mathrm{h}$	&	$v_\mathrm{total}$	&	$v_\mathrm{obs}$ &	$v_\mathrm{g}$, $v_{\mathrm{s}}^\Upsilon$, $v_\mathrm{h}$	&	$v_\mathrm{total}$ & 	$v_\mathrm{obs}$ & $R_\mathrm{flat}$ & $v_\mathrm{flat}$		\\
	& {km\,s$^{-1}$kpc$^{-1}$}& km\,s$^{-1}$& km\,s$^{-1}$& km\,s$^{-1}$& km\,s$^{-1}$& km\,s$^{-1}$& km\,s$^{-1}$ & kpc& km\,s$^{-1}$\\
$\,\,\,\,\,\,$(1)		&	(2)		&		(3)		&	(4)		&	(5)		&	(6)		&	(7)		&	(8)	& (9) &(10) \\
		\hline
		\endhead
		\hline
		\endfoot
D512-2	&	14.3&	7.2	,	18.5	,	29.6	&	35.7	&	36.4	&	8.2	,	17.9	,	31.9	&	37.5	&	35.7	&	3.8	&	35.9	\\
D564-8	&	13.9&	5.9	,	4.0	,	16.0		&	17.5	&	17.8	&	6.8	,	3.9	,	20.1		&	21.6	&	21.5	&	3.1	&	25.0	\\
D631-7	&	19.7&	17.8	,	10.1	,	19.8&	28.4	&	28.8	&	19.9	,	8.9	,	27.7	&	35.3	&	34.6	&	6.3	&	58.5	\\
DDO064	&	28.3&	13.7	,	8.2	,	34.5	&	38.1	&	40.6	&	20.6	,	8.4	,	41.9	&	47.4	&	46.5	&	2.1	&	46.4	\\
DDO154	&	23.7&	6.0	,	6.1	,	15.8		&	18.0	&	18.6	&	8.8	,	6.2	,	21.9		&	24.5	&	24.5	&	4.9	&	48.2	\\
DDO161	&	14.1&	22.0	,	18.3	,	24.1&	37.5	&	35.9	&	25.6	,	17.7	,	33.0&	45.4	&	42.7	&	11.0&	66.8	\\
DDO168	&	21.9&	22.2	,	7.6	,	38.2	&	44.9	&	46.6	&	22.8	,	7.7	,	48.7	&	54.3	&	54.6	&	3.3	&	54.8	\\
DDO170	&	11.8&	18.1	,	8.9	,	42.6	&	47.2	&	48.0	&	24.3	,	8.6	,	49.7	&	56.0	&	55.4	&	10.8&	60.0	\\
E444-84	&	42.1&	13.6	,	13.1	,	34.3&	39.3	&	40.6	&	17.4	,	12.7	,	44.2&	49.2	&	49.6	&	3.4	&	63.1	\\
F561-1	&	8.4	&	19.5	,	19.1	,	38.0&	46.8	&	48.2	&	27.8	,	21.9	,	37.9&	51.8	&	50.4	&	8.1	&	50.4	\\
F563-1	&	12.7&	19.7	,	13.1	,	98.6&	101.4	&	93.0	&	29.7	,	12.6	,	103.4&	108.3	&	103.5	&	14.6&	112.0	\\
F563-V1	&	5.3	&	19.2	,	17.7	,	17.8&	31.5	&	30.4	&								&			&			&	5.2	&	29.5	\\
F563-V2	&	22.9&	19.3	,	39.8	,	101.4&	110.8	&	114.3	&	33.8	,	36.9	,	107.8&	118.8	&	115.9	&	7.9	&	116.0	\\
F565-V2	&	14.4&	17.2	,	8.4	,	60.7	&	63.6	&	65.4	&	23.2	,	9.2	,	72.5	&	76.7	&	76.1	&		&		\\
F567-2	&	7.5	&	6.8	,	29.4	,	37.8	&	49.3	&	47.8	&	25.2	,	29.6	,	38.1&	54.3	&	52.8	&	9.6	&	52.2	\\
F574-2	&	4.9	&	21.5	,	15.4	,	22.4&	34.8	&	34.4	&								&			&			&		&		\\
F583-1	&	14.0&	9.2	,	11.7	,	65.6	&	67.2	&	68.9	&	22.3	,	10.8	,	74.6&	78.6	&	79.0	&	16.3&	85.8	\\
IC2574	&	7.9	&	14.5	,	24.3	,	38.4&	47.7	&	46.2	&	22.7	,	22.7	,	53.0&	61.9	&	59.5	&		&		\\
KK98-251&	13.7&	17.9	,	5.7	,	29.4	&	34.9	&	34.5	&								&			&			&	2.8	&	34.6	\\
NGC0055	&	7.6	&	34.2	,	20.9	,	78.3&	88.0	&	87.0	&								&			&			&	12.9&	87.4	\\
NGC1705	&	82.0&	15.0	,	39.7	,	48.1&	64.3	&	66.2	&	17.4	,	37.8	,	57.8&	71.3	&	72.7	&	2.0	&	72.9	\\
NGC2366	&	21.7&	10.9	,	10.3	,	27.3&	31.2	&	30.0	&	18.2	,	9.9	,	34.7	&	40.5	&	41.0	&	4.2	&	53.7	\\
NGC2915	&	36.9&	11.2	,	14.3	,	47.8&	51.2	&	43.5	&	12.4	,	12.7	,	60.8&	63.3	&	62.3	&	3.4	&	83.6	\\
NGC3109	&	15.5&	15.8	,	6.7	,	47.8	&	50.7	&	50.6	&	17.8	,	6.3	,	58.7	&	61.6	&	61.7	&		&		\\
NGC3741	&	39.2&	4.4	,	12.1	,	8.5		&	15.5	&	16.6	&	6.2	,	11.8	,	12.3	&	18.2	&	22.0	&	7.0	&	51.6	\\
NGC4068	&	20.7&	12.8	,	10.5	,	25.2&	30.1	&	30.7	&	19.3	,	10.2	,	33.8&	40.3	&	40.6	&		&		\\
NGC4214	&	63.6&	16.9	,	63.8	,	29.6&	72.4	&	67.1	&	16.8	,	62.2	,	40.5&	76.2	&	72.8	&	3.1	&	80.1	\\
NGC6789	&	98.5&	11.7	,	18.9	,	52.2&	56.7	&	57.4	&								&			&			&		&		\\
UGC00128&	9.2	&	15.4	,	25.0	,	111.6&	115.4	&	113.6	&	31.0	,	24.5	,	120.0&	126.3	&	126.3	&	31.3&	131.0	\\
UGC00191&	22.6&	15.8	,	18.1	,	66.3&	70.8	&	73.5	&	24.3	,	18.5	,	73.0&	79.1	&	76.8	&	10.0&	83.9	\\
UGC00634&	14.4&	21.8	,	19.2	,	69.0&	74.9	&	74.5	&	32.3	,	18.7	,	87.9&	95.5	&	95.0	&	13.5&	108.0	\\
UGC00731&	13.0&	17.1	,	13.1	,	60.7&	64.4	&	62.5	&	30.3	,	13.2	,	62.8&	71.0	&	71.7	&	9.1	&	74.0	\\
UGC00891&	14.8&	18.8	,	8.7	,	33.6	&	39.4	&	44.3	&	22.9	,	8.4	,	36.4	&	43.8	&	54.7	&		&		\\
UGC01230&	11.6&	25.8	,	23.5	,	102.1&	107.9	&	103.9	&	35.1	,	22.0	,	103.8&	111.8	&	110.6	&	14.9&	113.0	\\
UGC01281&	17.7&	18.3	,	8.6	,	48.8	&	52.8	&	54.2	&								&			&			&	5.0	&	56.9	\\
UGC02023&	17.9&	16.8	,	16.4	,	46.7&	52.3	&	52.3	&								&			&			&		&		\\
UGC02259&	24.5&	16.4	,	18.4	,	78.6&	82.4	&	81.6	&	22.1	,	17.5	,	80.7&	85.5	&	83.9	&	7.1	&	88.3	\\
UGC02455&	16.4&	16.4	,	15.2	,	27.7&	35.6	&	34.4	&	24.77	,	14.5	,	39.6&	49.0	&	46.6	&		&		\\
UGC04305&	14.8&	11.8	,	22.2	,	16.1&	29.8	&	34.9	&	23.1	,	23.0	,	15.5&	36.1	&	30.5	&	4.8	&	35.3	\\
UGC04325&	23.9&	26.0	,	22.2	,	85.5&	92.1	&	91.2	&	29.9	,	19.8	,	85.0&	92.2	&	90.9	&	4.9	&	92.7	\\
UGC04483&	44.4&	7.1	,	4.4	,	14.2		&	16.5	&	16.8	&	9.4	,	4.3	,	17.3		&	20.2	&	20.8	&	1.2	&	24.2	\\
UGC04499&	17.4&	22.7	,	17.1	,	57.4&	64.0	&	62.5	&	28.0	,	15.9	,	63.0&	70.8	&	70.8	&	8.2	&	74.3	\\
UGC05005&	9.7	&	17.0	,	31.3	,	55.	&	65.6	&	65.1	&	24.0	,	30.3	,	69.5&	79.5	&	78.8	&	22.7&	100.0	\\
UGC05414&	19.9&	21.2	,	26.1	,	43.8&	55.3	&	54.9	&								&			&			&	4.1	&	61.4	\\
UGC05716&	21.7&	8.8	,	24.4	,	40.2	&	48.0	&	51.4	&	13.1	,	23.3	,	50.5&	57.2	&	57.4	&	8.3	&	73.5	\\
UGC05750&	9.1	&	17.8	,	16.8	,	58.9&	63.8	&	65.9	&	22.9	,	17.8	,	68.3&	74.3	&	76.8	&	11.4&	77.6	\\
UGC05764&	25.3&	19.9	,	5.1	,	50.5	&	54.5	&	55.7	&								&			&			&	2.9	&	54.6	\\
UGC05829&	12.2&	20.1	,	19.1	,	45.7&	53.4	&	51.1	&	33.8	,	18.2	,	51.4&	64.1	&	66.0	&		&		\\
UGC05918&	13.4&	12.8	,	7.6	,	39.2	&	41.9	&	41.8	&								&			&			&		&		\\
UGC05986&	29.	&	14.8	,	49.8	,	85.8&	100.3	&	103.0	&	24.9	,	49.2	,	98.8&	113.1	&	115.5	&	6.3	&	116.0	\\
UGC05999&	12.1&	23.3	,	17.9	,	70.7&	76.5	&	80.9	&	31.2	,	17.8	,	82.	&	90.1	&	91.0	&		&		\\
UGC06399&	18.3&	16.7	,	19.8	,	72.6&	77.1	&	77.8	&	23.7	,	19.4	,	79.9&	85.6	&	84.9	&	7.9	&	87.6	\\
UGC06628&	8.0	&	23.0	,	21.0	,	29.5&	42.9	&	42.3	&								&			&			&	5.5	&	42.3	\\
UGC06818&	13.7&	8.6	,	18.4	,	39.4	&	44.4	&	41.3	&	13.7	,	17.9	,	54.	&	58.6	&	64.0	&		&		\\
UGC06917&	19.3&	29.5	,	36.9	,	96.2&	105.8	&	109.7	&	31.2	,	36.7	,	108.2&	117.7	&	114.0	&	10.5&	111.0	\\
UGC06923&	25.0&	22.8	,	25.0	,	63.8&	72.2	&	74.4	&	28.8	,	23.9	,	72.	&	81.1	&	79.5	&	5.2	&	81.1	\\
UGC07089&	13.1&	16.4	,	45.2	,	38.7&	61.8	&	62.4	&	24.2	,	43.3	,	53.1&	72.5	&	80.4	&		&		\\
UGC07125&	8.1	&	17.8	,	16.8	,	50.3&	56.0	&	56.5	&	27.8	,	15.9	,	52.8&	61.8	&	63.1	&	13.0&	65.6	\\
UGC07232&	64.9&	10.3	,	11.5	,	32.8&	36.2	&	35.4	&								&			&			&		&		\\
UGC07261&	26.5&	4.7	,	22.9	,	63.7	&	67.8	&	65.7	&	12.9	,	22.0	,	67.3&	72.0	&	69.7	&	5.7	&	74.7	\\
UGC07323&	19.0&	25.6	,	54.9	,	53.3&	80.7	&	80.7	&								&			&			&		&		\\
UGC07399&	27.8&	21.3	,	31.2	,	88.7&	96.4	&	94.2	&	24.4	,	28.8	,	91.4&	98.9	&	101.7	&	6.1	&	106.0	\\
UGC07524&	10.8&	46.4	,	16.0	,	54.8&	73.6	&	76.4	&	65.4	,	16.8	,	53.3&	86.0	&	73.6	&		&		\\
UGC07559&	19.3&	8.6	,	7.6	,	20.5		&	23.5	&	23.5	&	12.0	,	7.5	,	25.1	&	28.8	&	29.9	&		&		\\
UGC07577&	11.8&	10.3	,	6.9	,	14.3	&	18.8	&	19.9	&								&			&			&		&		\\
UGC07608&	18.4&	16.2	,	7.5	,	56.1	&	58.8	&	57.9	&	24.2	,	7.3	,	64.3	&	69.1	&	69.4	&		&		\\
UGC07690&	53.6&	12.7	,	30.4	,	49.7&	59.4	&	62.5	&	19.0	,	28.5	,	51.0&	61.2	&	59.8	&	1.8	&	60.7	\\
UGC07866&	20.3&	8.7	,	7.8	,	24.2		&	26.9	&	25.9	&	13.3	,	7.3	,	26.6	&	30.6	&	30.3	&		&		\\
UGC08490&	50.0&	12.7	,	41.3	,	51.5&	67.3	&	69.3	&	16.3	,	39.7	,	61.9&	75.4	&	75.7	&	3.1	&	78.5	\\
UGC08837&	12.9&	19.4	,	11.6	,	39.0&	45.1	&	46.6	&	30.3	,	10.6	,	58.5&	66.2	&	59.0	&		&		\\
UGC09992&	15.3&	14.0	,	10.3	,	28.1&	33.1	&	32.8	&	17.6	,	10.0	,	27.0&	33.7	&	33.6	&	3.9	&	34.3	\\
UGC10310&	18.2&	15.3	,	35.0	,	53.7&	66.0	&	67.5	&	27.6	,	34.8	,	57.2&	72.4	&	71.2	&	6.6	&	72.4	\\
UGC11557&	15.0&	17.7	,	42.0	,	59.5&	74.9	&	84.7	&	35.5	,	39.0	,	67.8	&	85.8	&	81.6	&		&		\\
UGC11820&	13.1&	14.5	,	45.5	,	29.2&	56.8	&	56.8	&	23.1	,	44.1	,	40.9	&	64.9	&	64.7	&		&		\\
UGC12632&	12.8&	14.7	,	26.2	,	57.3&	64.7	&	64.4	&	25.9	,	25.8	,	60.8	&	70.9	&	70.2	&	10.0	&	73.0	\\
UGC12732&	16.4&	9.5	,	31.3	,	58.8	&	67.3	&	67.3	&	17.3	,	30.9	,	69.4	&	77.9	&	75.2	&		&		\\
UGCA281	&	40.6&	8.8	,	7.9	,	24.2		&	27.0	&	27.1	&	10.0	,	7.5	,	26.8	&	29.6	&	29.3	&	1.1	&	29.5	\\
UGCA442	&	18.2&	21.0	,	12.0	,	38.8&	45.7	&	45.0	&	22.8	,	11.2	,	46.9	&	53.3	&	54.7	&	5.5	&	57.8	\\
UGCA444	&	18.8&	15.6	,	4.7	,	27.7	&	32.1	&	32.6	&	13.6	,	4.6	,	30.4	&	33.6	&	39.7	&	2.6	&	38.3 	
\label{TableB3}	
\end{longtable}

\bsp	
\label{lastpage}

\end{document}